\documentclass[twocolumn,superscriptaddress,nofootinbib,floatfix,aps,prb,citeautoscript,longbibliography]{revtex4-2}
\usepackage[utf8]{inputenc}
\usepackage[T1]{fontenc}
\usepackage{lineno}

\usepackage{graphicx}
\usepackage{color}
\usepackage{array}
\usepackage{dcolumn}
\usepackage{bm}
\usepackage{multirow}
\usepackage{chemformula}
\let\ce\ch
\usepackage{cleveref}
\usepackage{tabularx}
\usepackage{comment}
\usepackage{tikz}
\usepackage{xstring}
\usepackage{cleveref}

\usepackage{graphicx}
\usepackage{amsmath}
\usepackage{amsfonts}
\usepackage{float}
\usepackage{booktabs}
\usepackage{tabularray}

\begin{document}

\newcommand{\bochum}{Research Center Future Energy Materials and Systems of the University Alliance Ruhr and Interdisciplinary Centre for Advanced Materials Simulation, Ruhr University Bochum, Universitätsstraße 150, D-44801 Bochum, Germany}

\newcommand{\coimbra}{CFisUC, Department of Physics, University of Coimbra, Rua Larga, 3004-516 Coimbra, Portugal}
\newcommand{\mpi}{Max-Planck-Institut f\"ur Mikrostrukturphysik, Weinberg 2, D-06120 Halle, Germany}

\author{Thalis H. B. da Silva}
\affiliation{\coimbra}
\author{Tiago F. T. Cerqueira}
\affiliation{\coimbra}
\author{Hai-Chen Wang}
\affiliation{\bochum}
\author{Miguel A. L. Marques} 
\email{miguel.marques@rub.de}
\affiliation{\bochum} 
\date{\today}

\title{High-throughput study of kagome compounds in the \ce{AV3Sb5} family}

\begin{abstract}
The kagome lattice has emerged as a fertile ground for exotic quantum phenomena, including superconductivity, charge density wave, and topologically nontrivial states. While AV$_3$Sb$_5$ (A = K, Rb, Cs) compounds have been extensively studied in this context, the broader AB$_3$C$_5$ family remains largely unexplored. In this work, we employ machine-learning accelerated, high-throughput density functional theory calculations to systematically investigate the stability and electronic properties of kagome materials derived from atomic substitutions in the AV$_3$Sb$_5$ structure. We identify 36 promising candidates that are thermodynamically stable, with many more close to the convex hull. Stable compounds are not only found with a pnictogen (Sb or Bi) as the C atom, but also with Au, Hg, Tl, and Ce. This diverse chemistry opens the way to tune the electronic properties of the compounds. In fact, many of these compounds exhibit Dirac points, Van Hove singularities, or flat bands close to the Fermi level. Our findings provide an array of compounds for experimental synthesis and further theoretical exploration of kagome superconductors beyond the already known systems.
\end{abstract}

\maketitle

\section{Introduction}

Kagome materials have become a central focus in quantum materials research due to their distinct geometric, electronic, and topological properties. The kagome lattice, with its unique pattern of corner-sharing triangles and hexagons, is particularly intriguing as it holds the potential to exhibit quantum spin liquid phases~\cite{Balents2010}, flat bands~\cite{Lin2018}, and Dirac electronic states~\cite{Mazin2014}, which could lead to topological~\cite{Guo2009} and Chern insulating phases~\cite{Xu2015}. For kagome metals, calculations based on a simple tight-binding model with nearest-neighbor hopping already predict the topologically protected, linearly dispersive electronic bands near the Dirac point, as well as dispersionless flat bands. These electronic characteristics have been observed in the kagome magnet YMn$_6$Sn$_6$~\cite{Li2021}, which also demonstrates several nontrivial magnetic phases, with at least one phase exhibiting a large topological Hall effect~\cite{Ghimire2020}. 
An excellent example of kagome metals exhibiting these properties is the ``132'' kagome family, represented by compounds such as LaRu$_3$Si$_2$, YRu$_3$Si$_2$, ThRu$_3$Si$_2$, and LaIr$_3$Ga$_2$. These materials not only feature flat bands, Dirac cones, and non-trivial topological surface states~\cite{Liu2024,Mielke2021,Yin2022}, but also have attracted significant attention due to their rich interplay of topology, electronic correlations, and superconductivity. Another extensively studied family of kagome systems is the \ch{AV3Sb5} family (with A = K, Rb, and Cs), due to the interplay between topology, charge order, and superconductivity in these systems~\cite{Ortiz2019,Ortiz2021,Ortiz2020,Jiang2022,Xu2022}.

Since their discovery, \ch{AV3Sb5} compounds have been found to host a variety of interesting physical properties. For instance, KV$_3$Sb$_5$ single crystals exhibit a remarkably large and unconventional anomalous Hall effect~\cite{Yang2020}. CsV$_3$Sb$_5$ has been identified as a nonmagnetic $\mathbb{Z}_2$ topological metal, with protected surface states emerging close to the Fermi level~\cite{Ortiz2020}, and has recently been found to exhibit multi-band superconductivity, leading to two distinct superconducting regimes characterized by different transport and thermodynamic properties~\cite{Hossain2025}. Furthermore, studies using angle-resolved photoemission spectroscopy (ARPES) and density functional theory (DFT) have revealed that both KV$_3$Sb$_5$ and CsV$_3$Sb$_5$ possess multiple Dirac points near the Fermi level~\cite{Ortiz2019,Yang2020,Ortiz2020}.

An interesting structural feature in these compounds is the formation of ``Star of David'' and ``Inverse Star of David'' motifs~\cite{Ortiz2021_2}, a periodic lattice distortions closely linked to charge density wave (CDW) formation. These distortions, characterized by an in-plane $2\times2$ reconstruction, lead to exotic properties such as time-reversal symmetry breaking and rotational symmetry breaking~\cite{Hu2022}. The CDW transition happens below $T_{*} \sim 80-100$~K, and these compounds enter a superconducting state at lower temperatures, with critical temperatures ($T_c$) ranging from 0.3 to 3~K. The emergence of superconductivity in these kagome metals is closely linked to the sub-lattice interference mechanism, which drives a complex interplay between electronic correlations and lattice effects~\cite{Wu2021}.

In this work, we expand upon previous studies by exploring a broader class of kagome materials beyond the well-known \ch{AV3Sb5} family. Specifically, we investigate the more general \ch{AB3C5} family, where A, B, and C can be any element of the periodic table. We generate more than 300\,000 kagome structures based on the ``135'' kagome structural prototype through systematic atomic substitutions of the A, B, and C atoms. We employ machine learning models to efficiently screen and select promising kagome materials for further analysis using first-principles DFT calculations. We then investigate the electronic structure of the thermodynamically stable candidates. Our results indicate that many of these materials have a strong tendency to undergo structural distortions, likely stabilizing into the Star of David and Inverse Star of David deformation patterns.

\section{Results and Discussions}
\label{sec:results}

\begin{figure}[tbh]
    \centering
    \includegraphics[width=1\linewidth]{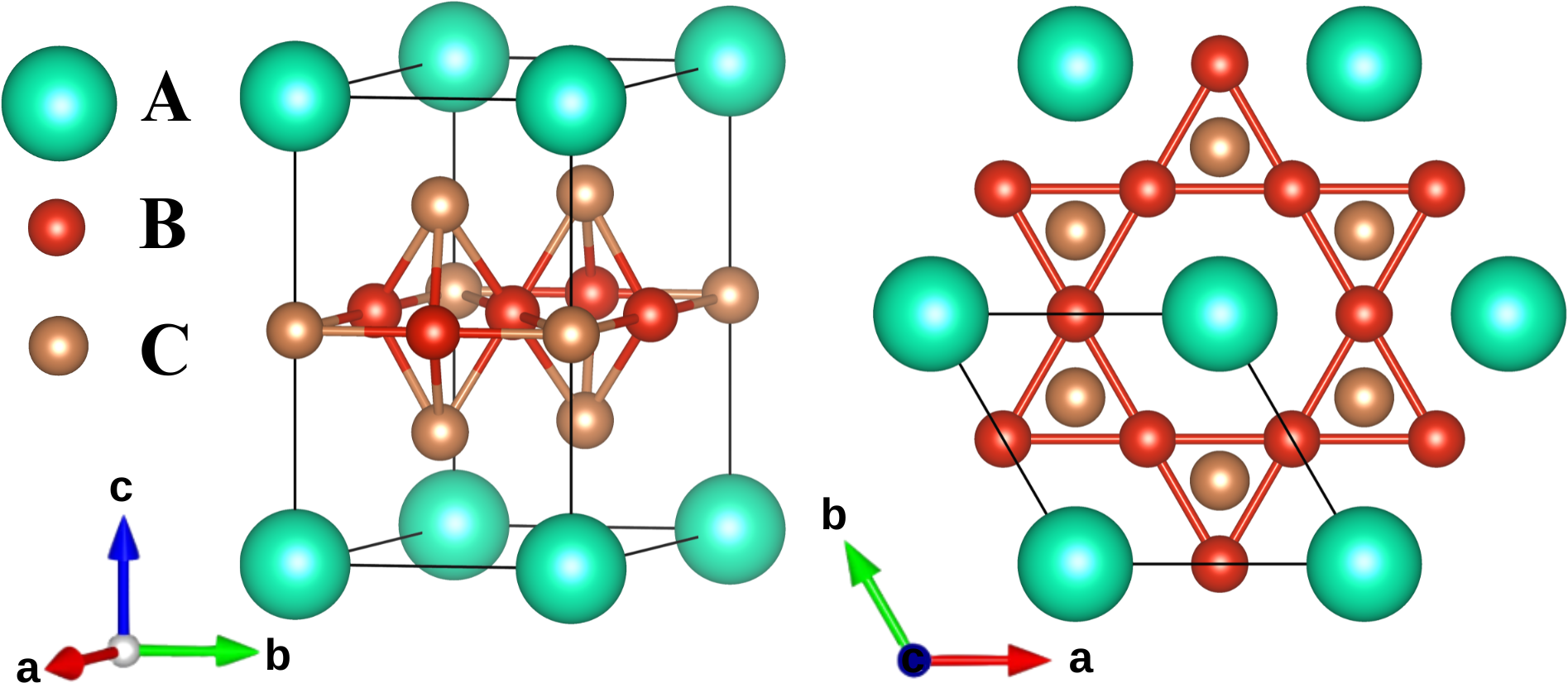}
        \caption{Illustration of the kagome structure, showing both side (left) and top (right) views. Image produced with \textsc{vesta}~\cite{vesta}.}
    \label{fig:kagome}
\end{figure}

The space of possible kagome compounds was systematically explored through atom substitution in the ``135'' Kagome structure \ch{AB3C5}~\cite{Ortiz2019} (see \cref{fig:kagome}). By substituting the 3 different chemical species in \ce{AB3C5} with all elements up to Bi, an extensive dataset of more than 300\,000 kagome structures was generated.

Due to the very large number of generated compounds, a direct DFT relaxation for all structures is unfeasible within a reasonable time frame. To accelerate the search, the geometry of each structure was first optimized using the universal machine-learning interatomic potential \textsc{m3gnet}~\cite{m3gnet}. Following this, thermodynamic stability was assessed by estimating the distance to the convex hull using an \textsc{alignn}~\cite{alignn} model. From the entire dataset, around 15\,000 compounds with the smallest distance to the convex hull were then selected for further analysis. Among these, 36 compounds were identified on the convex hull, as shown in Table~\ref{tab:convex_hull}. This set includes the well-known kagome superconductors \ce{KV3Sb5}, \ce{CsV3Sb5}, and \ce{RbV3Sb5} as well as other compounds which were also previously reported, but also a series of other thermodynamically stable materials that had been overlooked in the literature. Furthermore, a total of 269 compounds were found within 50~meV/atom and 1386 were identified within 100~meV/atom above the hull.


\begin{table}[h]
\centering
\caption{List of 36 compounds found on the convex hull, with calculated lattice constants $a$ and $c$ (in \AA), and magnetization (in $\mu_B$/formula).}
\label{tab:convex_hull}
\renewcommand{\arraystretch}{1.0}
\begin{tabular}{@{}c l c c c@{}}
\toprule
\textbf{C} & \textbf{Formula} & \textbf{$a$} & \textbf{$c$} & \textbf{Mag.} \\
\midrule
\multirow{3}{*}{Group 11} 
  & \ce{CaBe3Au5}  & 5.03 & 6.74  \\
  & \ce{SrBe3Au5}  & 5.12 & 6.76  \\
  & \ce{PmBe3Au5}  & 5.05 & 6.75  \\
\midrule
\multirow{2}{*}{Group 12} 
  & \ce{KPd3Hg5}  & 5.75 & 7.47  \\
  & \ce{RbPd3Hg5} & 5.78 & 7.52  \\
\midrule
\multirow{5}{*}{Group 13} 
  & \ce{RbPd3Tl5} & 5.76 & 8.33  \\
  & \ce{CsPd3Tl5} & 5.69 & 9.44  \\
  & \ce{KPt3Tl5}  & 5.67 & 8.79 \\
  & \ce{RbPt3Tl5} & 5.67 & 9.32  \\
  & \ce{CsPt3Tl5} & 5.66 & 9.94  \\
\midrule
\multirow{16.5}{*}{Group 15} 
  & \ce{CsTi3Sb5}~\cite{Jiang2022b}   & 5.68 & 9.80  \\
  & \ce{KTi3Bi5}~\cite{Jiang2022b}    & 5.80 & 9.44  \\
  & \ce{RbTi3Bi5}~\cite{Yi2022,Yi2023}   & 5.82 & 9.66  \\
  & \ce{CsTi3Bi5}~\cite{Yi2022,Yi2023}   & 5.83 & 9.92  \\
  & \ce{CsHf3Bi5}~\cite{Jiang2022b}   & 6.11 & 9.63  \\
  & \ce{BaTi3Bi5}   & 5.85 & 8.89  \\
  & \ce{KV3Sb5}~\cite{Ortiz2019}     & 5.48 & 9.31  \\
  & \ce{RbV3Sb5}~\cite{Ortiz2019}    & 5.49 & 9.55  \\
  & \ce{CsV3Sb5}~\cite{Ortiz2019}    & 5.51 & 9.82  \\
  & \ce{IV3Sb5}   & 5.45 & 8.78  \\
  & \ce{HgV3Sb5}  & 5.44 & 8.79  \\
  & \ce{RbNb3Bi5}~\cite{Jiang2022b}  & 5.90 & 9.51  \\
  & \ce{CsNb3Bi5}~\cite{Jiang2022b}  & 5.91 & 9.77  \\
  & \ce{INb3Bi5}   & 5.85 & 9.03  \\
  & \ce{KMn3Sb5}~\cite{Jiang2022b}   & 5.43 & 9.26 & 7.75 \\
  & \ce{RbMn3Sb5}~\cite{Jiang2022b}  & 5.44 & 9.53 & 7.76 \\
\midrule
\multirow{11}{*}{Ce} 
  & \ce{PbRu3Ce5}  & 5.84 & 7.34 & \\
  & \ce{InOs3Ce5}  & 5.81 & 7.45 & \\
  & \ce{TlOs3Ce5}  & 5.82 & 7.44 & \\
  & \ce{PbOs3Ce5}  & 5.84 & 7.44 & \\
  & \ce{BiOs3Ce5}  & 5.87 & 7.41 & \\
  & \ce{CdCo3Ce5}  & 5.46 & 7.34 & \\
  & \ce{HgCo3Ce5}  & 5.45 & 7.34 & \\
  & \ce{InCo3Ce5}  & 5.49 & 7.32 & \\
  & \ce{TlCo3Ce5}  & 5.51 & 7.32 & \\
  & \ce{PbCo3Ce5}  & 5.56 & 7.28 & 0.91 \\
\bottomrule
\end{tabular}
\end{table}

We divided the set of stable systems by the group in the periodic table of the C atom. We find that most stable compounds contain a group 15 atom in the C position. This family includes all the materials that have been synthesized experimentally or proposed theoretically in the literature~\cite{Ortiz2019, Jiang2022b, Jiang2022b, Yi2023}. In the B position we find a +4 metal, specifically Ti or Mn, or the +5 metals V or Nb. The A position is predominantly occupied by an alkali metal. However, we also find Ba, commonly found in the +2 oxidation state, in \ch{BaTi3Bi5}, and I in \ch{IV3Sb5} and \ch{INbSb5}. The existence of stable materials with A atoms in different charge states opens opportunities to tune the Fermi energy of the compounds through substitution or alloying.
Interestingly, we also find several stable systems with other C atoms, specifically Au (group 11), Hg (group 12), Tl (group 13), Sn (group 14), and Ce. The period 6 metals are not entirely surprising, as they are neighboring elements of Bi and exhibit a large degree of chemical similarity to it~\cite{Glawe2016}. Ce, on the other hand, always appears combined with Co, Ru, and Os, and a heavy metal around Pb.

\begin{figure*}[tbh]
    \centering
    \includegraphics[width=0.98\linewidth]{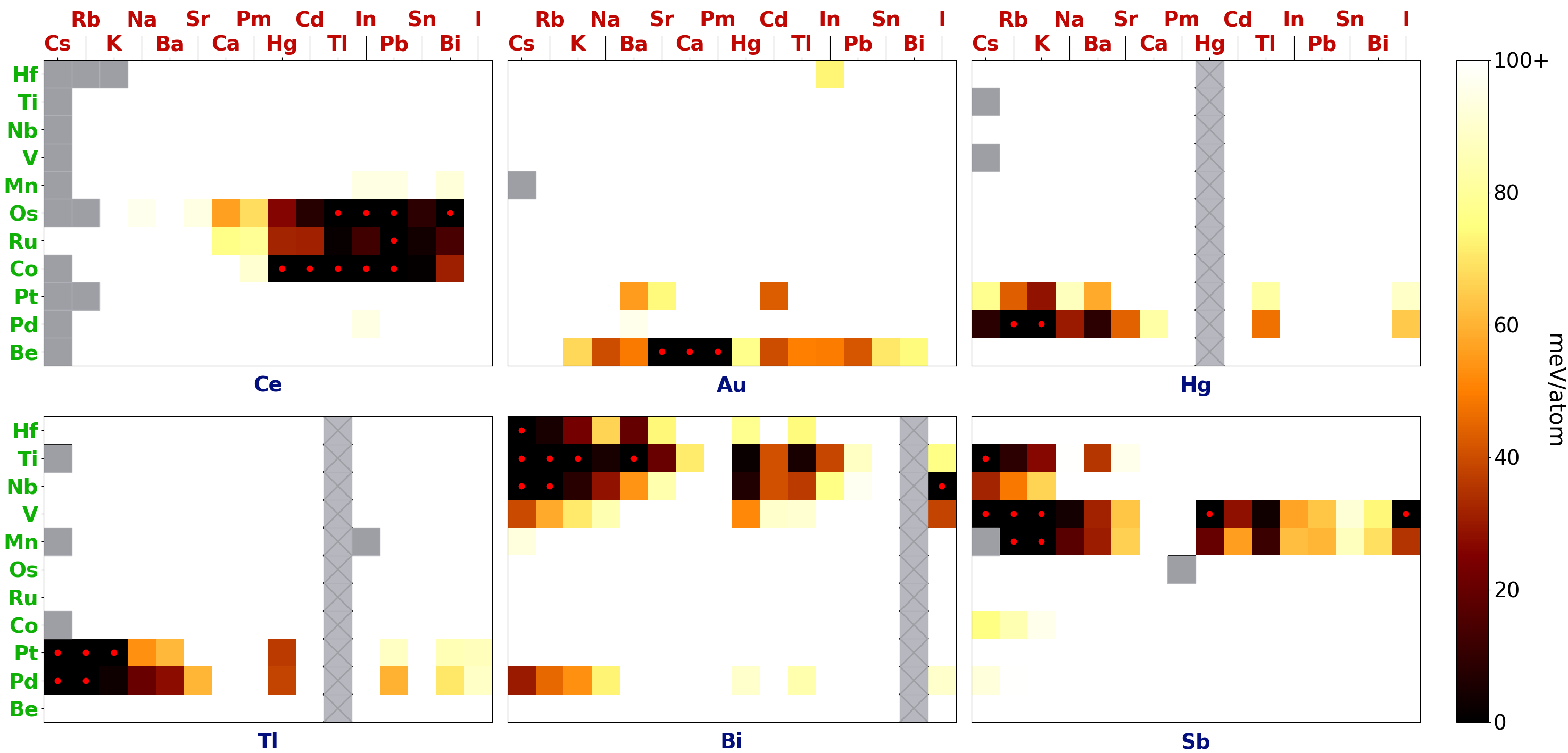}
        \caption{Distance above the convex hull (meV/atom) for candidate kagome compounds with composition \textbf{\textcolor[HTML]{BF0704}{A}}\textbf{\textcolor[HTML]{10B107}{B}}$_3$\textbf{\textcolor[HTML]{020F7C}{C}}$_5$ (note the color code). Each subplot represents a different \textbf{\textcolor[HTML]{020F7C}{C}} element, while the x-axis and y-axis correspond to \textbf{\textcolor[HTML]{BF0704}{A}} and \textbf{\textcolor[HTML]{10B107}{B}} elements, respectively. In order to put into evidence chemical similarity, we ordered the elements using the modified Pettifor scale~\cite{Glawe2016}. The 36 structures found on the convex hull are marked with a red dot and the grey cells are the compounds for which the calculations did not converge.}
    \label{fig:stability}
\end{figure*}

To gain a better overview of the possible chemistry of this family of kagome compounds, we performed an additional DFT screening of an extra $\sim800$ materials constructed from the chemical elements that were found to favor stability (see \cref{fig:stability}). Among these, no new structures were found to lie on the convex hull, which confirms the soundness of our machine-learning accelerated approach to find thermodynamically stable compounds. We do find, however, several other compounds very close to the convex hull. A complete list of all materials we found within 50~meV/atom from the convex hull can be found in the supplementary information (SI).

Each element placed in the $C$ position in the kagome structure exhibits a distinct stability pattern, while the dependence on the A atom is less pronounced. This can be understood from the fact that the A atoms are typically alkali and alkali earths whose main role is to donate electrons to the kagome \ce{B3C5} layers.

The Au, Hg, and Tl subplots in \cref{fig:stability} show that kagomes with these elements tend to be less stable overall than with Ce, Sn, Sb or Bi. Nevertheless, it is still possible to identify that all subplots show ``islands'' or regions of higher stability surrounding a cluster of stable compounds. If Au is in the C position, we find that the large majority of the most stable kagome include Be, the smallest alkali earth, in the B site. On the other hand, Hg and Tl prefer to combine with the noble metals Pd and Pt in the B-site. 

Materials including Sb, and Bi on the C site exhibit a broader stability range, accepting a larger set of chemical elements both in the A and B sites. While in the B site we find that stable compounds include a series of transition metals (Hf, Ti, Nb, V, and Mn), the A site accepts a series of alkali, alkali earths and post-transition metals.

Compounds with Ce occupying the C position show large stability islands, with elements near the late transition metals Cd and Hg, as well as post-transition metals In, Tl, Sn, Pb, and Bi in the A site, and transition metals Os, Ru and and Co in the B site. The discovery of thermodynamically stable Ce-based kagome compounds represents a significant and unexpected finding in our systematic investigation. Ce, being a rare earth element with partially filled 4$f$ orbitals, exhibits fundamentally different chemical properties compared to the $p$-block elements Sb and Bi of previously studied kagomes. 

The remarkable contrast between the high stability of Ce systems and the comparatively lower stability of other kagome compounds featuring other lanthanides highlights the complex and often non-intuitive nature of thermodynamic stability in these materials. This observation reinforces the importance of allowing computational algorithms to systematically survey the vast chemical space without imposing restrictions based on traditional chemical similarity or previously successful compositions. Such an approach maximizes the potential for discovering unexpected yet promising candidate materials that can expand both our fundamental understanding of kagome physics and the range of potentially synthesizable compounds with novel properties.

\begin{figure}[bh]
    \hspace{0cm}
    \includegraphics[width=1\linewidth]{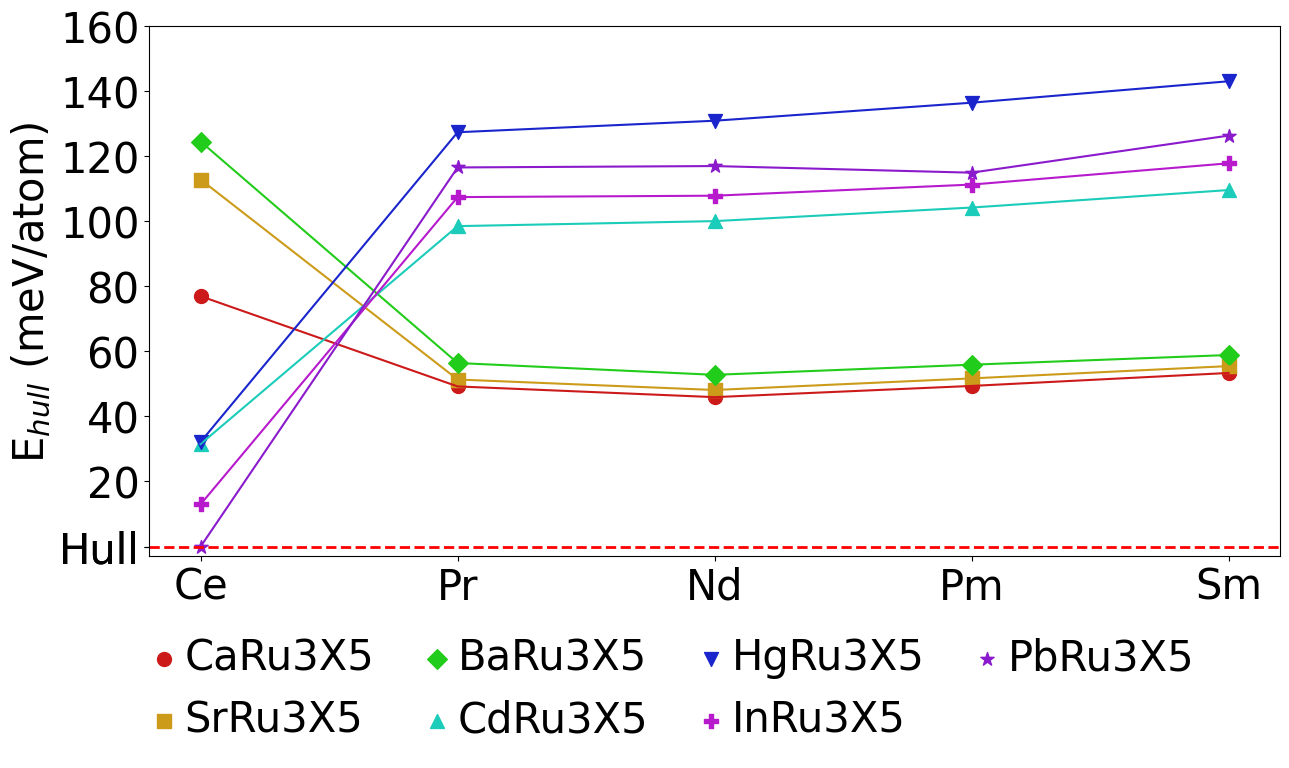}
        \caption{Stability profile of kagome compounds containing a lanthanide in the C position.}
    \label{fig:lanthanides}
\end{figure}

To better understand the preference for Ce, \cref{fig:lanthanides} presents the stability profile of various compounds with different lanthanides in the C position. It is evident that Ce behaves differently from other lanthanides. As a result, compounds such as \ce{CdRu3Ce5}, \ce{InRu3Ce5}, \ce{HgRu3Ce5}, and \ce{PbRu3Ce5} exhibit higher stability, whereas \ce{CaRu3Ce5}, \ce{SrRu3Ce5}, and \ce{BaRu3Ce5} display lower stability compared to their Pr/Nd/Pm/Sm counterparts. The latter, in all cases, show little to no change in stability when compared with each other, reflecting the chemical similarity of these lanthanides. This variation in stability for compounds with Ce likely arises from its ability to exhibit both the +3 and +4 oxidation states, unlike other lanthanides which predominantly adopt the +3 state.

\begin{figure*}[tbh]
    \centering
    \begin{tabular}{ccc}
        \includegraphics[width=0.33\textwidth]{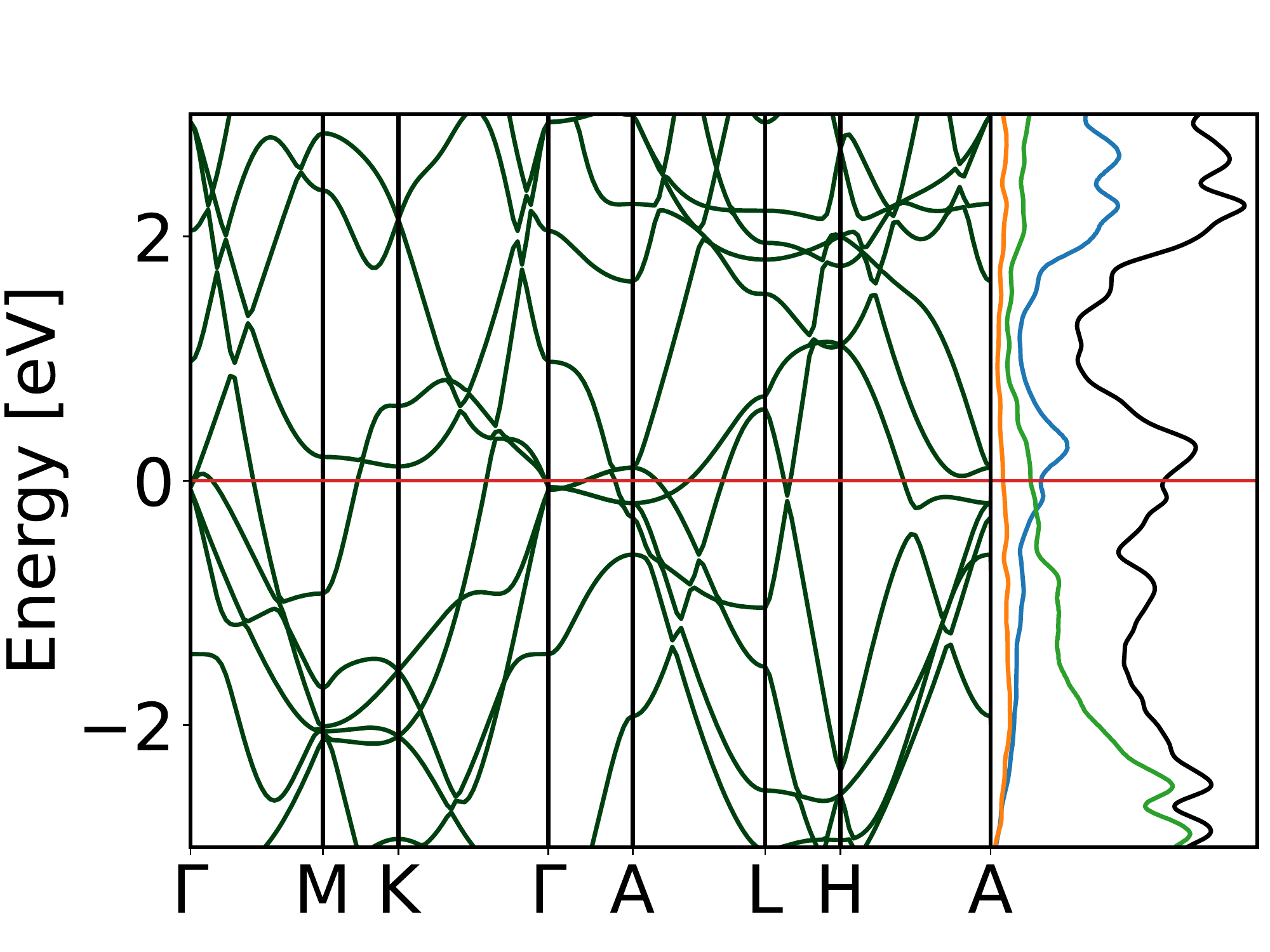} & 
        \includegraphics[width=0.33\textwidth]{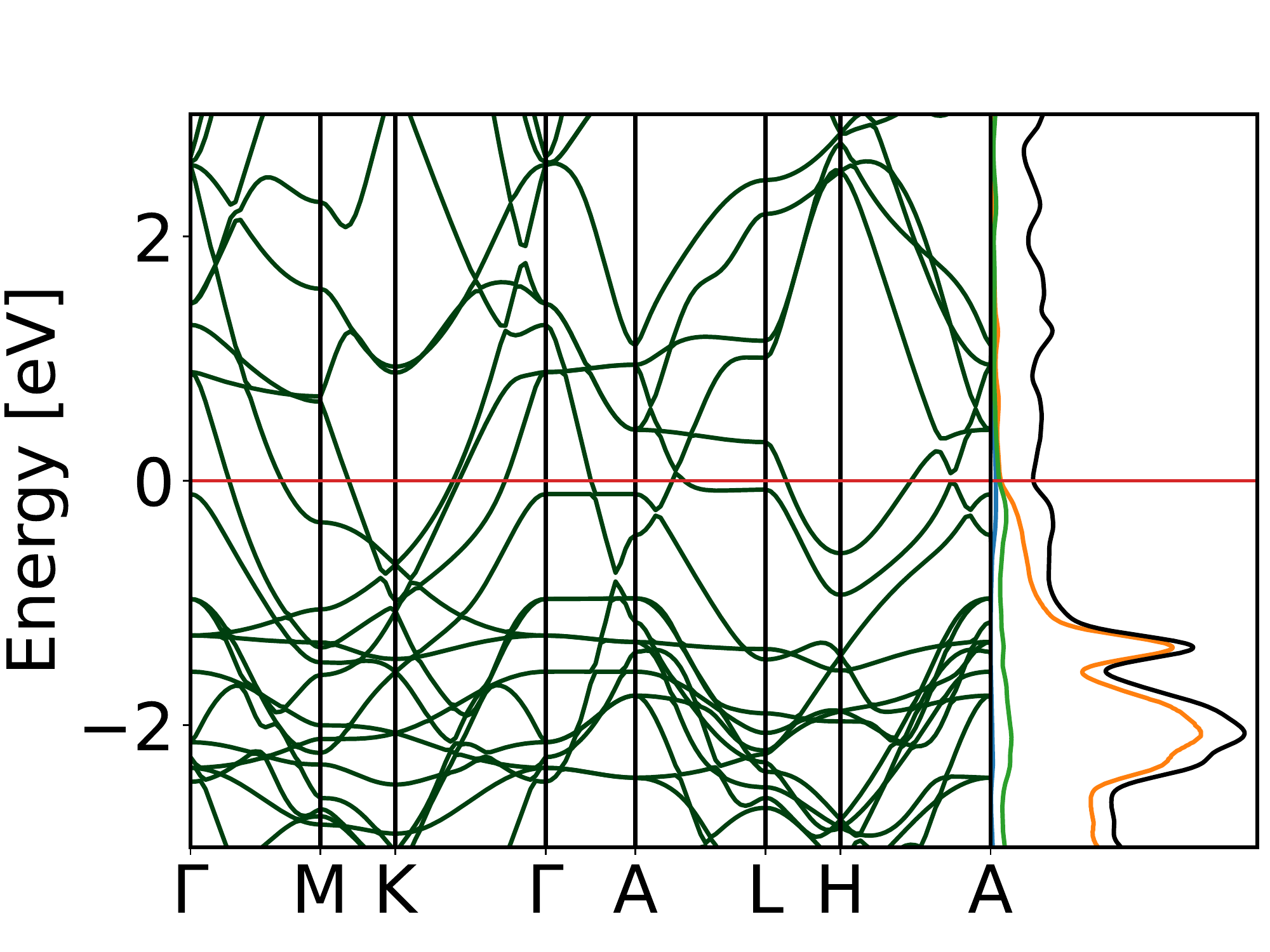} & 
        \includegraphics[width=0.33\textwidth]{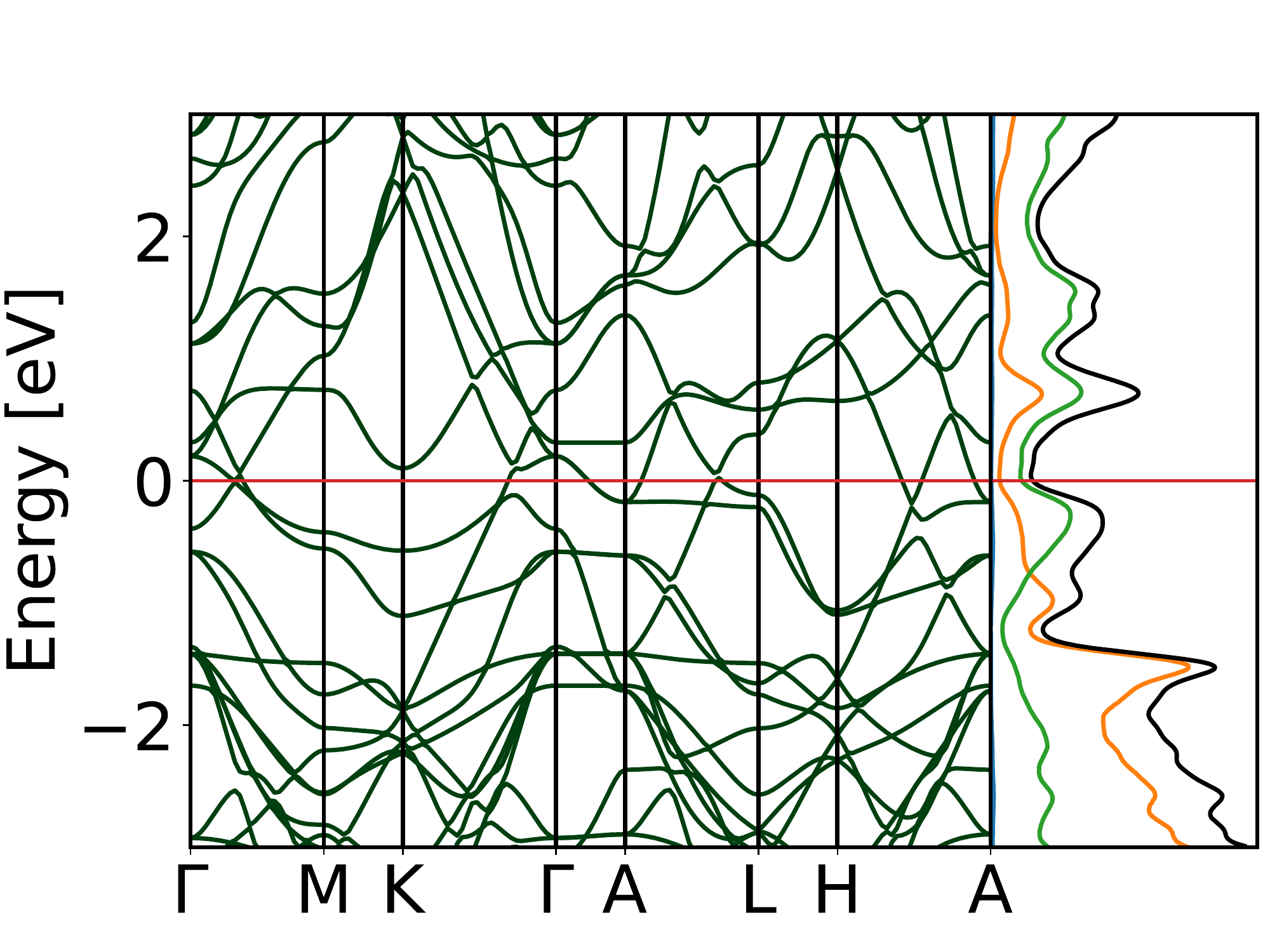} \\
        \ce{PmBe3Au5} (Group 11) & \ce{KPd3Hg5} (Group 12) & \ce{CsPt3Tl5} (Group 13) \\
        \\[-1em]
        \includegraphics[width=0.33\textwidth]{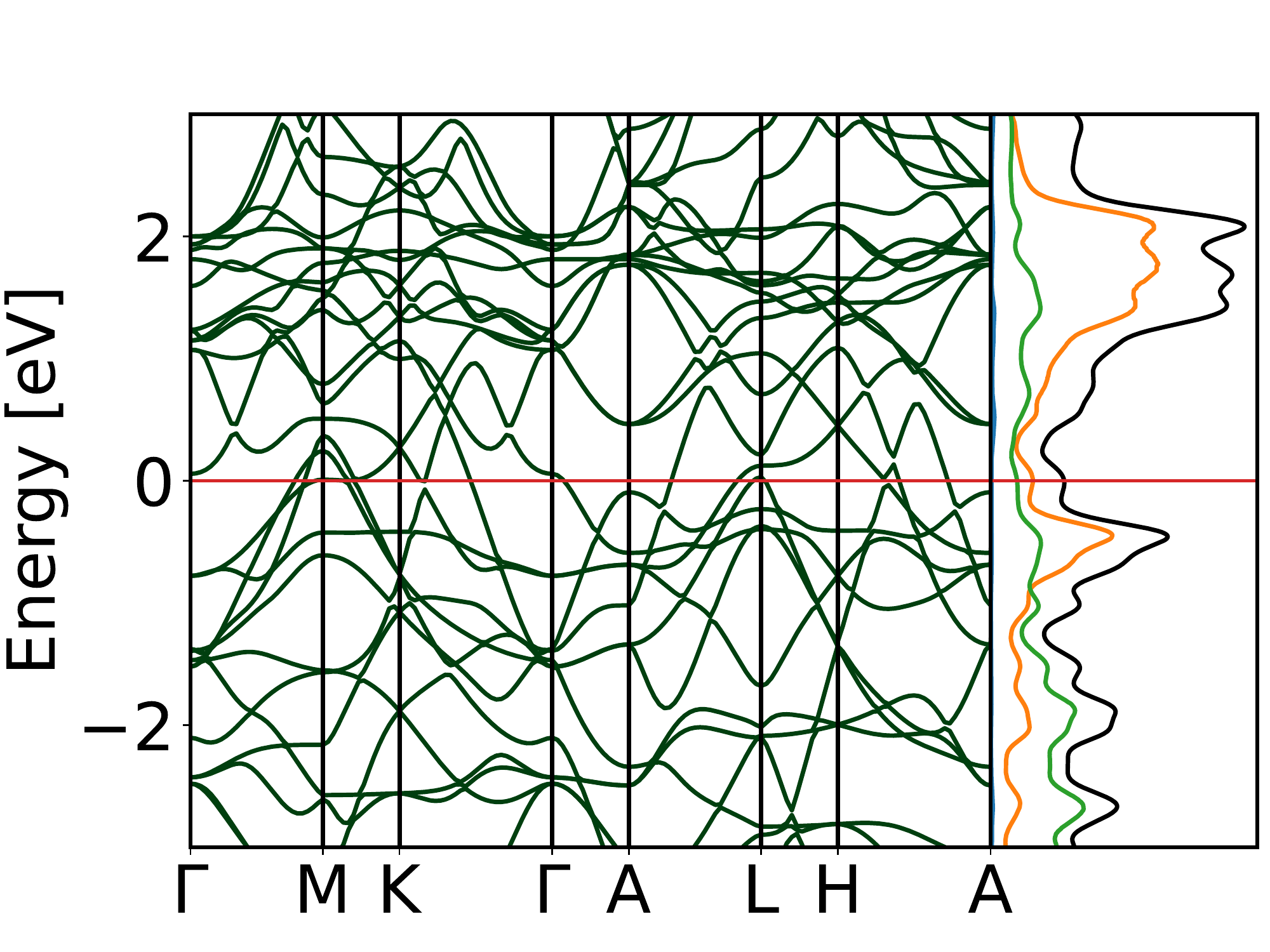} & 
        \includegraphics[width=0.33\textwidth]{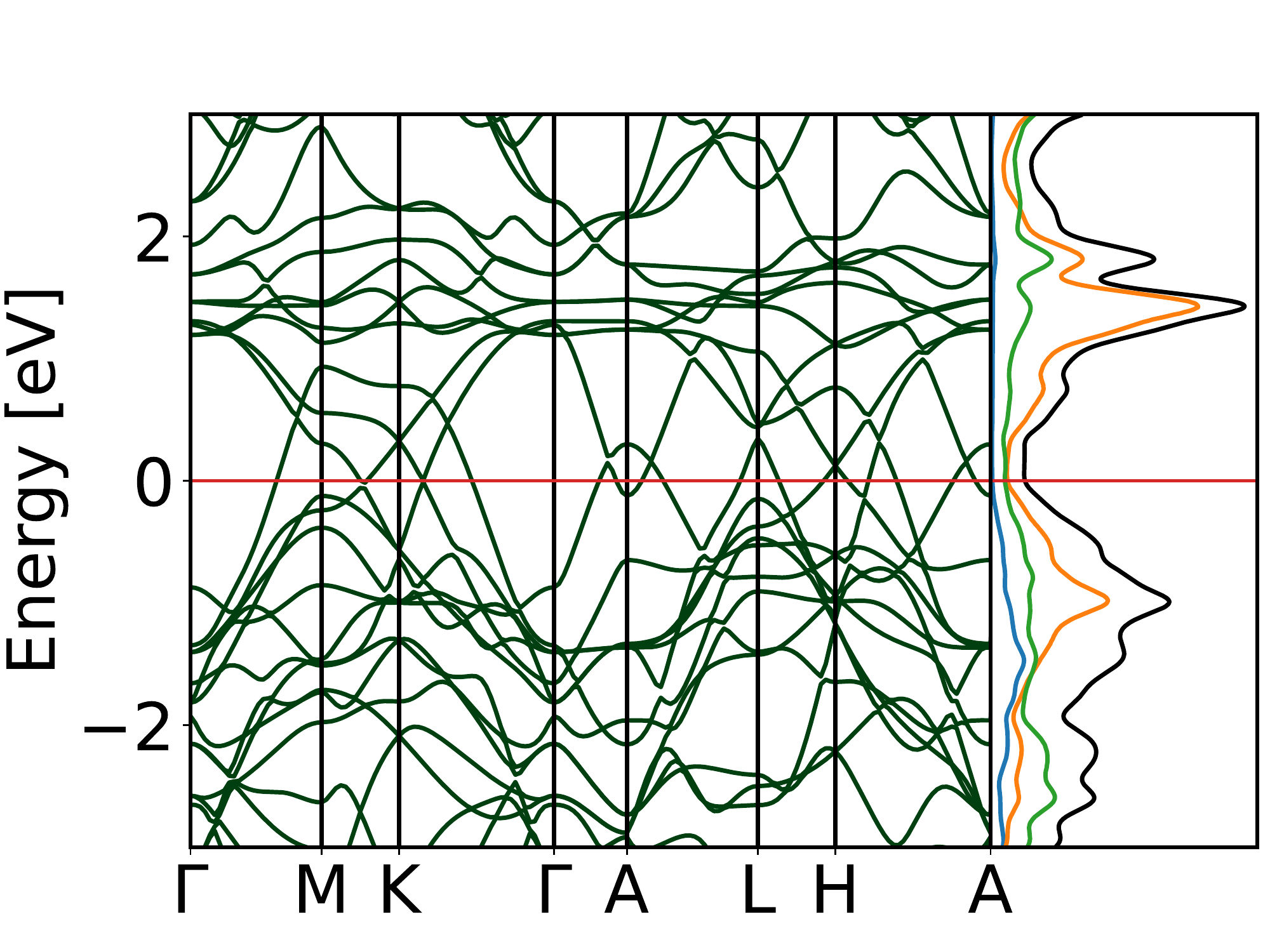} & 
        \includegraphics[width=0.33\textwidth]{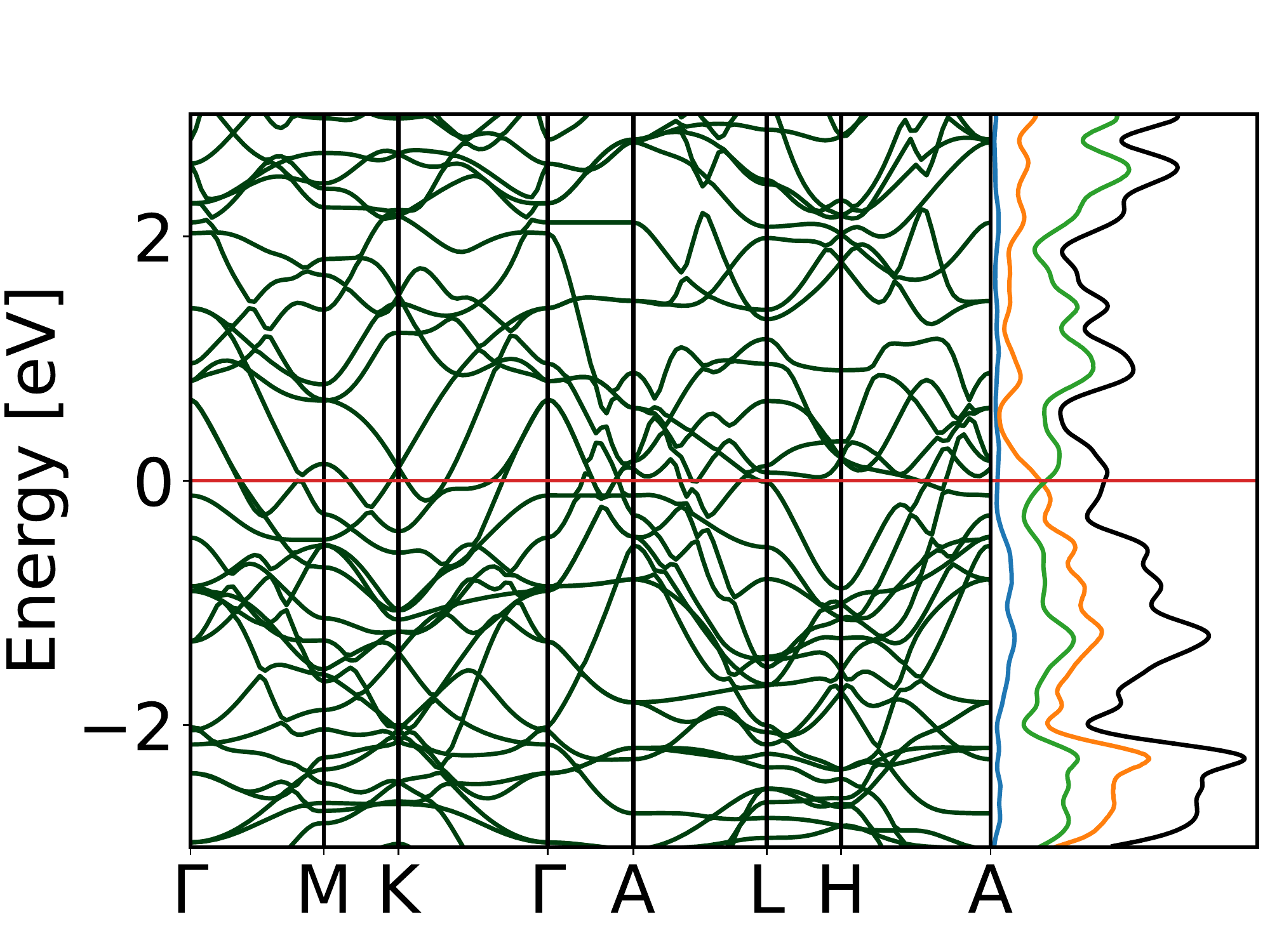} \\
        \ce{BaTi3Bi5} (Group 15) & \ce{IV3Sb5} (Group 15) & \ce{PbOs3Ce5} (Ce) \\
    \end{tabular}
    \caption{Electronic band structures of six newly proposed compounds. The black curve represents the total density of states (DOS), while the projected atomic density of states is shown in blue (atom A), orange (atom B), and green (atom C).}
    \label{fig:bandstructures}
\end{figure*}

In \cref{fig:bandstructures}, we show the band structures of a representative compound from each group that is not yet known experimentally or studied theoretically. In general, the band structures of compounds within the same group are similar due to the analogous chemical properties of the elements occupying the A and B positions. The complete set of band structures of all compounds on the convex hull can be found in the SI.

We can see that states stemming from the A atom are mostly absent from the bands close to the Fermi level. This is to be expected when A is a alkali or Ba that are fully ionized, but we witness a similar behavior also for I and Pb. The only exception is the Au compounds, where the lanthanide in the A position contributes significantly to the density of states at the Fermi level. We always find a sizeable contribution of the B (that forms the kagome sub-lattice) and C metals to the states surrounding the Fermi level.

A key feature of these band structures, due to the well-known properties of the kagome lattice, is the presence of Van Hove singularities, Dirac points, and flat bands close to the Fermi level, all of which play a crucial role in the exotic electronic behavior of these compounds. In fact, we see that all band structures exhibit Dirac points at the high-symmetry K point and Van Hove singularities along the high-symmetry M point, both below the Fermi level. Regarding the flat bands, groups 12 and 13, along with Ce-containing compounds, display flat bands very close to the Fermi level specifically along the $\Gamma$--A--L path. Group 15 compounds, however, exhibit more extensive flat bands that are not confined to the $\Gamma$--A--L trajectory.

Finally, we turn our attention to the phonon dispersions. It turns out that 35 out of the 36 stable compounds identified in this work exhibit imaginary phonon frequencies in their pristine phase, suggesting that these structures tend to collapse into the Star of David or Inverse Star of David configurations, as expected from previous studies on kagome superconductors of this family. In fact, \ce{INb3Bi5} is the only compound found to be stable in the pristine phase, exhibiting an electron-phonon coupling constant $\lambda = 0.28$ and a logarithmic average frequency of $\omega_\text{log}=141$~K. This leads to a critical transition temperature ($T_\text{c}$) of 0.03~K (calculated with the Allen-Dynes~\cite{Allen1975} correction to the McMillan formula~\cite{McMillan1968} and a typical $\mu^*$ of 0.1). For the dynamically unstable compounds, we found that structural distortions are minimal, of the order of a small fraction of an \AA, resulting in energy changes of only a few meV/atom~\cite{Tan2021}. Such small deformations do not impose any significant impact on the stability of the compounds or their electronic band structures.

In conclusion, our high-throughput study has significantly expanded the landscape of potential kagome materials. Through systematic computational screening of over 300\,000 structures using machine learning-accelerated approaches, we identified 36 thermodynamically stable kagome compounds, including several previously unreported materials. Remarkably, stable compounds were discovered not only with pnictogens (Sb, Bi) in the C position, but also with Au, Hg, Tl, and Ce, demonstrating unexpected chemical diversity in this materials class. The electronic band structures of these compounds consistently exhibit key kagome lattice features such as Dirac points, Van Hove singularities, and flat bands near the Fermi level, which are essential for exotic quantum phenomena. The majority of these materials show phonon instabilities, suggesting they likely adopt Star of David configurations similar to known kagome superconductors. Our work provides a comprehensive roadmap for experimental synthesis efforts and opens numerous opportunities to explore and tune the electronic properties of kagome materials, potentially leading to the discovery of new superconductors with higher critical temperatures and unique properties.

\section{Methods}
\label{sec:methods}

\subsection{Machine learning}

Starting from the initial geometry, we performed a preliminary geometry optimisation using the \textsc{m3gnet}~\cite{m3gnet} universal machine learning interatomic potential. The choice of force field was mainly dictated by the superior numerical efficiency and stability of \textsc{m3gnet}. Unfortunately, the total energies calculated with \textsc{m3gnet} do not have the accuracy necessary for the reliable prediction of the distance to the convex hull. Therefore we used an \textsc{alignn}~\cite{alignn} model to predict this crucial property.

The \textsc{alignn} model, that predicts DFT distances to the hull from \textsc{m3gnet} geometries, was trained in a dataset of 4.4 million data points with a range from 0 to 8~eV with a mean of 0.39~eV. From this dataset we split 300\,000 entries for validation and 300\,000 for testing, while the remaining 3.8 million entries were used for training. We performed 200 training epochs, and the best model according to the validation error was selected. The MAE of the \textsc{alignn} model was 33~meV/atom, considerably worse than the 16~meV/atom of the same model trained with relaxed PBE structures~\cite{alex2}. This is due to blind spots in the training data of the \textsc{m3gnet} model, leading to a larger number of outliers with unphysical geometries.

\subsection{DFT calculations}

Around 16\,000 compounds with the smallest distance to the convex hull or in the relevant chemical systems were selected to perform geometry relaxations and total energy calculations using the \textsc{vasp} code~\cite{vasp1,vasp2}.  All parameters, including pseudopotentials, were set to ensure compatibility with the data available in the Materials Project database~\cite{materialsproject}. Calculations were performed with the Perdew-Burke-Ernzerhof approximation~\cite{pbe} to the exchange-correlation functional. To sample the Brillouin zones we used a 6$\times$6$\times$4 $\Gamma$-centred k-point grids. Spin-polarised calculations were started from a ferromagnetic configuration. We utilised the projector augmented wave (PAW) setup~\cite{paw,paw2} within \textsc{vasp} version 5.2, applying a cutoff of 520~eV. We set the convergence criteria of the forces to be less than 0.005~eV/\AA. Distances to the convex hull where calculated against the convex hull of the Alexandria database~\cite{alex1, alex2}. We note that this is the largest convex hull freely available, considerably larger than the one of the Materials Project database~\cite{materialsproject}.

Phonon calculations were performed using version 7.1 of \textsc{Quantum Espresso}~\cite{qe1,qe2} with the Perdew-Burke-Ernzerhof functional for solids (PBEsol)~\cite{pbe-sol} generalized gradient approximation. We used the PBEsol pseudopotentials from the \textsc{pseudodojo} project~\cite{pseudodojo}, specifically the stringent, scalar-relativistic norm-conserving set. Geometry optimizations were performed using uniform 6$\times$6$\times$4 $\Gamma$-centered $k$-point grids. Convergence thresholds for energies, forces, and stresses were set to 10$^{-8}$~a.u., $10^{-6}$~a.u., and 0.05~kbar, respectively. Dynamical matrix were calculated on a 3$\times$3$\times$2 $q$-points grid, and a finer 24$\times$24$\times$16 grid was used for the double-grid method. 


\section{Data and code availability}

All DFT calculations will appear in the next release of the Alexandria database~\cite{alex1, alex2} through \url{https://alexandria.icams.rub.de/}. The \textsc{alignn} model can be downloaded from \url{https://github.com/hyllios/utils/tree/main/models/alexandria_v2/alignn}.

\section{Acknowledgements}
T.F.T.C. acknowledges the financial support from FCT - Fundação para a Ciência e Tecnologia, I.P. through the projects UIDB/04564/2020 and CEECINST/00152/2018/CP1570/CT0006, with DOI identifiers 10.54499/UIDB/04564/2020 and 10.54499/CEECINST/00152/2018/CP1570/CT0006 respectively, and computing resources provided by the project Advanced Computing Project 2023.14294.CPCA.A3, platform Deucalion.
M.A.L.M. was supported by a grant from the Simons Foundation (SFI-MPS-NFS-00006741-12, P.T.) in the Simons Collaboration on New Frontiers in Superconductivity and by the Keele and the Klaus Tschira foundations as a part of the SuperC collaboration.
The authors thank the Gauss Centre for Supercomputing e.V. (www.gauss-centre.eu) for funding this project by providing computing time on the GCS supercomputer SUPERMUC-NG at the Leibniz Supercomputing Centre (www.lrz.de) under the project pn25co. H.C.W and M.A.L.M. would like to thank the NHR Centre PC2 for providing computing time on the Noctua 2 supercomputers under project hpc-prf-vibff and hpc-prf-pbeml.

\section{Author  Contributions}
T.H.B.S. and H.-C.W. developed the high-throughput workflow and performed the analysis and band structure calculations. T.F.T.C. and M.A.L.M. trained the machine learning model, performed the predictions and the response calculations, as well as directed the research. All authors participated equally in the interpretation of the results and in the writing of the manuscript.

\section{Competing  Interests}
The authors declare that they have no competing interests.

\bibliography{bib.bib, cal.bib}
\end{document}


\newcommand{\bochum}{Research Center Future Energy Materials and Systems of the University Alliance Ruhr and Interdisciplinary Centre for Advanced Materials Simulation, Ruhr University Bochum, Universitätsstraße 150, D-44801 Bochum, Germany}

\newcommand{\coimbra}{CFisUC, Department of Physics, University of Coimbra, Rua Larga, 3004-516 Coimbra, Portugal}
\newcommand{\mpi}{Max-Planck-Institut f\"ur Mikrostrukturphysik, Weinberg 2, D-06120 Halle, Germany}

\author{Thalis H. B. da Silva}
\affiliation{\coimbra}
\author{Tiago F. T. Cerqueira}
\affiliation{\coimbra}
\author{Hai-Chen Wang}
\affiliation{\bochum}
\author{Miguel A. L. Marques} 
\email{miguel.marques@rub.de}
\affiliation{\bochum} 
\date{\today}

\title{Supplementary Information for: High-throughput study of kagome compounds in the \ce{AV3Sb5} family}
\maketitle

\begin{longtable}{p{2cm} p{2.2cm} p{.9cm} p{.9cm} p{1cm} p{1cm}}
\caption{List of compounds with E$_{hull}$ < 50 meV/atom. Lattice constants $a$ and $c$ in \AA, magnetization in $\mu_B$/formula, and distances to the hull in meV/atom. The material ID can be used to extract the compounds from the Alexandria database~\cite{alex1,alex2}}\\
Formula & ID &    a &    c & Mag.   &   E$_{hull}$ \\
\hline \\
\ce{KPt3Tl5 }    & agm072675582         & 5.67 &  8.79 &                 &   0    \\
\ce{KPd3Hg5 }    & agm072481683         & 5.75 &  7.47 &                 &   0    \\
\ce{RbMn3Sb5}    & agm072236772         & 5.44 &  9.53 & 7.76            &   0    \\
\ce{CsPt3Tl5}    & agm072528460         & 5.66 &  9.94 &                 &   0    \\
\ce{CsPd3Tl5}    & agm072529313         & 5.69 &  9.44 &                 &   0    \\
\ce{PbCo3Ce5}    & agm072489927         & 5.56 &  7.28 & 0.91            &   0    \\
\ce{TlOs3Ce5}    & agm072201098         & 5.82 &  7.44 &                 &   0    \\
\ce{PbRu3Ce5}    & agm072676997         & 5.84 &  7.34 &                 &   0    \\
\ce{CsTi3Bi5}    & agm072458489         & 5.83 &  9.92 &                 &   0    \\
\ce{RbTi3Bi5}    & agm072188670         & 5.82 &  9.66 &                 &   0    \\
\ce{KMn3Sb5 }    & agm072397024         & 5.43 &  9.26 & 7.75            &   0    \\
\ce{CsTi3Sb5}    & agm072443170         & 5.68 &  9.80 &                 &   0    \\
\ce{INb3Bi5 }    & agm072657566         & 5.85 &  9.03 &                 &   0    \\
\ce{CsNb3Bi5}    & agm072821533         & 5.91 &  9.77 &                 &   0    \\
\ce{RbV3Sb5 }    & agm072640091         & 5.49 &  9.55 &                 &   0    \\
\ce{RbPt3Tl5}    & agm072576688         & 5.67 &  9.32 &                 &   0    \\
\ce{InCo3Ce5}    & agm072574409         & 5.49 &  7.32 &                 &   0    \\
\ce{RbPd3Tl5}    & agm072537459         & 5.76 &  8.33 &                 &   0    \\
\ce{CaBe3Au5}    & agm072537467         & 5.03 &  6.74 &                 &   0    \\
\ce{IV3Sb5  }    & agm072419781         & 5.45 &  8.78 &                 &   0    \\
\ce{PmBe3Au5}    & agm072317599         & 5.05 &  6.75 &                 &   0    \\
\ce{KV3Sb5  }    & agm072182393         & 5.48 &  9.31 &                 &   0    \\
\ce{RbNb3Bi5}    & agm072613991         & 5.90 &  9.51 &                 &   0    \\
\ce{KTi3Bi5 }    & agm072606978         & 5.80 &  9.44 &                 &   0    \\
\ce{CsHf3Bi5}    & agm072522722         & 6.11 &  9.63 &                 &   0    \\
\ce{PbOs3Ce5}    & agm072220509         & 5.84 &  7.44 &                 &   0    \\
\ce{InOs3Ce5}    & agm072319909         & 5.81 &  7.45 &                 &   0    \\
\ce{SrBe3Au5}    & agm072283714         & 5.12 &  6.76 &                 &   0    \\
\ce{BaTi3Bi5}    & agm072475240         & 5.85 &  8.89 &                 &   0    \\
\ce{RbPd3Hg5}    & agm072750532         & 5.78 &  7.52 &                 &   0    \\
\ce{HgV3Sb5 }    & agm072298328         & 5.44 &  8.79 &                 &   0    \\
\ce{HgCo3Ce5}    & agm072326417         & 5.45 &  7.34 &                 &   0    \\
\ce{CsV3Sb5 }    & agm072439004         & 5.51 &  9.82 &                 &   0    \\
\ce{BiOs3Ce5}    & agm072628310         & 5.87 &  7.41 &                 &   0    \\
\ce{TlCo3Ce5}    & agm072764022         & 5.51 &  7.32 &                 &   0    \\
\ce{CdCo3Ce5}    & agm072591502         & 5.46 &  7.34 &                 &   0    \\
\ce{SnCo3Ce5}    & agm072645880         & 5.52 &  7.28 & 0.80            &   1.16 \\
\ce{TlRu3Ce5}    & agm072752275         & 5.82 &  7.35 &                 &   1.79 \\
\ce{HgTi3Bi5}    & agm072656815         & 5.74 &  9.19 &                 &   2.28 \\
\ce{CeBe3Au5}    & agm072229360         & 5.06 &  6.76 & 1.13            &   2.72 \\
\ce{KPd3Tl5 }    & agm072409398         & 5.75 &  8.13 &                 &   2.80 \\
\ce{TlV3Sb5 }    & agm072347547         & 5.46 &  8.87 &                 &   3.36 \\
\ce{SnRu3Ce5}    & agm072299097         & 5.81 &  7.34 &                 &   3.55 \\
\ce{CsTa3Sb5}    & agm072580330         & 5.82 &  9.55 &                 &   3.99 \\
\ce{NaV3Sb5 }    & agm072642576         & 5.46 &  8.85 &                 &   4.28 \\
\ce{RbHf3Bi5}    & agm072523454         & 6.10 &  9.45 &                 &   4.91 \\
\ce{NaTi3Bi5}    & agm072645720         & 5.77 &  9.13 &                 &   4.92 \\
\ce{TlTi3Bi5}    & agm072727614         & 5.78 &  9.09 &                 &   4.98 \\
\ce{SmBe3Au5}    & agm072657102         & 5.04 &  6.75 &                 &   5.38 \\
\ce{NdBe3Au5}    & agm072662188         & 5.06 &  6.75 &                 &   5.72 \\
\ce{PrBe3Au5}    & agm072313974         & 5.08 &  6.75 &                 &   6.31 \\
\ce{HgNb3Bi5}    & agm072690233         & 5.85 &  9.15 &                 &   6.46 \\
\ce{CsRh3Tl5}    & agm072821795         & 5.62 &  9.76 &                 &   7.34 \\
\ce{CdOs3Ce5}    & agm072234423         & 5.81 &  7.42 &                 &   7.80 \\
\ce{BrV3Sb5 }    & agm072848030         & 5.43 &  8.42 &                 &   7.82 \\
\ce{KNb3Bi5 }    & agm072836551         & 5.89 &  9.39 &                 &   7.83 \\
\ce{CsPd3Hg5}    & agm072833881         & 5.83 &  7.57 &                 &   8.42 \\
\ce{BaPd3Hg5}    & agm072714296         & 5.73 &  7.58 &                 &   8.63 \\
\ce{SnOs3Ce5}    & agm072723504         & 5.81 &  7.43 &                 &   8.71 \\
\ce{RbTi3Sb5}    & agm072656534         & 5.67 &  9.51 &                 &   8.88 \\
\ce{MgCo3Ce5}    & agm072299589         & 5.49 &  7.32 &                 &   9.75 \\
\ce{TlMn3Sb5}    & agm072182829         & 5.41 &  8.67 & 7.68            &  11.58 \\
\ce{CsNi3Bi5}    & agm072743807         & 5.66 &  9.74 &                 &  12.38 \\
\ce{LaBe3Au5}    & agm072535796         & 5.10 &  6.76 &                 &  12.49 \\
\ce{YBe3Au5 }    & agm072663828         & 5.00 &  6.76 &                 &  12.57 \\
\ce{TbBe3Au5}    & agm072301026         & 5.00 &  6.76 &                 &  12.77 \\
\ce{InRu3Ce5}    & agm072164726         & 5.80 &  7.36 &                 &  12.99 \\
\ce{AgCo3Ce5}    & agm072366793         & 5.38 &  7.38 &                 &  14.22 \\
\ce{CsRh3Pb5}    & agm072760982         & 5.72 &  9.43 &                 &  14.47 \\
\ce{BaFe3Sn5}    & agm072773683         & 5.26 &  9.09 & 6.62            &  14.77 \\
\ce{BiRu3Ce5}    & agm072315847         & 5.87 &  7.31 & 0.19            &  14.83 \\
\ce{RbRh3Pb5}    & agm072495227         & 5.71 &  8.94 &                 &  14.84 \\
\ce{CsNi3Sn5}    & agm072715406         & 5.23 & 10.25 &                 &  15.05 \\
\ce{KFe3Sn5 }    & agm072574330         & 5.23 &  9.54 & 6.17            &  16.55 \\
\ce{DyBe3Au5}    & agm072682561         & 4.99 &  6.76 &                 &  17.15 \\
\ce{NaMn3Sb5}    & agm072303247         & 5.40 &  8.75 & 7.64            &  17.37 \\
\ce{MgOs3Ce5}    & agm072223355         & 5.84 &  7.41 &                 &  17.86 \\
\ce{BaMg3Hg5}    & agm072323083         & 6.09 &  7.43 &                 &  18.71 \\
\ce{RbNi3Sn5}    & agm072823042         & 5.22 &  9.93 &                 &  18.83 \\
\ce{BaHf3Bi5}    & agm072552349         & 6.15 &  8.74 &                 &  19.59 \\
\ce{CsCu3In5}    & agm072735402         & 5.27 & 10.53 &                 &  19.62 \\
\ce{HgMn3Sb5}    & agm072519029         & 5.38 &  8.60 & 7.04            &  20.02 \\
\ce{NaPd3Tl5}    & agm072846939         & 5.71 &  7.99 &                 &  20.36 \\
\ce{SrTi3Bi5}    & agm072336379         & 5.82 &  8.83 &                 &  20.36 \\
\ce{RbFe3Sn5}    & agm072647995         & 5.24 &  9.90 & 6.16            &  20.44 \\
\ce{ZnCo3Ce5}    & agm072736265         & 5.37 &  7.37 &                 &  20.49 \\
\ce{CsAu3Hg5}    & agm072212901         & 5.72 &  9.79 &                 &  20.91 \\
\ce{RbTi3Te5}    & agm072667428         & 6.08 &  8.51 &                 &  20.94 \\
\ce{RbTa3Sb5}    & agm072268217         & 5.81 &  9.35 &                 &  20.95 \\
\ce{BaNi3In5}    & agm072672867         & 5.26 &  8.80 &                 &  20.97 \\
\ce{HoBe3Au5}    & agm072568987         & 4.98 &  6.76 &                 &  21.50 \\
\ce{KRh3Tl5 }    & agm072518035         & 5.77 &  7.67 &                 &  22.62 \\
\ce{KHf3Bi5 }    & agm072415006         & 6.09 &  9.34 &                 &  22.79 \\
\ce{RbRh3Tl5}    & agm072410940         & 5.74 &  8.07 &                 &  23.07 \\
\ce{GaCo3Ce5}    & agm072415285         & 5.40 &  7.34 &                 &  24.37 \\
\ce{CsFe3Sn5}    & agm072390997         & 5.25 & 10.25 & 6.15            &  24.72 \\
\ce{CsNi3In5}    & agm072815241         & 5.23 & 10.22 &                 &  24.86 \\
\ce{CsZr3Bi5}    & agm072306615         & 6.15 &  9.66 &                 &  25.22 \\
\ce{InRh3Ce5}    & agm072362201         & 5.73 &  7.50 & 1.42            &  25.44 \\
\ce{HgOs3Ce5}    & agm072619086         & 5.80 &  7.42 &                 &  25.85 \\
\ce{ErBe3Au5}    & agm072431603         & 4.98 &  6.76 &                 &  25.98 \\
\ce{BaCu3In5}    & agm072544508         & 5.30 &  8.99 &                 &  26.00 \\
\ce{KTi3Sb5 }    & agm072504011         & 5.66 &  9.30 &                 &  26.43 \\
\ce{KAg3Cd5 }    & agm072701291         & 5.81 &  7.60 &                 &  26.48 \\
\ce{CsCo3Sn5}    & agm072515930         & 5.26 &  9.94 &                 &  27.11 \\
\ce{BaRh3Tl5}    & agm072232389         & 5.67 &  8.27 &                 &  27.34 \\
\ce{RbCu3In5}    & agm072402702         & 5.24 & 10.15 &                 &  27.57 \\
\ce{BaPd3Tl5}    & agm072831371         & 5.75 &  8.13 &                 &  27.73 \\
\ce{CdV3Sb5 }    & agm072668146         & 5.44 &  8.64 &                 &  28.31 \\
\ce{RbNi3Bi5}    & agm072187342         & 5.65 &  9.34 &                 &  28.35 \\
\ce{RbFe3Sb5}    & agm072727305         & 5.42 &  9.42 & 6.74            &  28.36 \\
\ce{PbIr3Ce5}    & agm072174948         & 5.79 &  7.52 & 0.93            &  28.46 \\
\ce{KPt3Hg5 }    & agm072415202         & 5.70 &  7.52 &                 &  28.86 \\
\ce{NaNb3Bi5}    & agm072489894         & 5.87 &  9.12 &                 &  28.88 \\
\ce{KNi3Sn5 }    & agm072252701         & 5.21 &  9.57 &                 &  29.00 \\
\ce{RbCd3Hg5}    & agm072248864         & 6.11 &  8.13 &                 &  29.07 \\
\ce{RbNi3In5}    & agm072394679         & 5.22 &  9.83 &                 &  29.35 \\
\ce{PbRh3Ce5}    & agm072482342         & 5.77 &  7.48 & 1.92            &  29.67 \\
\ce{KCd3Hg5 }    & agm072452219         & 6.05 &  8.14 &                 &  29.75 \\
\ce{CsRh3Bi5}    & agm072299161         & 5.83 &  9.70 &                 &  29.93 \\
\ce{InFe3Ce5}    & agm072418033         & 5.58 &  7.16 &                 &  30.04 \\
\ce{TlFe3Ce5}    & agm072212220         & 5.59 &  7.16 &                 &  30.16 \\
\ce{NaPd3Hg5}    & agm072688703         & 5.67 &  7.44 &                 &  30.28 \\
\ce{CsPd3Bi5}    & agm072779238         & 5.88 &  9.71 &                 &  30.38 \\
\ce{BaV3Sn5 }    & agm072838061         & 5.41 &  8.94 &                 &  30.48 \\
\ce{TlIr3Ce5}    & agm072600097         & 5.77 &  7.53 & 0.47            &  30.51 \\
\ce{BaMn3Sb5}    & agm072676666         & 5.52 &  8.65 & 8.16            &  30.60 \\
\ce{BrAu3Hg5}    & agm072792588         & 5.78 &  7.83 &                 &  30.61 \\
\ce{TmBe3Au5}    & agm072241394         & 4.97 &  6.76 &                 &  30.92 \\
\ce{BiCo3Ce5}    & agm072679021         & 5.59 &  7.25 & 1.24            &  31.02 \\
\ce{NaCu3In5}    & agm072804053         & 5.37 &  7.85 &                 &  31.22 \\
\ce{SrFe3Ga5}    & agm072491376         & 5.16 &  6.59 & 5.63            &  31.47 \\
\ce{RbCo3Sn5}    & agm072283258         & 5.25 &  9.59 &                 &  31.47 \\
\ce{CdRu3Ce5}    & agm072293082         & 5.80 &  7.33 &                 &  31.53 \\
\ce{BaV3Sb5 }    & agm072361037         & 5.53 &  8.79 &                 &  31.73 \\
\ce{KCu3In5 }    & agm072318826         & 5.22 &  9.79 &                 &  31.81 \\
\ce{CsNb3Sb5}    & agm072582356         & 5.83 &  9.54 &                 &  32.08 \\
\ce{RbMg3Tl5}    & agm072522643         & 6.24 &  8.25 &                 &  32.11 \\
\ce{HgRu3Ce5}    & agm072420474         & 5.79 &  7.34 &                 &  32.17 \\
\ce{TlRh3Ce5}    & agm072751532         & 5.74 &  7.50 & 1.47            &  32.31 \\
\ce{NaFe3Sn5}    & agm072159573         & 5.23 &  8.47 & 5.99            &  32.52 \\
\ce{CsMg3Tl5}    & agm072240218         & 6.28 &  8.32 &                 &  32.76 \\
\ce{SrCu3In5}    & agm072664035         & 5.27 &  8.62 &                 &  32.98 \\
\ce{RbAg3Cd5}    & agm072599221         & 5.85 &  7.62 &                 &  33.20 \\
\ce{CsMn3Se5}    & agm072491687         & 5.98 &  7.56 & 10.38           &  33.36 \\
\ce{PbRu3La5}    & agm072460592         & 6.16 &  7.79 &                 &  33.39 \\
\ce{CsFe3Ge5}    & agm072534460         & 4.88 &  9.71 & 4.90            &  33.63 \\
\ce{BaNi3Ga5}    & agm072683373         & 4.93 &  8.33 &                 &  34.65 \\
\ce{TlFe3Sn5}    & agm072602894         & 5.24 &  8.51 & 6.02            &  34.78 \\
\ce{TlCu3In5}    & agm072163172         & 5.32 &  8.25 &                 &  34.93 \\
\ce{InIr3Ce5}    & agm072775519         & 5.76 &  7.53 & 0.32            &  35.01 \\
\ce{IMn3Sb5 }    & agm072353950         & 5.40 &  8.67 & 5.75            &  35.18 \\
\ce{CsHf3Te5}    & agm072513153         & 6.36 &  8.66 &                 &  35.29 \\
\ce{SrRu3La5}    & agm072594917         & 6.35 &  8.03 &                 &  35.37 \\
\ce{NaFe3Ga5}    & agm072202010         & 5.08 &  6.56 & 5.58            &  35.55 \\
\ce{BaTi3Sb5}    & agm072723172         & 5.72 &  8.70 &                 &  35.70 \\
\ce{INi3Bi5 }    & agm072731967         & 5.63 &  8.56 &                 &  35.80 \\
\ce{KRh3Pb5 }    & agm072269153         & 5.71 &  8.56 &                 &  36.17 \\
\ce{BaZn3Au5}    & agm072731187         & 5.46 &  7.22 &                 &  36.31 \\
\ce{HgPt3Tl5}    & agm072239199         & 5.67 &  8.35 &                 &  36.50 \\
\ce{TlNb3Bi5}    & agm072379406         & 5.85 &  9.09 &                 &  36.53 \\
\ce{RbRh3Bi5}    & agm072701693         & 5.83 &  9.25 &                 &  37.10 \\
\ce{ClAu3Hg5}    & agm072458539         & 5.75 &  7.82 &                 &  37.27 \\
\ce{SrZn3Au5}    & agm072352873         & 5.37 &  7.21 &                 &  37.34 \\
\ce{SbOs3Ce5}    & agm072744365         & 5.83 &  7.40 &                 &  37.38 \\
\ce{BrNi3Bi5}    & agm072211668         & 5.62 &  8.15 &                 &  37.51 \\
\ce{KNi3Bi5 }    & agm072233956         & 5.63 &  9.07 &                 &  37.63 \\
\ce{CsCd3Hg5}    & agm072478938         & 6.24 &  8.06 &                 &  37.77 \\
\ce{BaFe3Ge5}    & agm072291283         & 4.90 &  8.60 & 5.79            &  37.88 \\
\ce{KTi3Te5 }    & agm072709657         & 6.06 &  8.42 &                 &  37.95 \\
\ce{BaCd3Hg5}    & agm072763456         & 6.01 &  8.01 &                 &  37.98 \\
\ce{SbCo3Ce5}    & agm072347155         & 5.54 &  7.24 & 0.80            &  38.06 \\
\ce{BaLi3Hg5}    & agm072820030         & 5.77 &  7.49 &                 &  38.09 \\
\ce{IAu3Hg5 }    & agm072581438         & 5.84 &  7.82 &                 &  38.18 \\
\ce{HgPd3Tl5}    & agm072196882         & 5.72 &  7.98 &                 &  38.35 \\
\ce{IV3Bi5  }    & agm072545395         & 5.60 &  9.11 &                 &  38.36 \\
\ce{RbFe3Ge5}    & agm072455319         & 4.87 &  9.36 & 4.86            &  38.42 \\
\ce{CaRu3La5}    & agm072344483         & 6.32 &  7.92 &                 &  38.48 \\
\ce{AlCo3Ce5}    & agm072691832         & 5.41 &  7.34 &                 &  38.65 \\
\ce{InTi3Bi5}    & agm072190376         & 5.77 &  9.00 &                 &  38.88 \\
\ce{KCu3Sn5 }    & agm072330846         & 5.35 &  9.19 &                 &  38.98 \\
\ce{CaCu3In5}    & agm072716265         & 5.24 &  8.41 &                 &  39.03 \\
\ce{CsAg3Cd5}    & agm072528597         & 5.93 &  7.61 &                 &  39.03 \\
\ce{BaMn3Ge5}    & agm072720672         & 5.02 &  8.26 & 4.88            &  39.20 \\
\ce{HgNi3Bi5}    & agm072612843         & 5.57 &  8.47 &                 &  39.31 \\
\ce{SrCu3Au5}    & agm072301595         & 5.23 &  7.22 &                 &  39.40 \\
\ce{KMg3Tl5 }    & agm072267263         & 6.19 &  8.23 &                 &  39.49 \\
\ce{SrFe3Sn5}    & agm072723713         & 5.25 &  8.75 & 6.59            &  39.57 \\
\ce{KTa3Sb5 }    & agm072803141         & 5.80 &  9.24 &                 &  39.61 \\
\ce{NaCu3Sn5}    & agm072359213         & 5.34 &  8.31 &                 &  39.64 \\
\ce{SnIr3Ce5}    & agm072344551         & 5.77 &  7.51 & 0.85            &  39.72 \\
\ce{NaNi3In5}    & agm072701483         & 5.38 &  7.29 &                 &  39.75 \\
\ce{CsV3Bi5 }    & agm072721025         & 5.64 & 10.06 &                 &  39.82 \\
\ce{CsNa3K5 }    & agm072452948         & 7.94 & 10.98 &                 &  39.88 \\
\ce{HgFe3Sn5}    & agm072169817         & 5.27 &  7.98 & 5.86            &  39.92 \\
\ce{CdBe3Au5}    & agm072737287         & 4.92 &  6.81 &                 &  40.04 \\
\ce{NaBe3Au5}    & agm072466972         & 5.01 &  6.69 &                 &  40.05 \\
\ce{BaAg3Cd5}    & agm072846324         & 5.82 &  7.59 &                 &  40.52 \\
\ce{SrCu3Zn5}    & agm072208318         & 5.15 &  6.65 &                 &  40.63 \\
\ce{CdNb3Bi5}    & agm072845152         & 5.85 &  9.08 &                 &  40.65 \\
\ce{CdTi3Bi5}    & agm072566306         & 5.74 &  9.07 &                 &  40.76 \\
\ce{SrAl3Au5}    & agm072283613         & 5.48 &  7.01 &                 &  41.02 \\
\ce{NaCu3Cd5}    & agm072430321         & 5.36 &  7.37 &                 &  41.04 \\
\ce{BaZn3Hg5}    & agm072592110         & 5.69 &  7.75 &                 &  41.13 \\
\ce{PbCu3In5}    & agm072658363         & 5.32 &  8.17 &                 &  41.32 \\
\ce{RbCu3Sn5}    & agm072552419         & 5.34 &  9.77 &                 &  41.42 \\
\ce{BaRu3La5}    & agm072420357         & 6.36 &  8.19 &                 &  41.47 \\
\ce{CdRh3Ce5}    & agm072381891         & 5.71 &  7.51 & 0.84            &  41.59 \\
\ce{PbBe3Au5}    & agm072833491         & 5.07 &  6.76 &                 &  42.01 \\
\ce{NaRu3La5}    & agm072464654         & 6.32 &  7.93 &                 &  42.17 \\
\ce{KCo3Sn5 }    & agm072457299         & 5.24 &  9.17 &                 &  42.26 \\
\ce{TlAu3Hg5}    & agm072497801         & 5.70 &  8.19 &                 &  42.28 \\
\ce{HgRh3Pb5}    & agm072820223         & 5.71 &  7.99 &                 &  42.49 \\
\ce{PbTc3Ce5}    & agm072194225         & 5.92 &  7.33 &                 &  42.87 \\
\ce{BaNi3Sn5}    & agm072564473         & 5.23 &  9.05 &                 &  42.91 \\
\ce{CsNa3Rb5}    & agm072196480         & 8.25 & 11.69 &                 &  43.48 \\
\ce{RbPt3Hg5}    & agm072352936         & 5.74 &  7.55 &                 &  43.63 \\
\ce{MgRu3Ce5}    & agm072803376         & 5.83 &  7.32 &                 &  43.77 \\
\ce{RbNa3K5 }    & agm072521388         & 7.81 & 11.02 &                 &  43.82 \\
\ce{RbMn3Se5}    & agm072844540         & 5.98 &  7.32 & 10.40           &  43.86 \\
\ce{CdRu3La5}    & agm072320856         & 6.19 &  7.74 &                 &  44.10 \\
\ce{SrPd3Hg5}    & agm072528517         & 5.66 &  7.57 &                 &  44.40 \\
\ce{SnRh3Ce5}    & agm072640588         & 5.74 &  7.47 & 1.77            &  44.55 \\
\ce{BaMn3Sn5}    & agm072538329         & 5.35 &  8.90 & 6.39            &  44.86 \\
\ce{SbRu3Ce5}    & agm072344071         & 5.83 &  7.30 &                 &  44.88 \\
\ce{BaCu3Zn5}    & agm072707379         & 5.24 &  6.70 &                 &  45.15 \\
\ce{RbPd3Bi5}    & agm072835197         & 5.88 &  9.19 &                 &  45.58 \\
\ce{CaFe3Ga5}    & agm072536404         & 5.10 &  6.55 & 5.40            &  45.73 \\
\ce{KAg3Hg5 }    & agm072385436         & 5.85 &  7.75 &                 &  45.73 \\
\ce{BaAg3Hg5}    & agm072371502         & 5.86 &  7.68 &                 &  45.74 \\
\ce{SrNi3In5}    & agm072583847         & 5.28 &  8.15 &                 &  45.75 \\
\ce{CaRu3Nd5}    & agm072154850         & 6.23 &  7.76 &                 &  45.92 \\
\ce{KRh3Bi5 }    & agm072601612         & 5.83 &  8.76 &                 &  45.98 \\
\ce{ClNi3Bi5}    & agm072383330         & 5.60 &  8.04 &                 &  46.39 \\
\ce{KFe3Ge5 }    & agm072214090         & 4.87 &  9.01 & 4.81            &  46.41 \\
\ce{TlRu3La5}    & agm072234018         & 6.17 &  7.76 &                 &  46.62 \\
\ce{HgTi3Te5}    & agm072496332         & 5.98 &  8.56 &                 &  46.68 \\
\ce{TlPd3Hg5}    & agm072258239         & 5.67 &  7.53 &                 &  46.91 \\
\ce{BaLi3Tl5}    & agm072615104         & 5.91 &  7.82 &                 &  47.33 \\
\ce{KHg3Cs5 }    & agm072652514         & 7.99 & 11.02 &                 &  47.44 \\
\ce{KFe3Sb5 }    & agm072438417         & 5.41 &  9.13 & 6.70            &  47.45 \\
\ce{BaAl3Au5}    & agm072782269         & 5.60 &  6.96 &                 &  47.50 \\
\ce{KNi3In5 }    & agm072691309         & 5.37 &  7.74 &                 &  47.60 \\
\ce{LaFe3Ga5}    & agm072264041         & 5.15 &  6.57 & 6.05            &  47.80 \\
\ce{TlNi3Bi5}    & agm072826218         & 5.58 &  8.50 &                 &  47.82 \\
\ce{AgOs3Ce5}    & agm072788037         & 5.80 &  7.39 &                 &  47.88 \\
\ce{CeCu3Au5}    & agm072444976         & 5.16 &  7.23 & 1.07            &  47.90 \\
\ce{InFe3Sn5}    & agm072269071         & 5.24 &  8.22 & 5.88            &  48.07 \\
\ce{SrRu3Nd5}    & agm072773426         & 6.26 &  7.88 &                 &  48.11 \\
\ce{BaTi3Sn5}    & agm072224773         & 5.87 &  7.47 &                 &  48.31 \\
\ce{BrAg3Hg5}    & agm072409879         & 5.83 &  7.76 &                 &  48.50 \\
\ce{HgCu3In5}    & agm072215657         & 5.40 &  7.76 &                 &  48.54 \\
\ce{BaCu3Sn5}    & agm072648583         & 5.36 &  9.15 &                 &  48.62 \\
\ce{RbNb3Sb5}    & agm072272354         & 5.81 &  9.33 &                 &  48.62 \\
\ce{RbAu3Hg5}    & agm072182010         & 5.90 &  7.74 &                 &  48.66 \\
\ce{BaBe3Au5}    & agm072709309         & 5.24 &  6.79 &                 &  48.87 \\
\ce{CsAg3Br5}    & agm072311859         & 6.69 &  7.35 &                 &  48.99 \\
\ce{CaRu3Pr5}    & agm072726265         & 6.27 &  7.83 &                 &  49.21 \\
\ce{RbMn3As5}    & agm072527570         & 5.18 &  8.67 & 5.70            &  49.21 \\
\ce{CaRu3Pm5}    & agm072584623         & 6.19 &  7.71 &                 &  49.36 \\
\ce{KV3As5  }    & agm072168418         & 5.21 &  8.47 &                 &  49.38 \\
\ce{PmCu3Au5}    & agm072639973         & 5.15 &  7.22 &                 &  49.44 \\
\ce{InBe3Au5}    & agm072644968         & 4.96 &  6.81 &                 &  49.49 \\
\ce{CsZr3Te5}    & agm072201186         & 6.42 &  8.69 &                 &  49.58 \\
\ce{InRu3La5}    & agm072588814         & 6.14 &  7.77 &                 &  49.66 \\
\ce{CaCu3Au5}    & agm072842371         & 5.16 &  7.21 &                 &  49.81 \\
\ce{RbHf3Te5}    & agm072183333         & 6.35 &  8.58 &                 &  49.85 \\
\ce{NaNi3Bi5}    & agm072179318         & 5.60 &  8.50 &                 &  49.86 \\
\ce{PbFe3Ce5}    & agm072718574         & 5.62 &  7.15 &                 &  49.92 \\
\end{longtable}

\newcommand{\bandfigure}[4]{%
  \begin{minipage}[t]{0.45\textwidth}  
    \centering  
    \textbf{\Large \ce{#1}}\\[0.3cm]  
    \textbf{\large a = #2, c = #4}\\[1ex]  
    \includegraphics[width=\linewidth]{img/all_band_structures/el-bs-#1.pdf}  
  \end{minipage}%
}

\newcommand{\removetag}{%
  \renewcommand{\thefigure}{}%
  \renewcommand{\figurename}{}%
  \renewcommand{\caption}[1]{\par\centering ##1\par}%
}

\begin{figure*}[ht]
\centering
\caption{Electronic band structures of the 36 stable compounds. Lattice constants $a$ and $c$ in \AA. The black curve is the total DOS, while the projected atomic DoS is in blue (atom A), orange (atom B), and green (atom C). For magnetic materials \ce{RbMn3Sb5}, \ce{KMn3Sb5} and \ce{PbCo3Ce5} spin-projected DOS is shown, positive values for spin-up and negative values for spin-down. Spin up band structure is shown in green and spin down in orange.}
\removetag
\caption{\large \textbf{\underline{Group 11}}}
\begin{tabular}{cc}
  \bandfigure{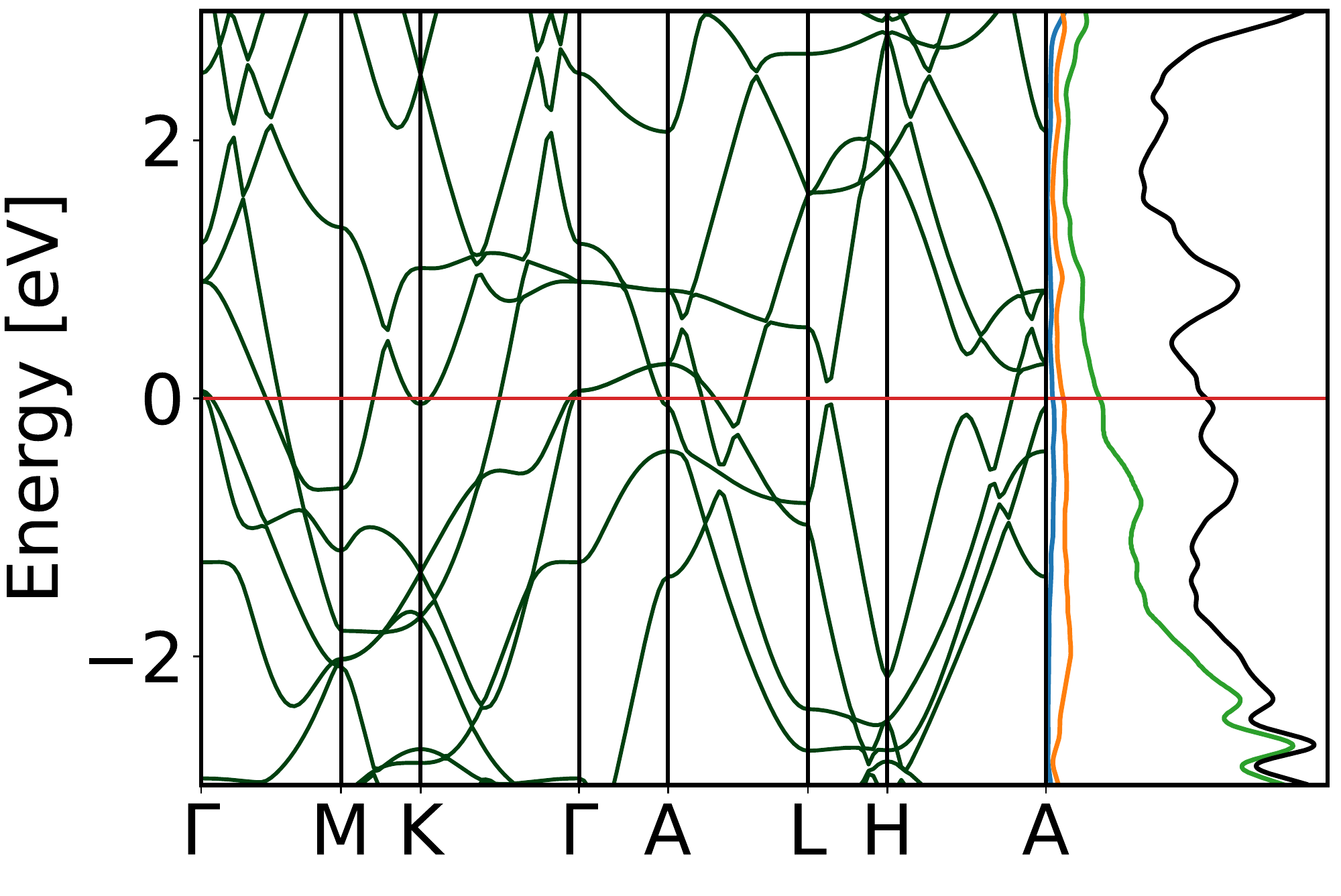}{5.03}{5.03}{6.74}    & \bandfigure{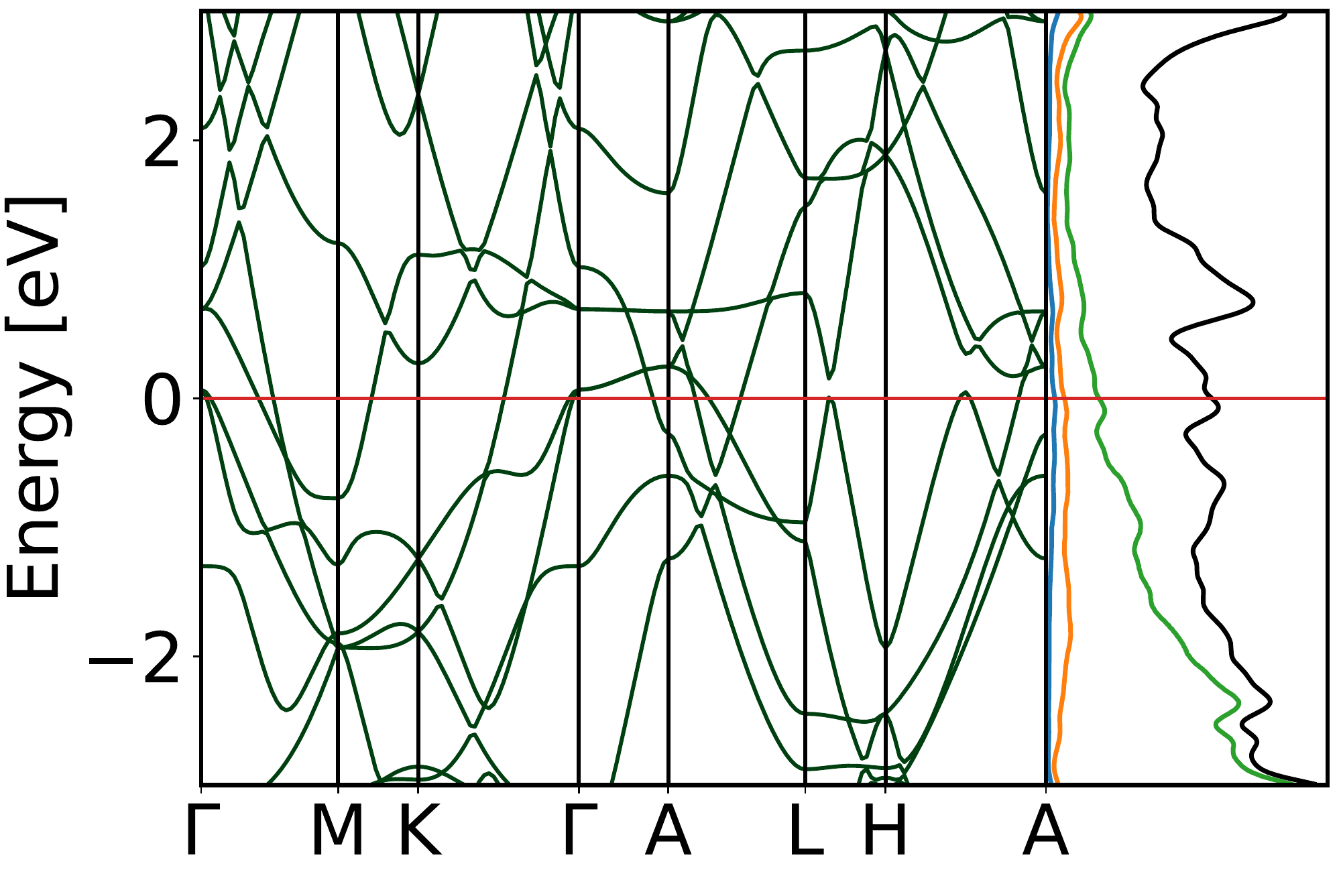}{5.12}{5.12}{6.76}    \\[5cm]
  \bandfigure{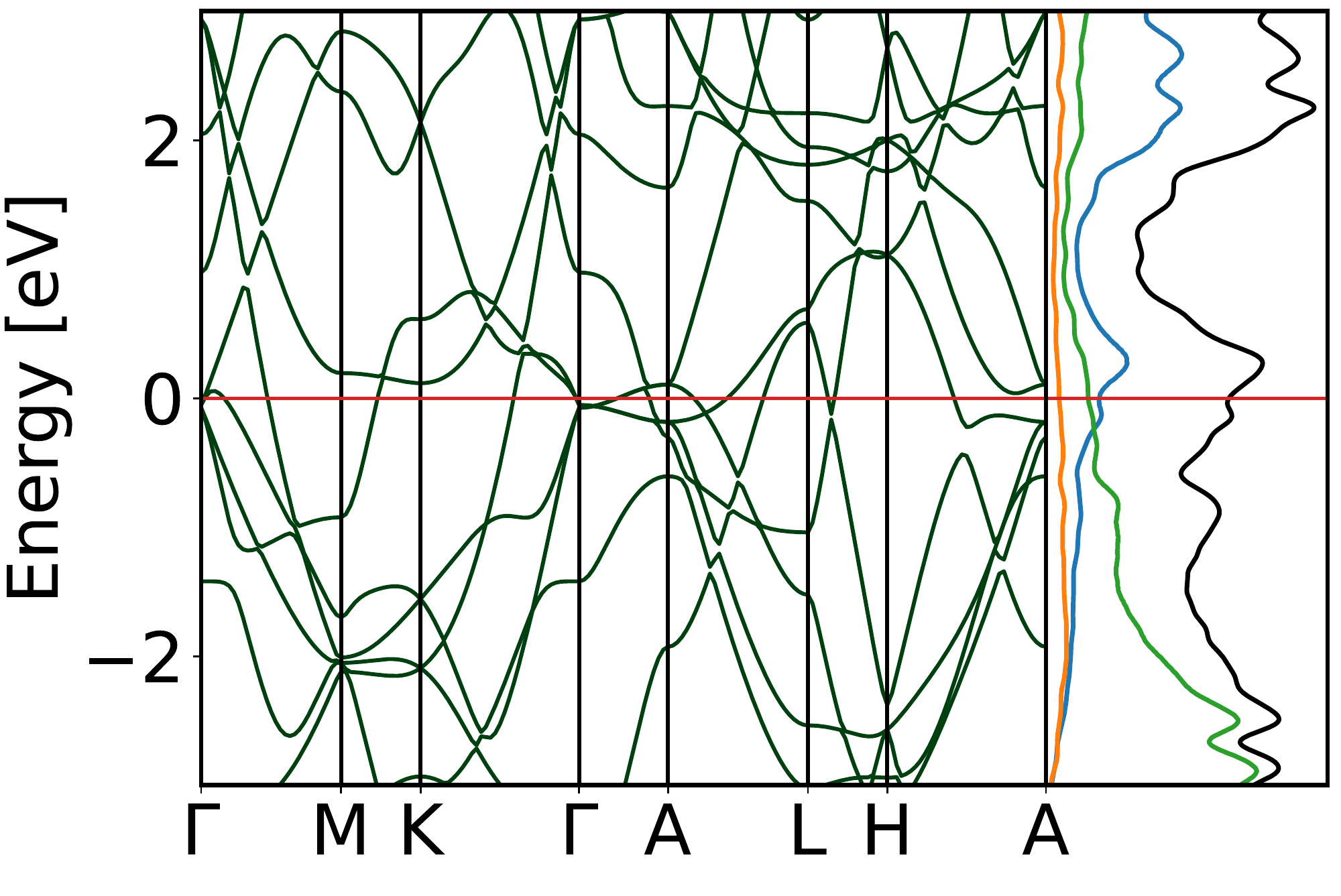}{5.05}{5.05}{6.75}    & {} \\[5cm]
\end{tabular}
\end{figure*}

\begin{figure*}[ht]
\centering
\removetag
\caption{\large \textbf{\underline{Group 12}}}
\begin{tabular}{cc}
  \bandfigure{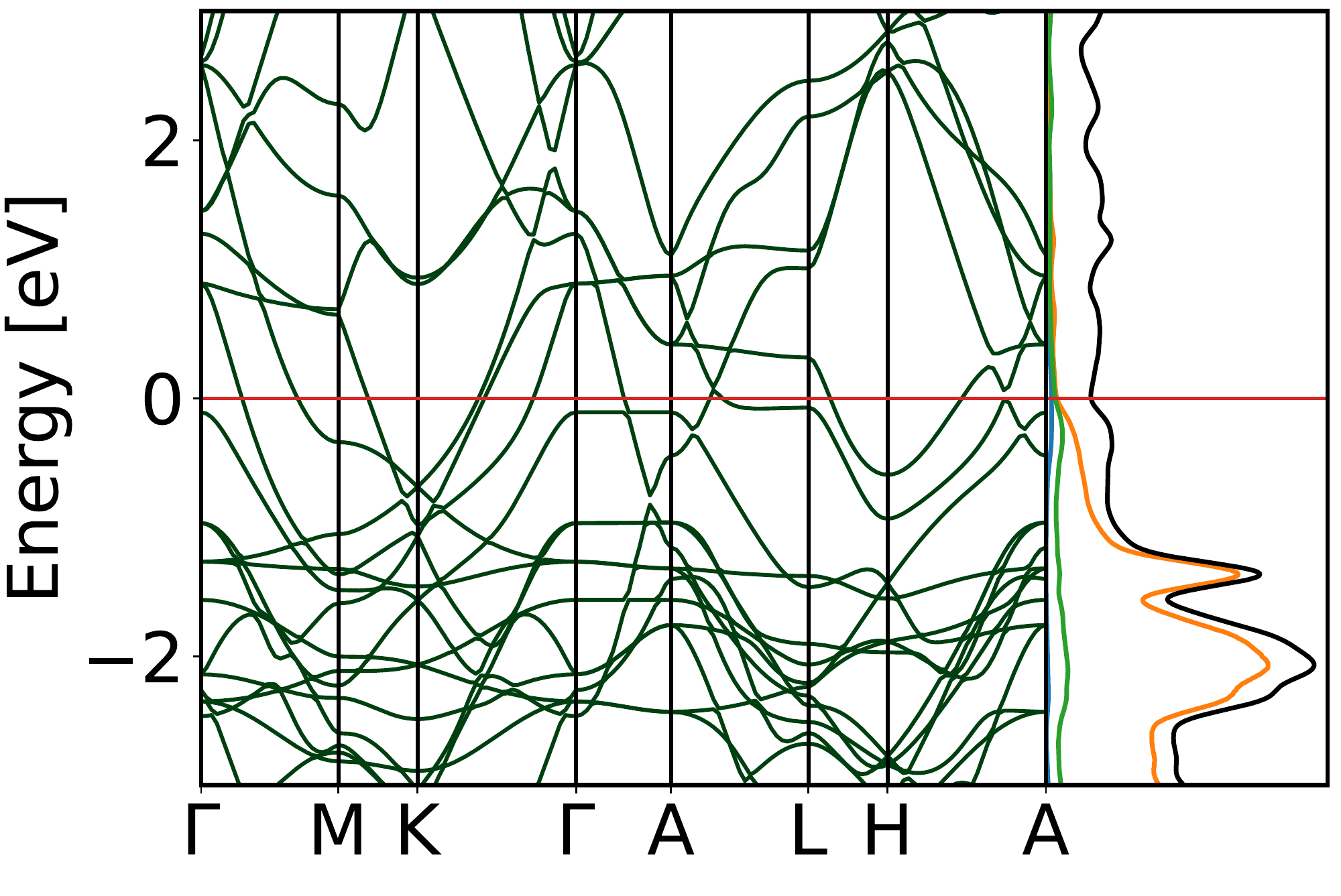}{5.75}{5.75}{7.47}    & \bandfigure{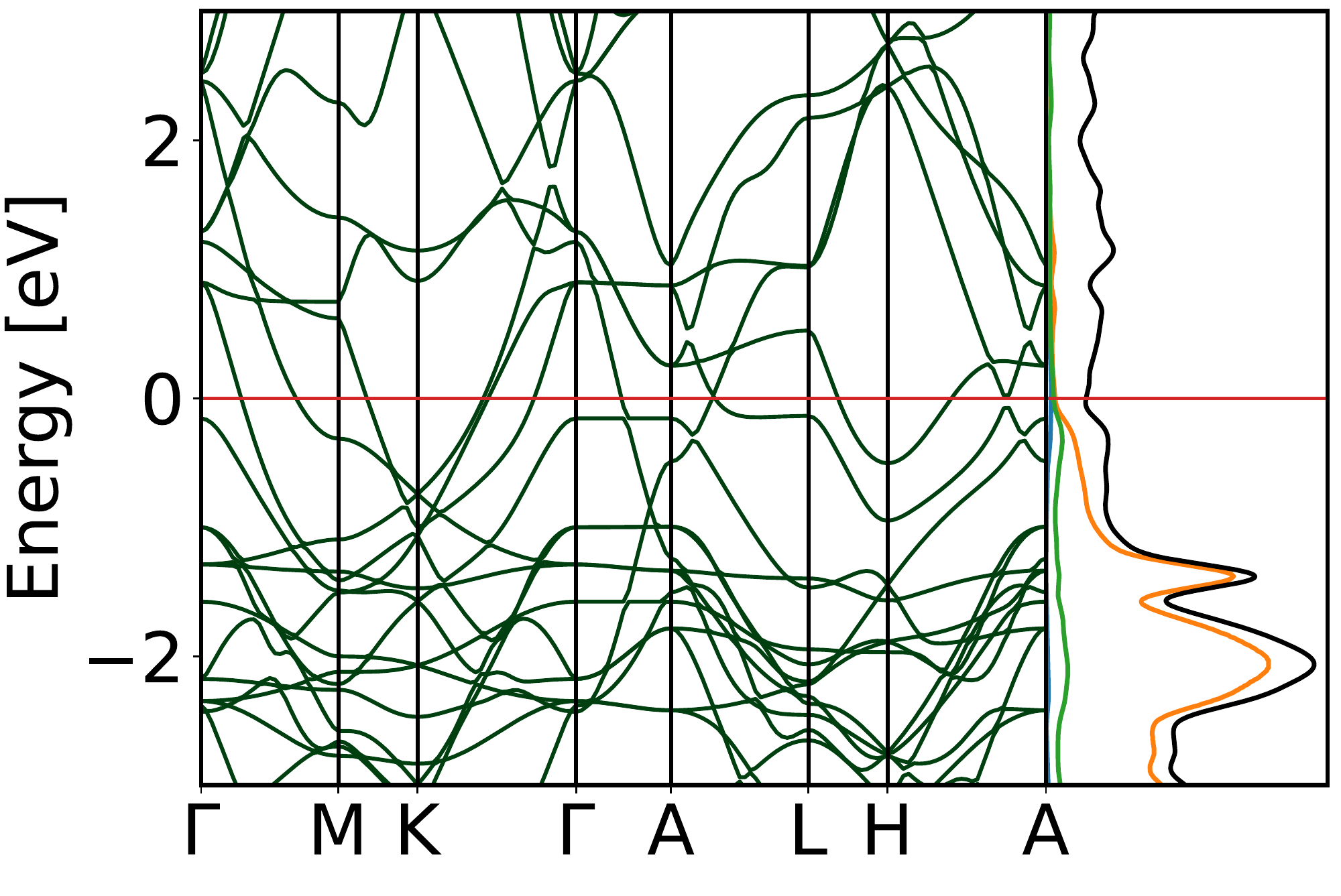}{5.78}{5.78}{7.52}    \\[5cm]
\end{tabular}
\end{figure*}

\begin{figure*}[ht]
\centering
\removetag
\caption{\large \textbf{\underline{Group 13}}}
\begin{tabular}{cc}
  \bandfigure{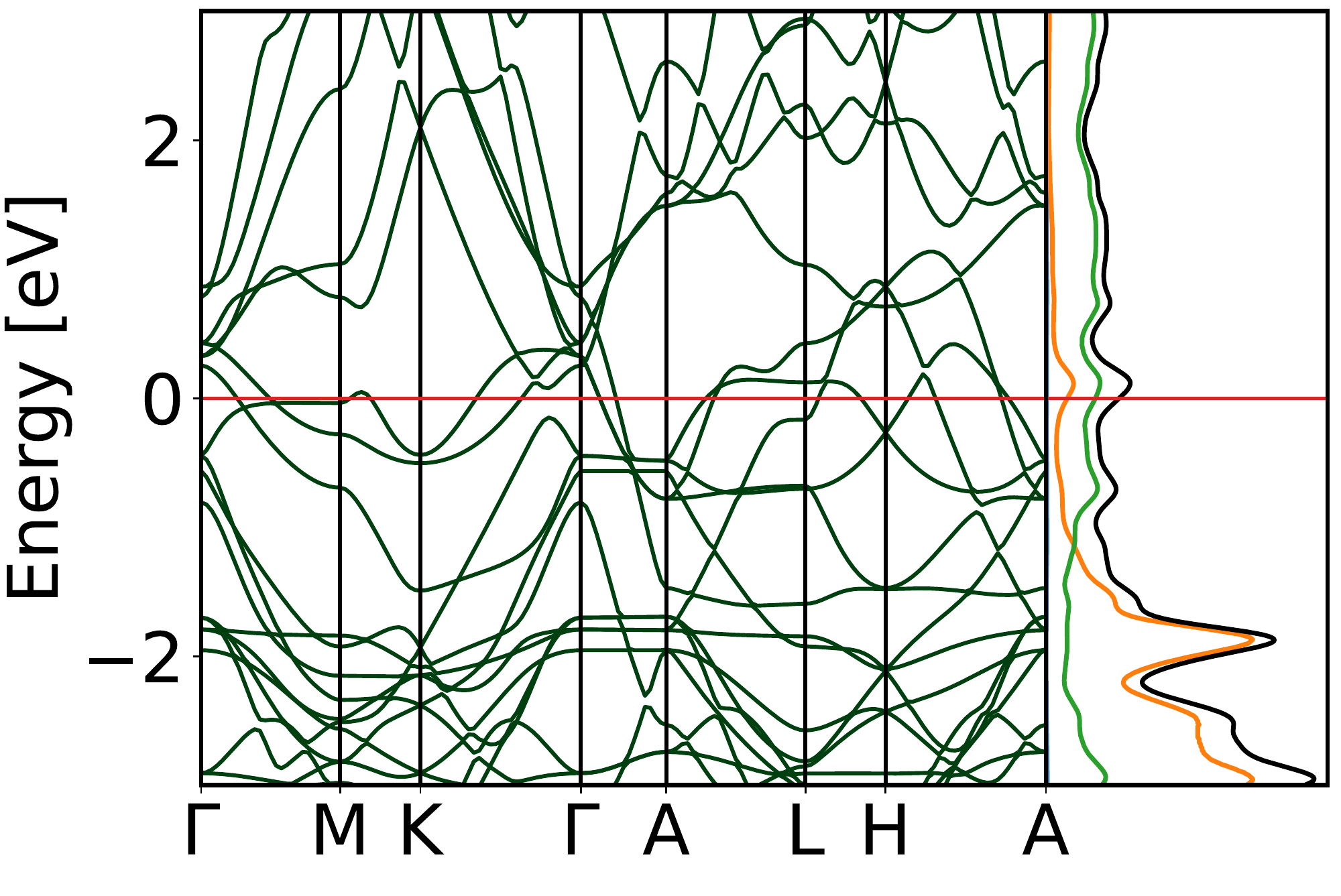}{5.76}{5.76}{8.33}    & \bandfigure{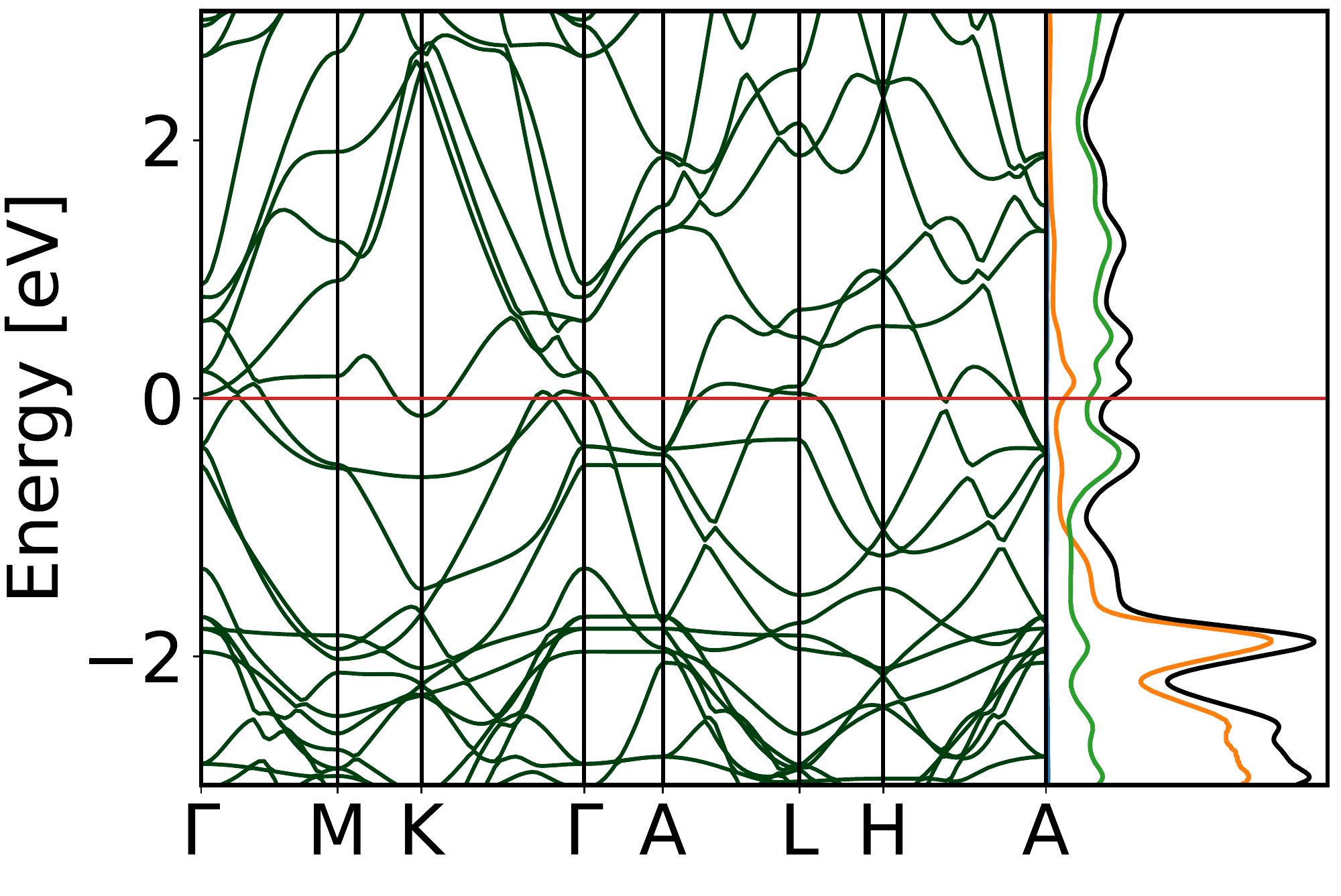}{5.69}{5.69}{9.44}    \\[5cm]
  \bandfigure{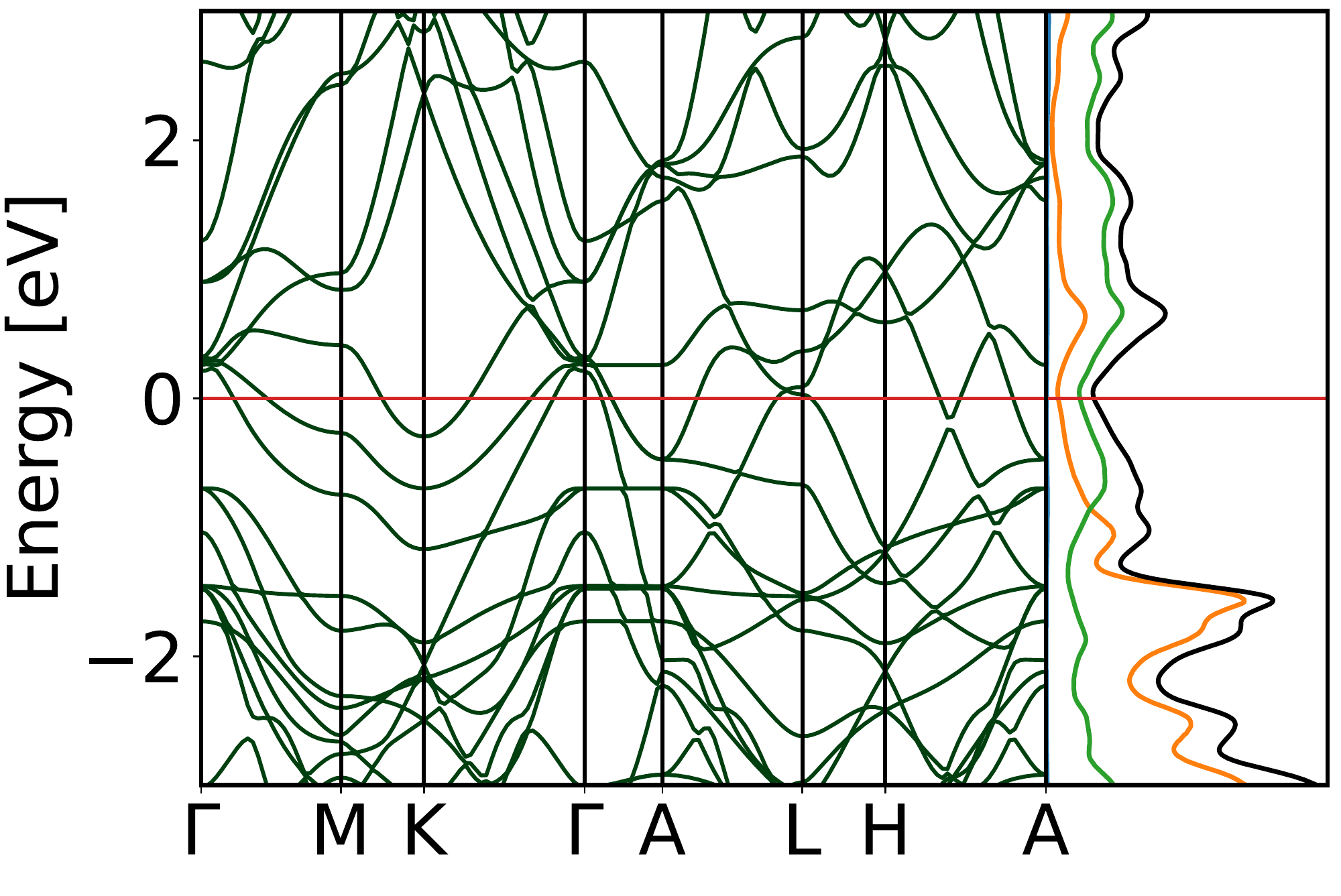}{5.67}{5.67}{8.79}    & \bandfigure{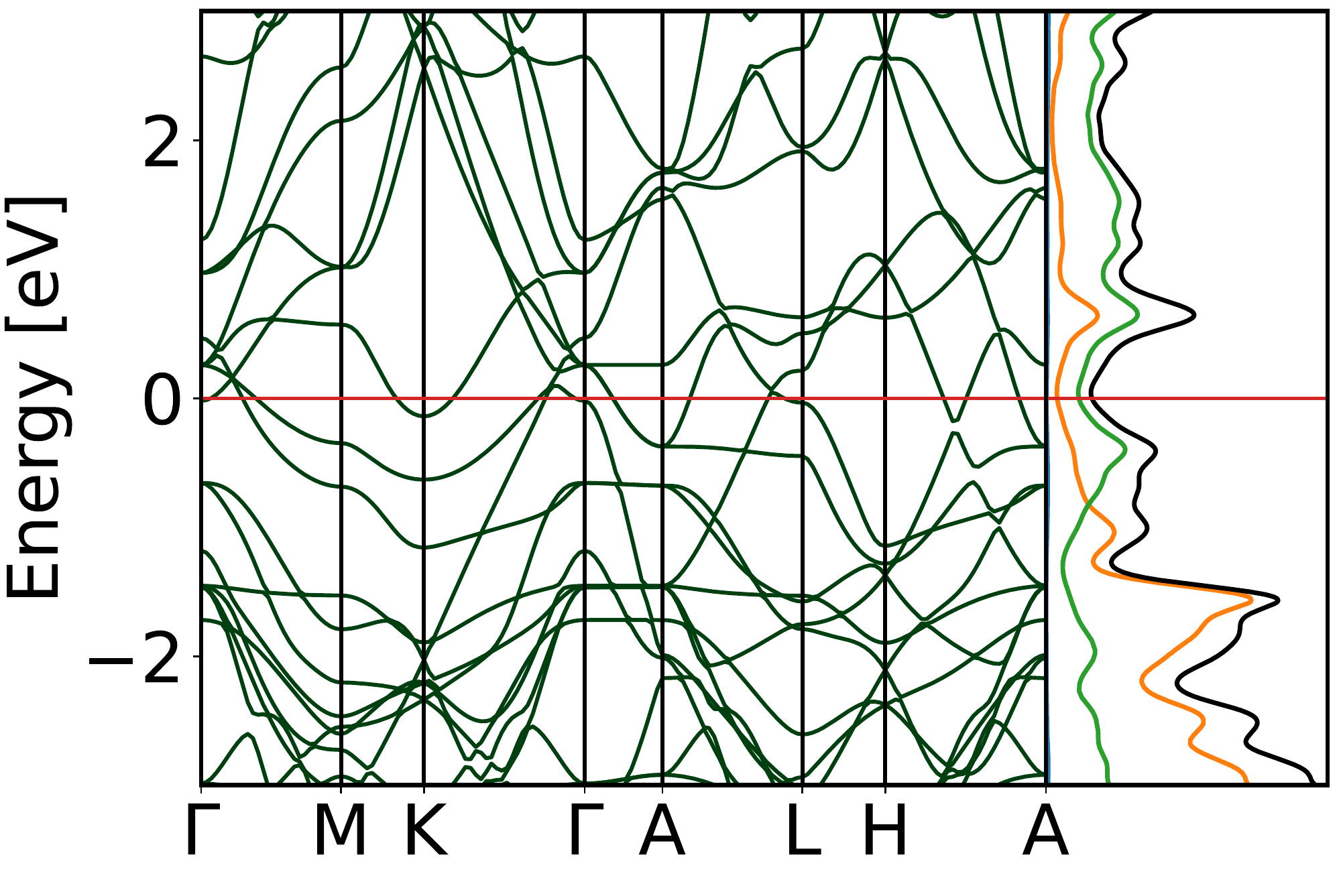}{5.67}{5.67}{9.32}    \\[5cm]
  \bandfigure{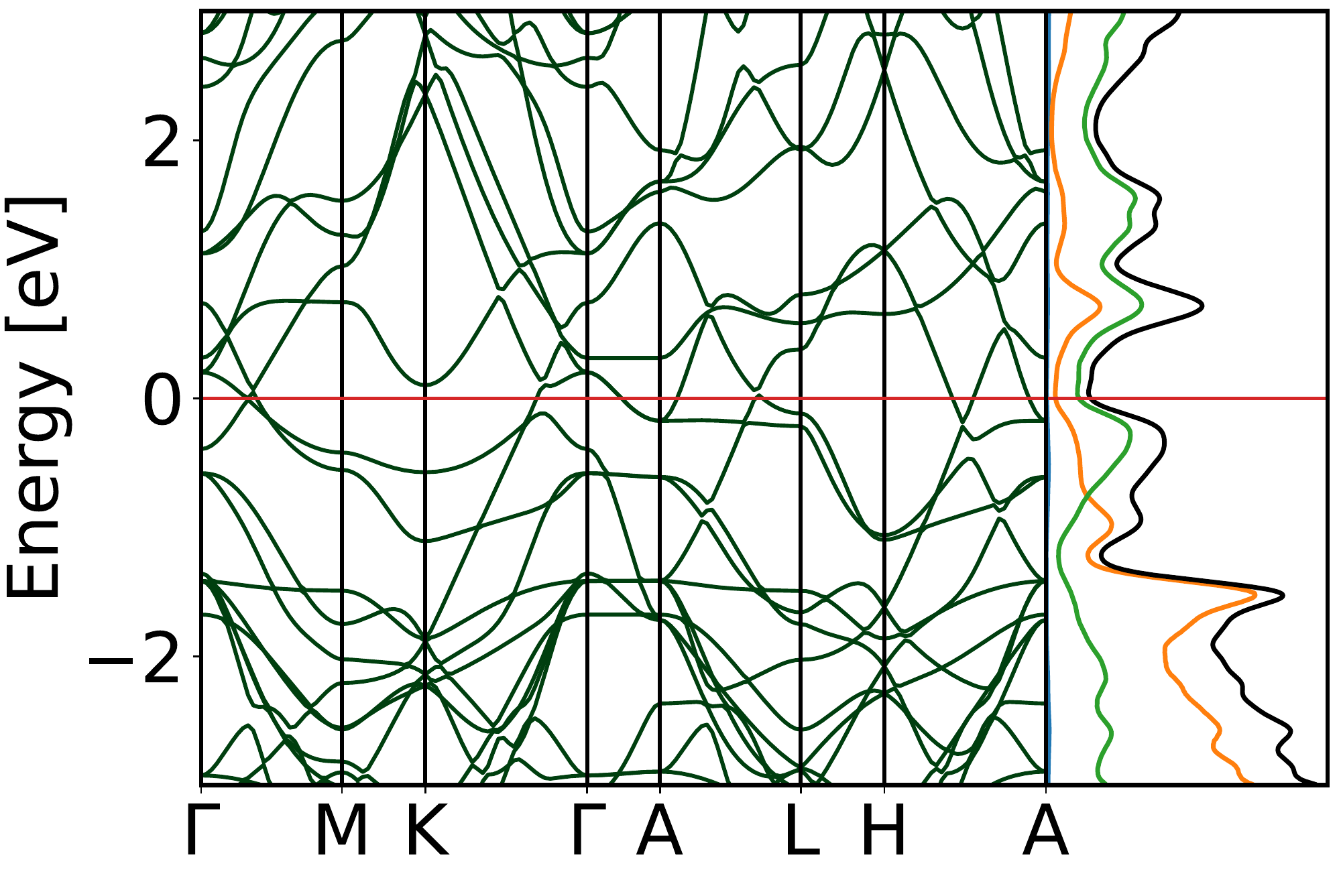}{5.66}{5.66}{9.94}    & {} \\[5cm]
\end{tabular}
\end{figure*}

\begin{figure*}[ht]
\centering
\removetag
\caption{\large \textbf{\underline{Group 15}}}
\begin{tabular}{cc}
  \bandfigure{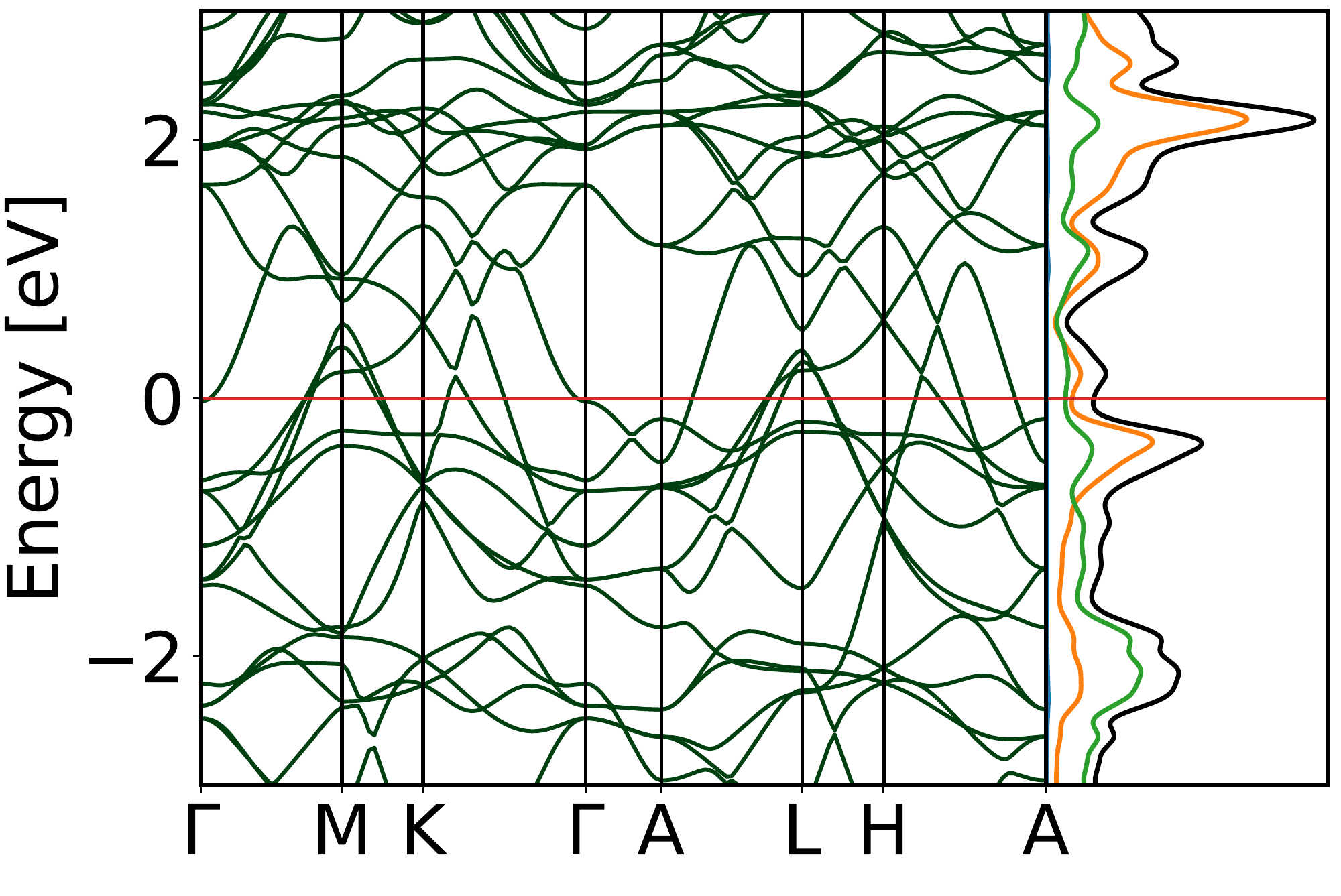}{5.68}{5.68}{9.80}    & \bandfigure{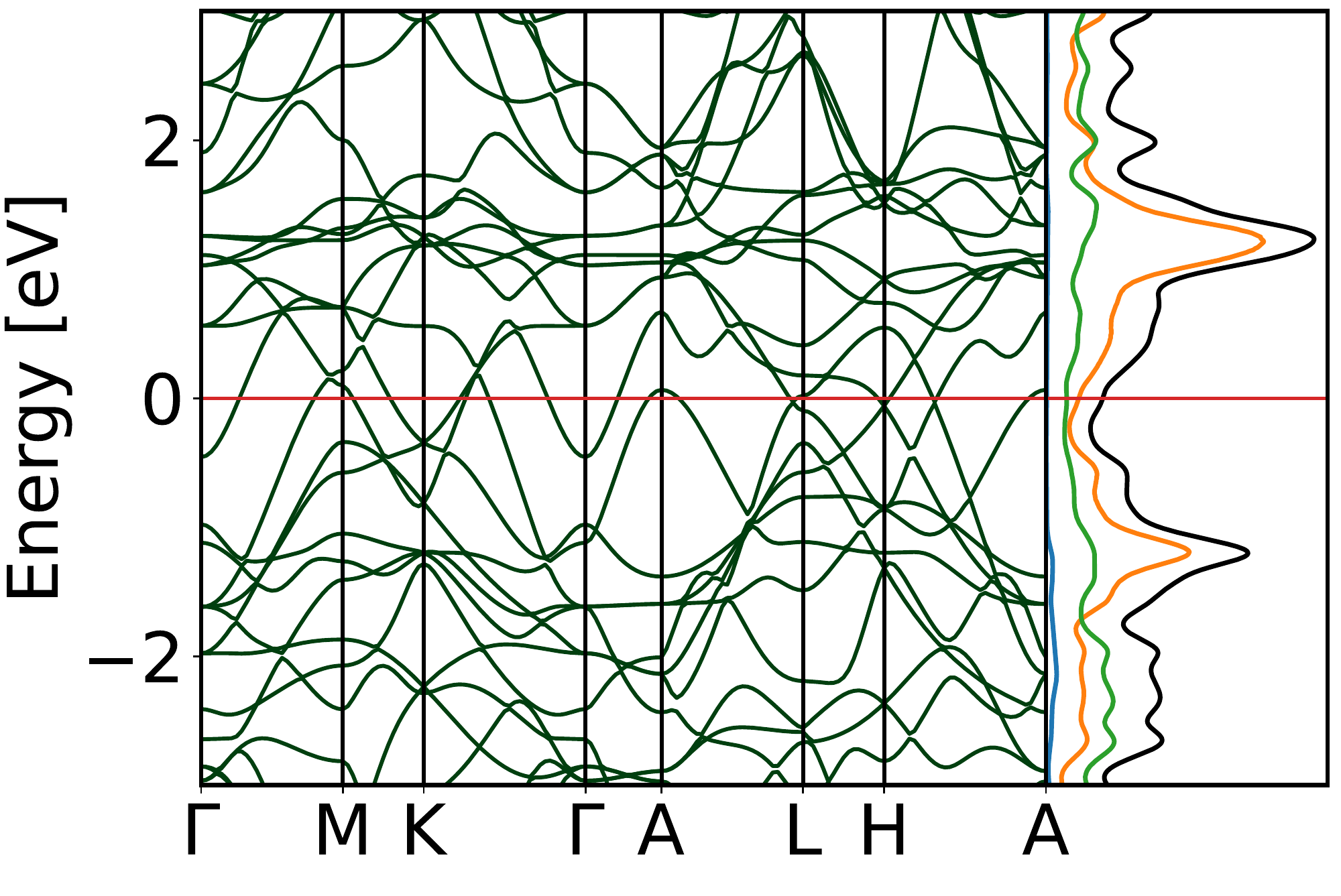}{5.44}{5.44}{8.79}    \\[5cm]
  \bandfigure{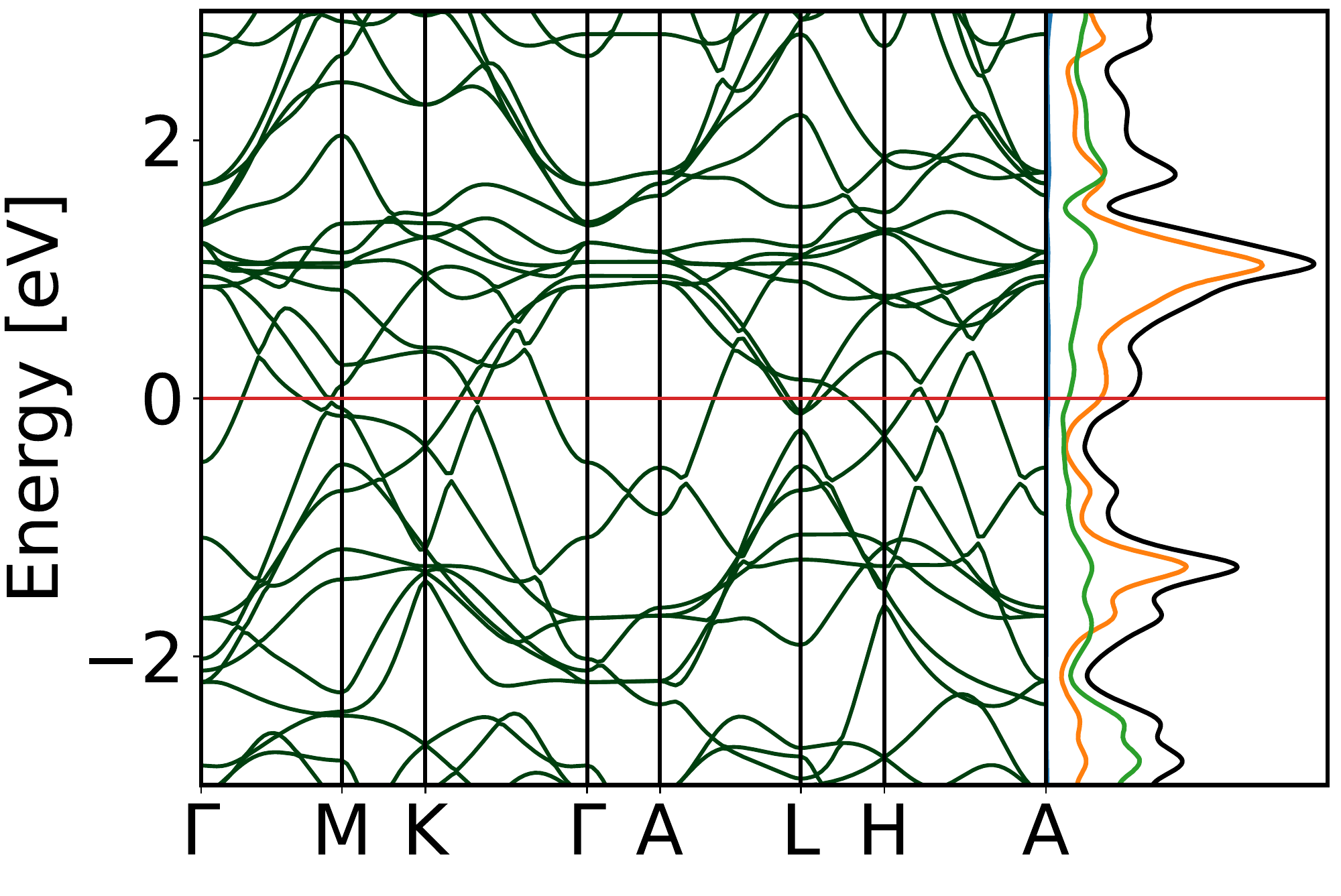}{5.48}{5.48}{9.31}      & \bandfigure{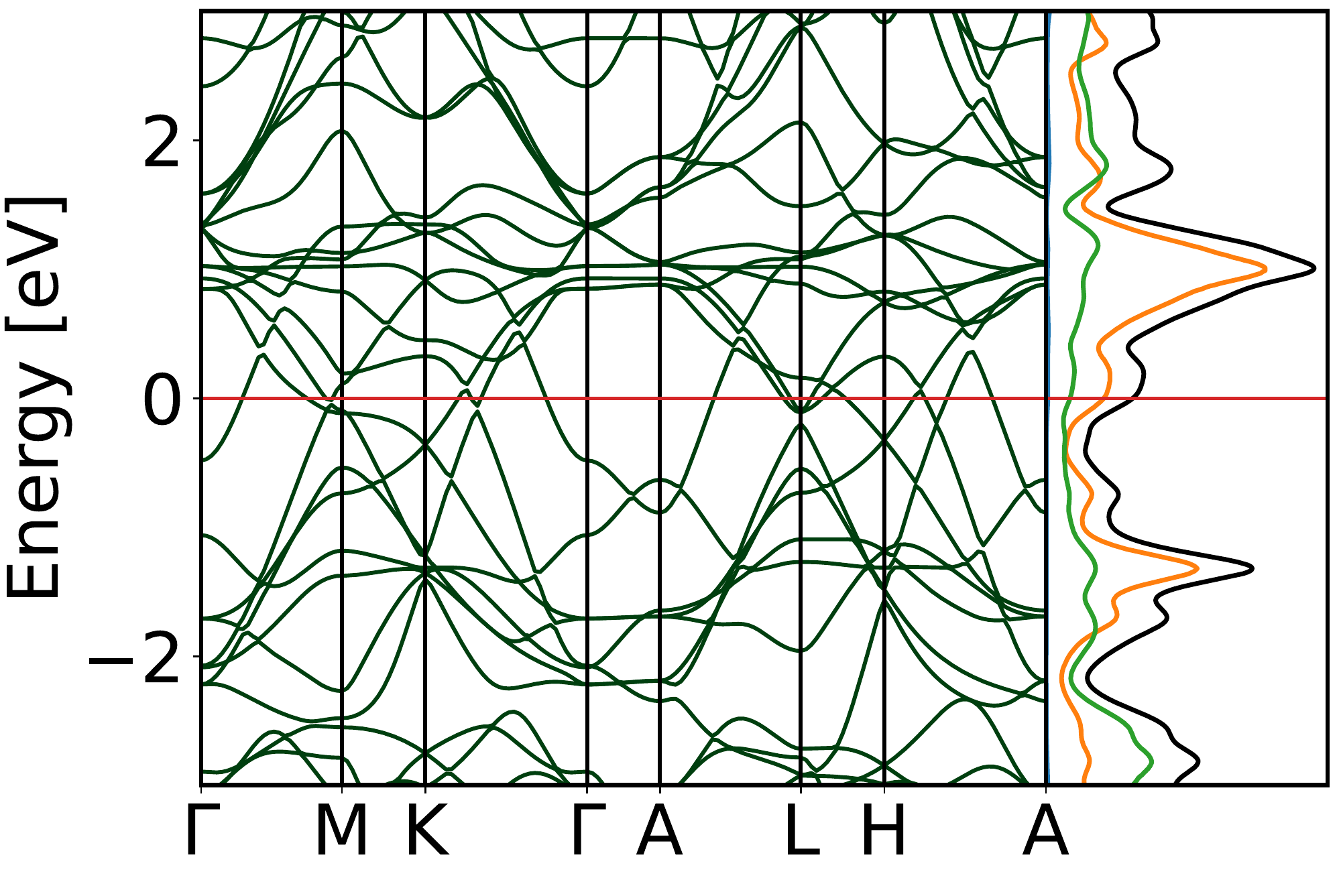}{5.49}{5.49}{9.55}    \\[5cm]
  \bandfigure{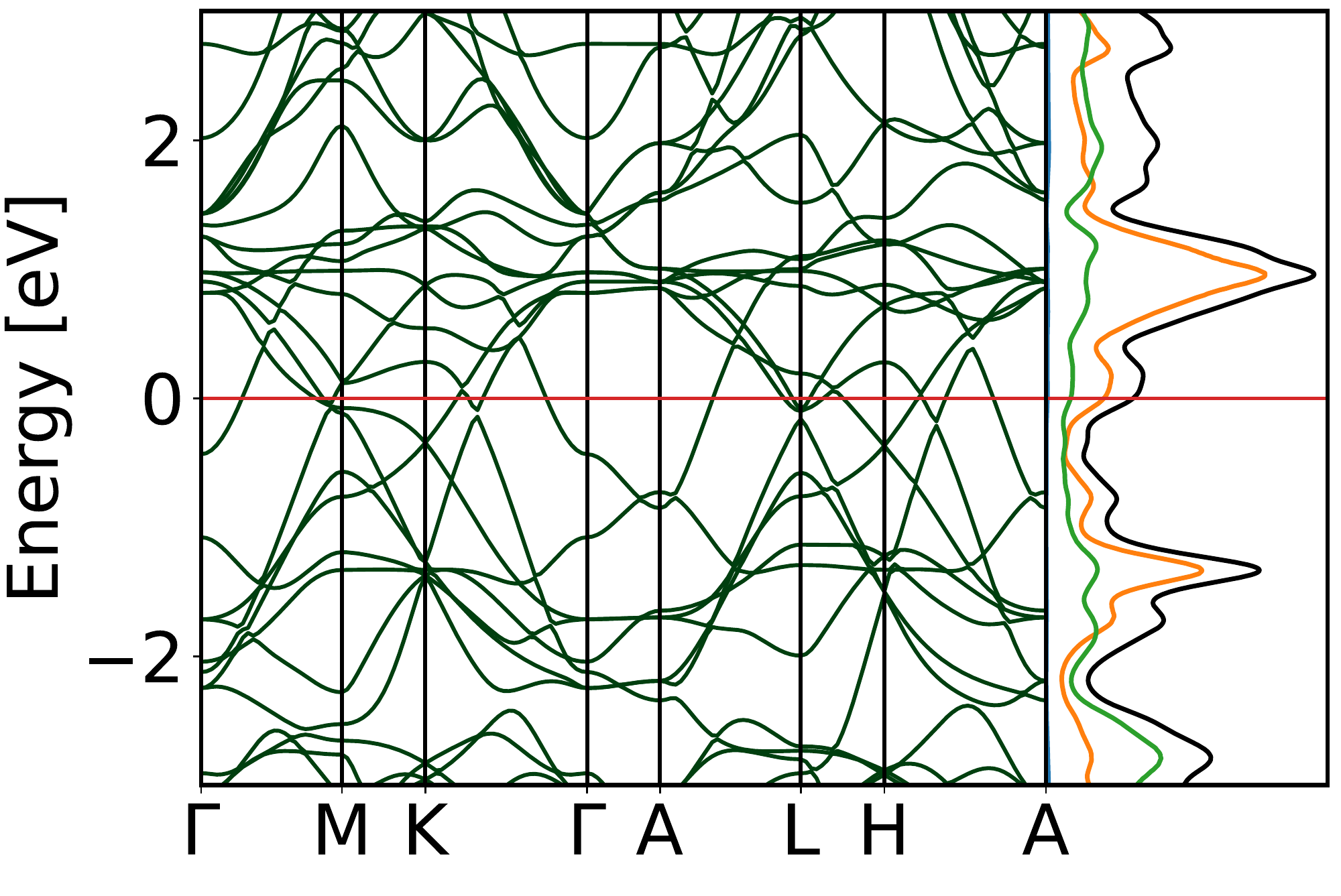}{5.51}{5.51}{9.82}     & \bandfigure{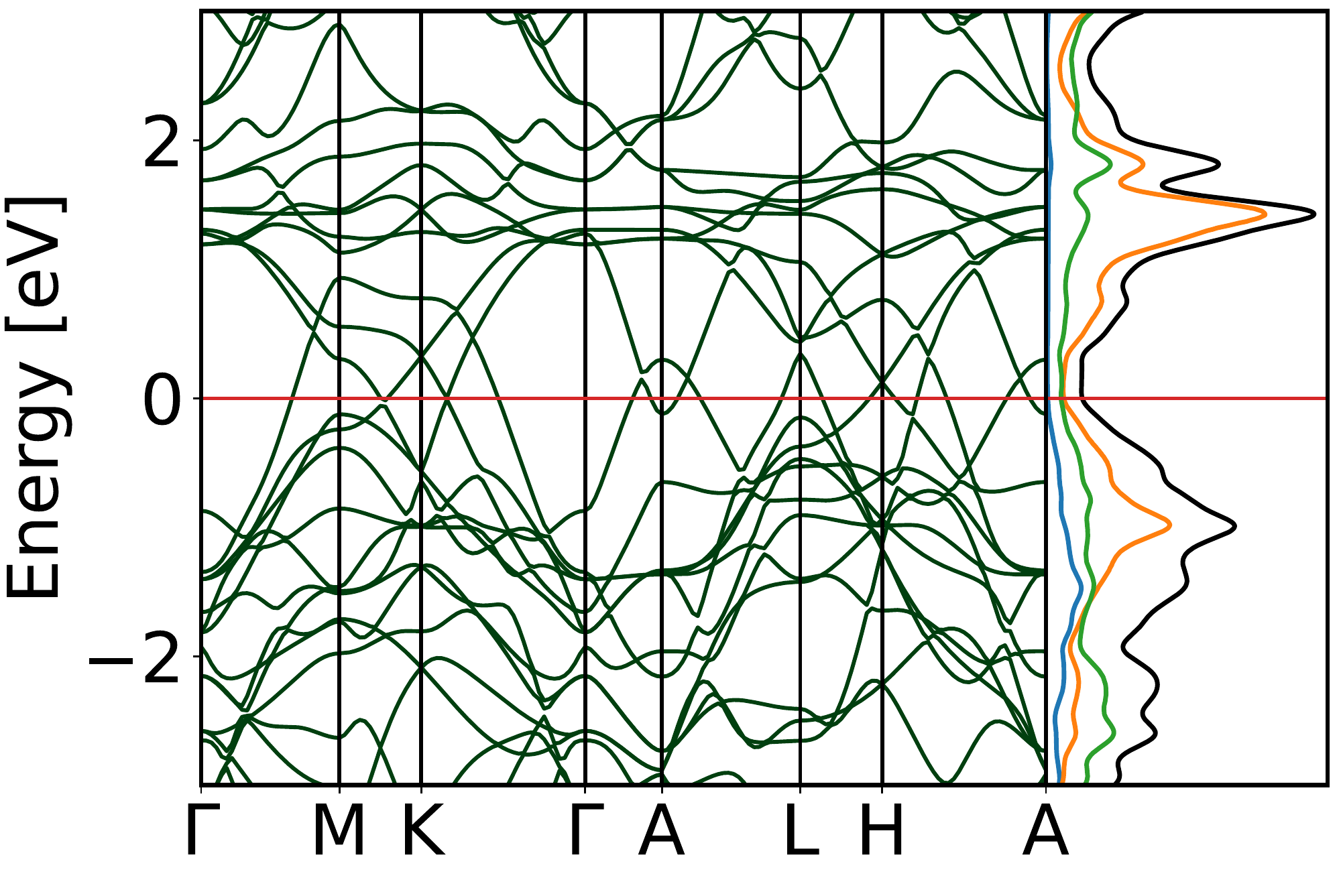}{5.45}{5.45}{8.78}     \\[5cm]
\end{tabular}
\end{figure*}

\begin{figure*}[ht]
\centering
\begin{tabular}{cc}
  \bandfigure{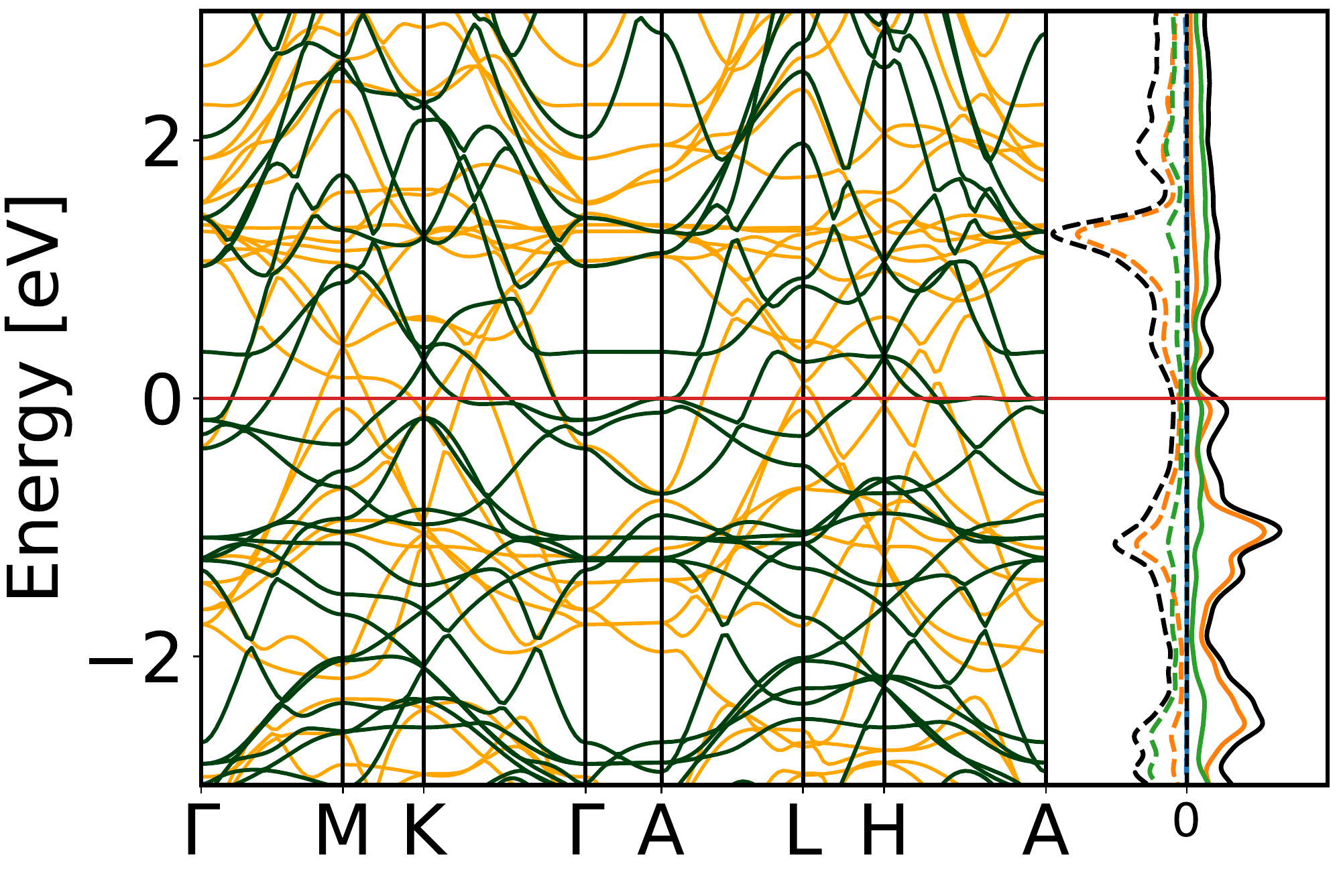}{5.43}{5.43}{9.26}     & \bandfigure{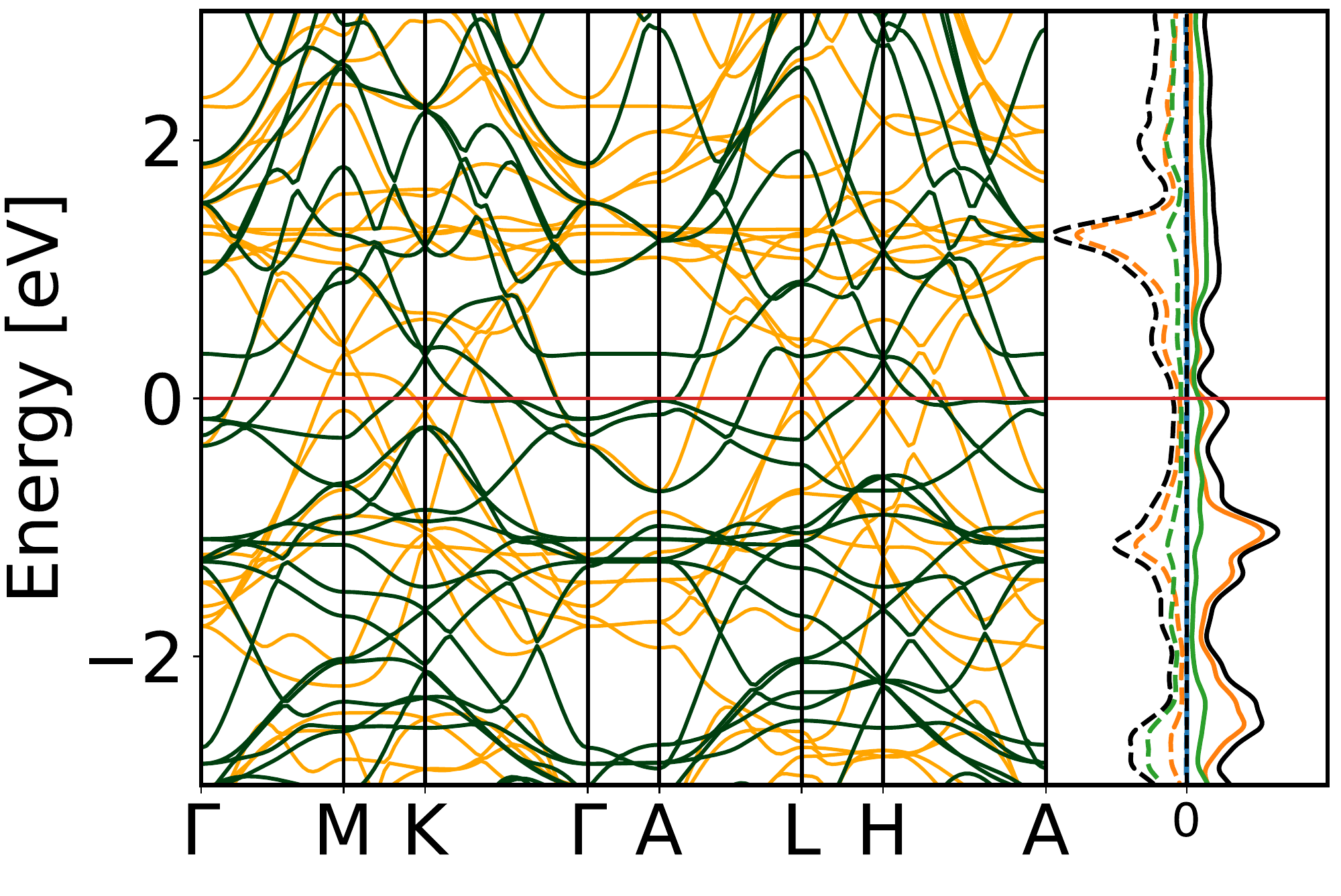}{5.44}{5.44}{9.53}    \\[5cm]
  \bandfigure{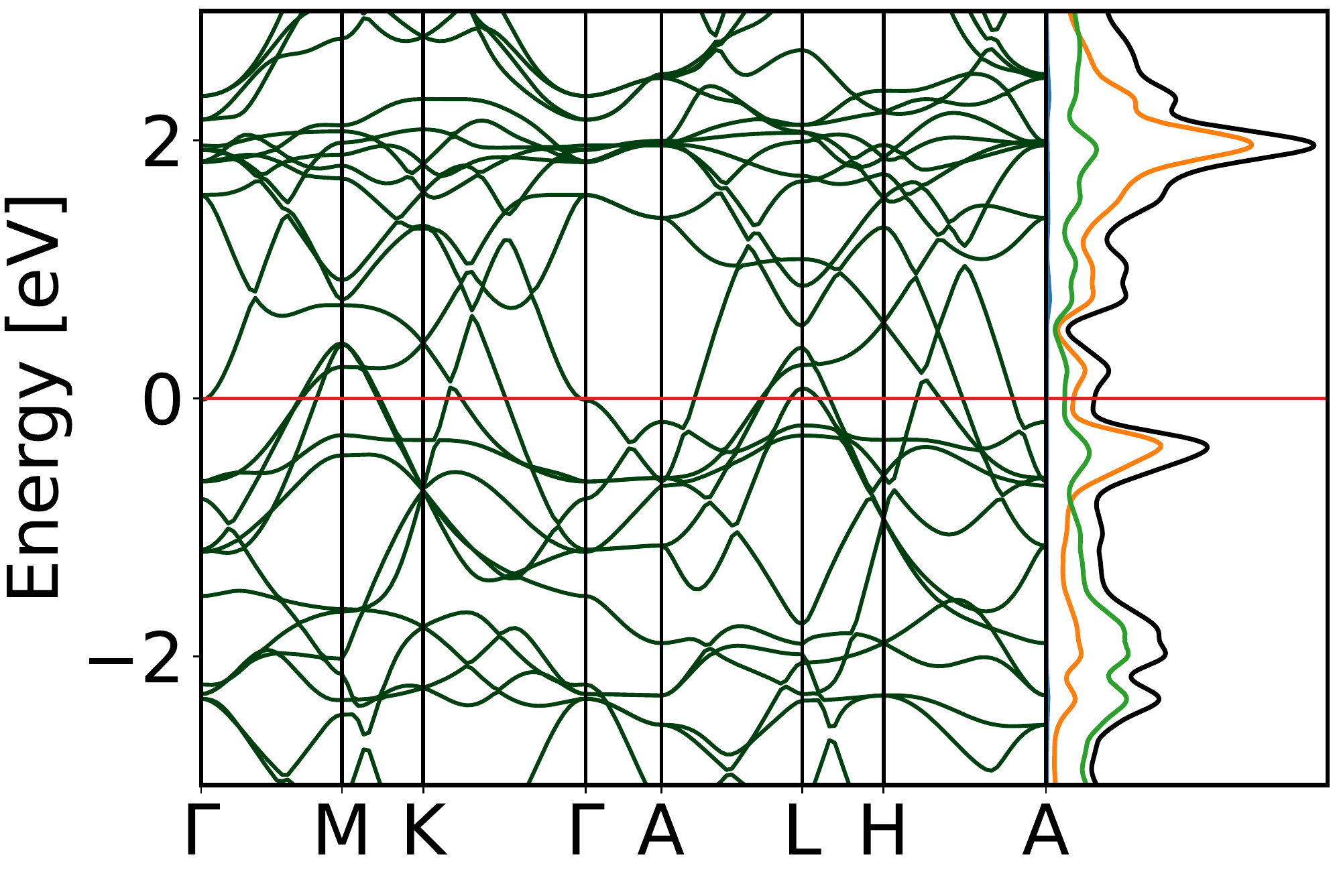}{5.82}{5.82}{9.66}    & \bandfigure{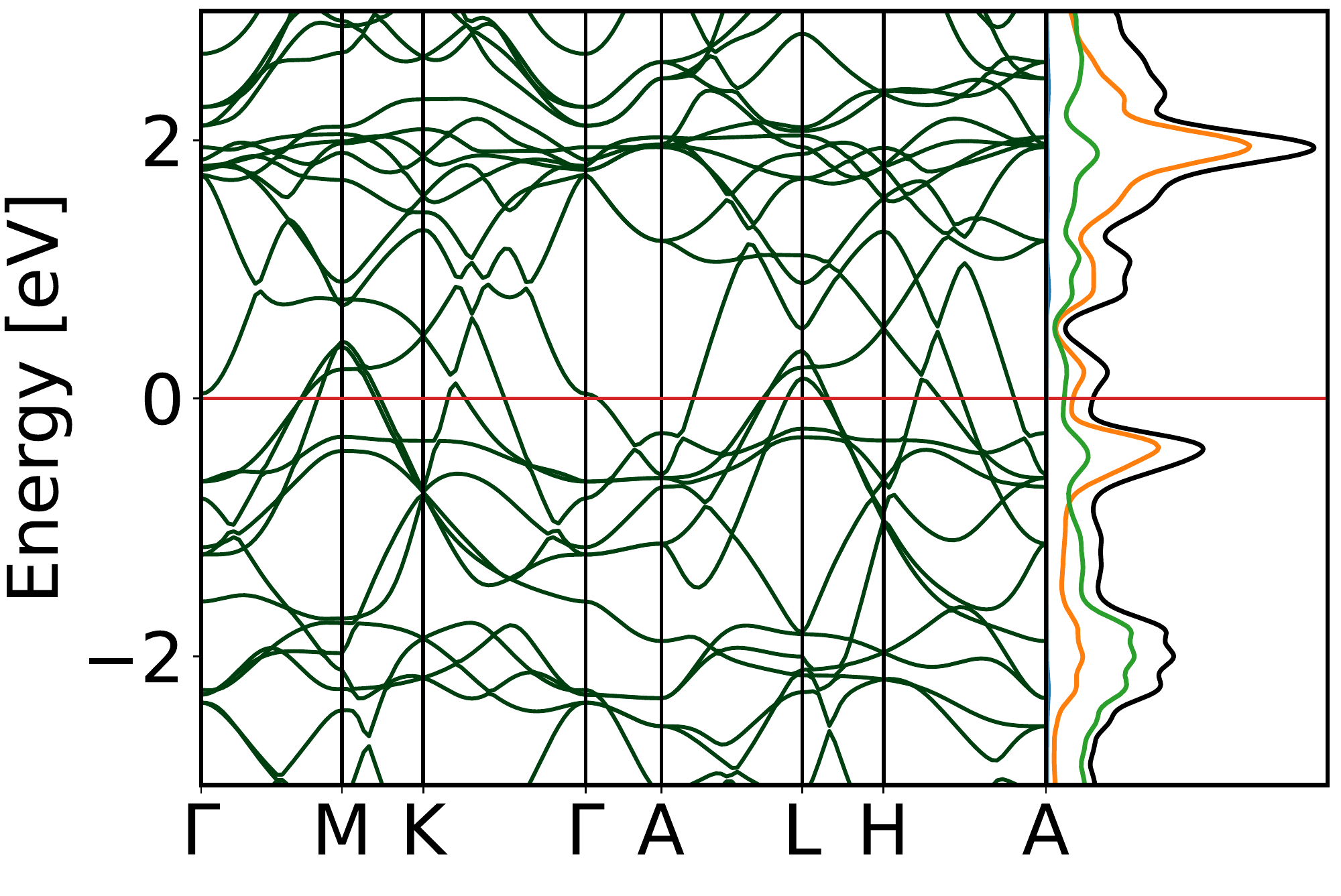}{5.83}{5.83}{9.92}    \\[5cm]
  \bandfigure{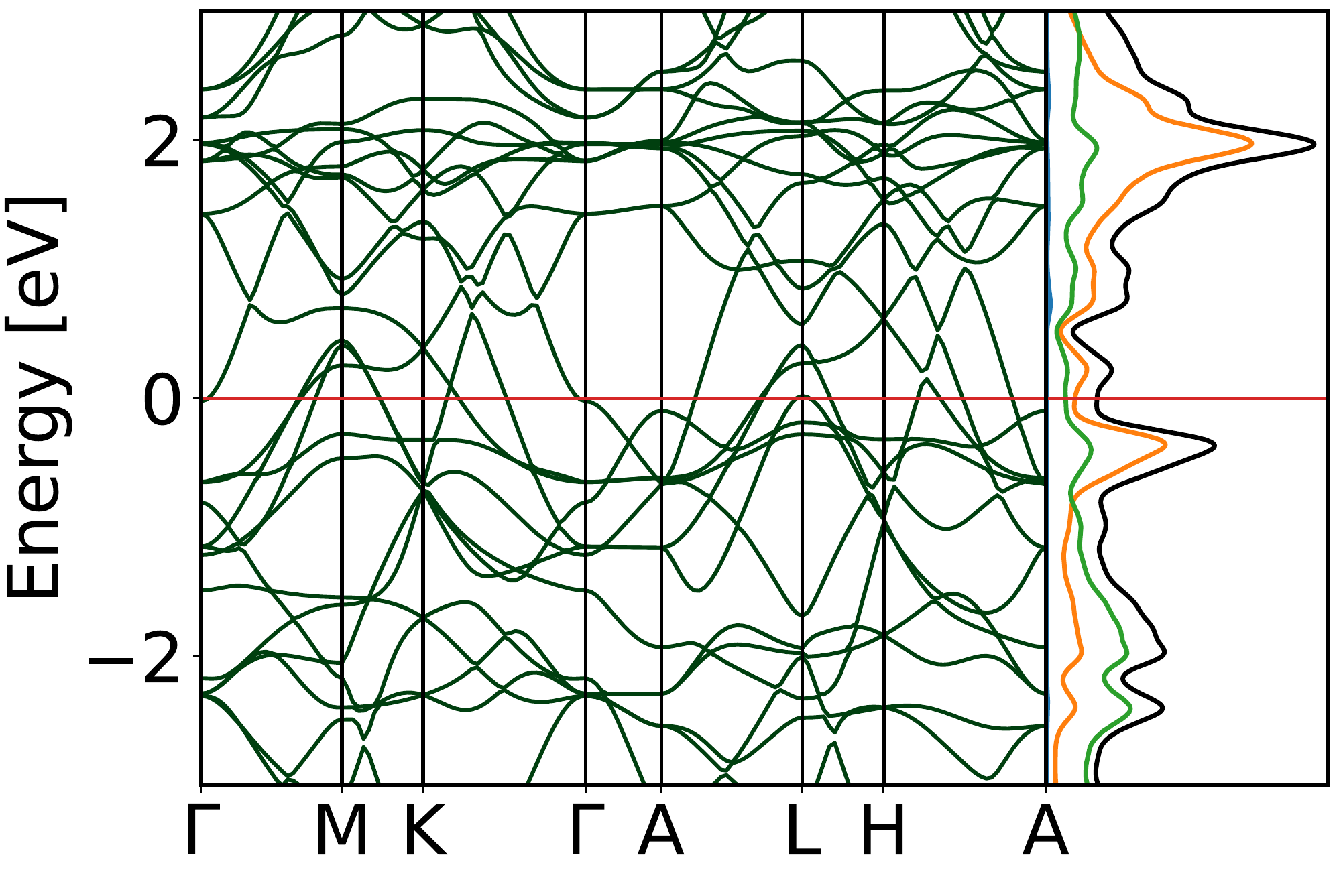}{5.89}{5.89}{9.44}     & \bandfigure{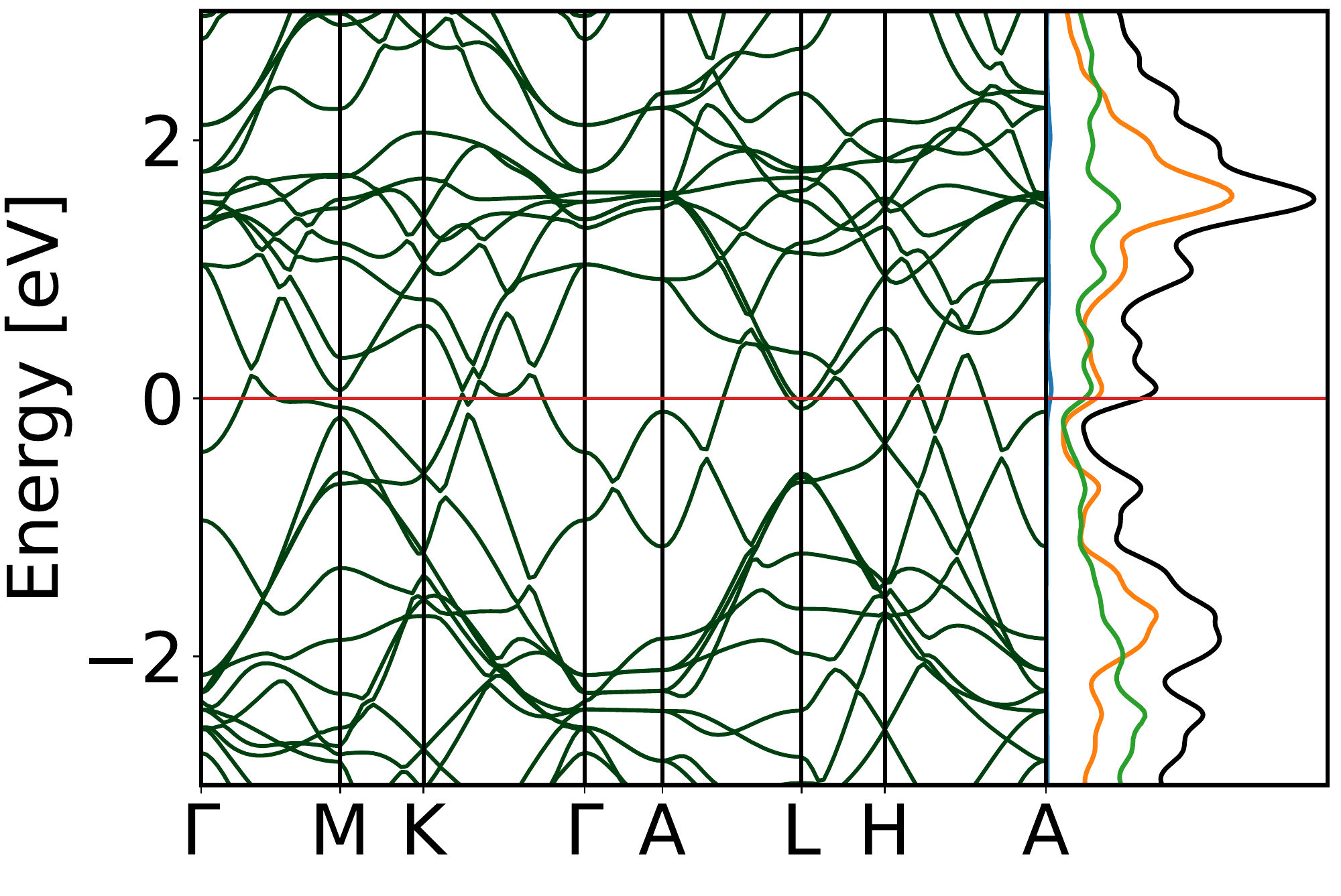}{5.99}{5.99}{9.51}    \\[5cm]
\end{tabular}
\end{figure*}

\begin{figure*}[ht]
\centering
\begin{tabular}{cc}
  \bandfigure{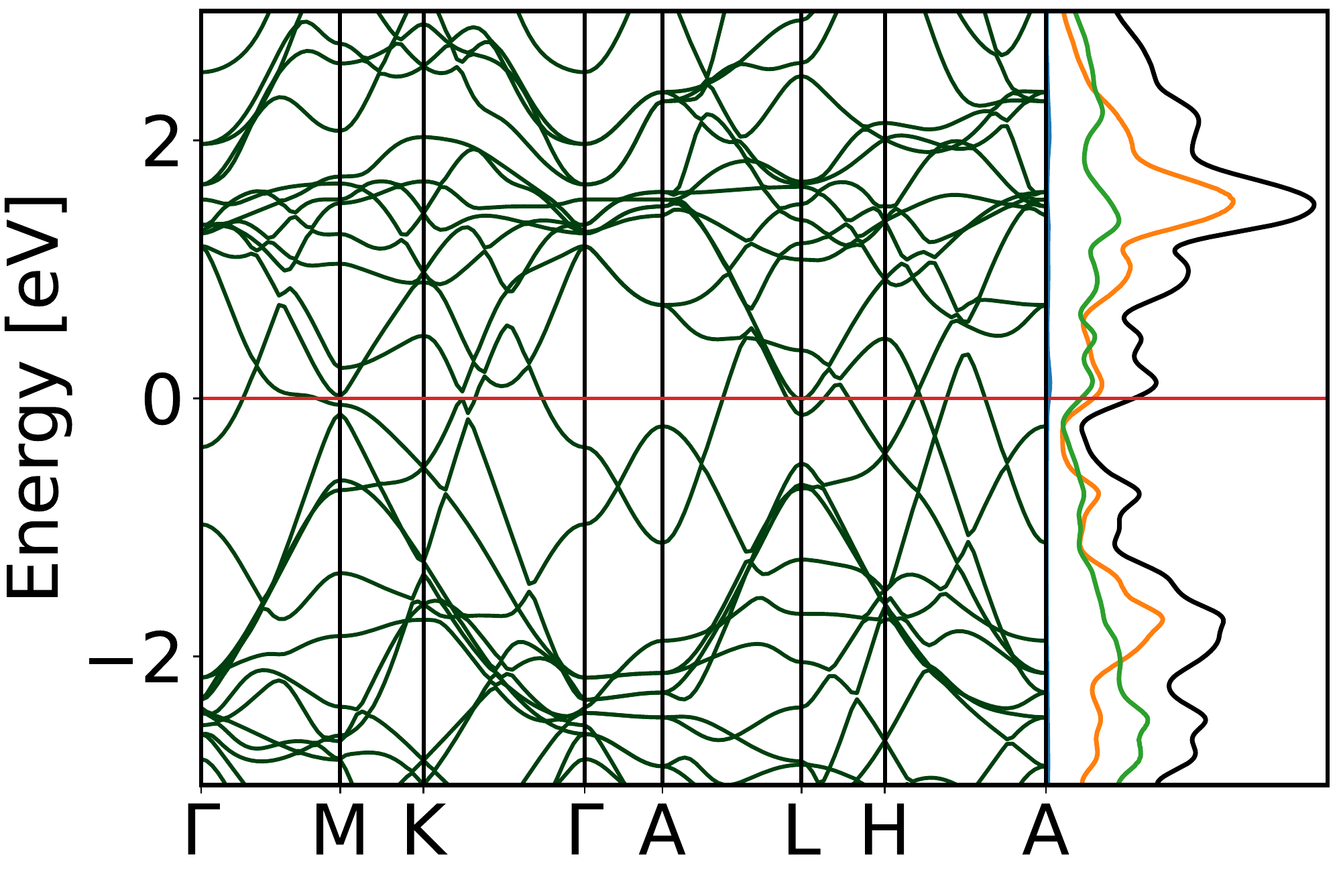}{5.91}{5.91}{9.77}    & \bandfigure{INb3Bi5}{5.85}{5.85}{9.03}     \\[5cm]
  \bandfigure{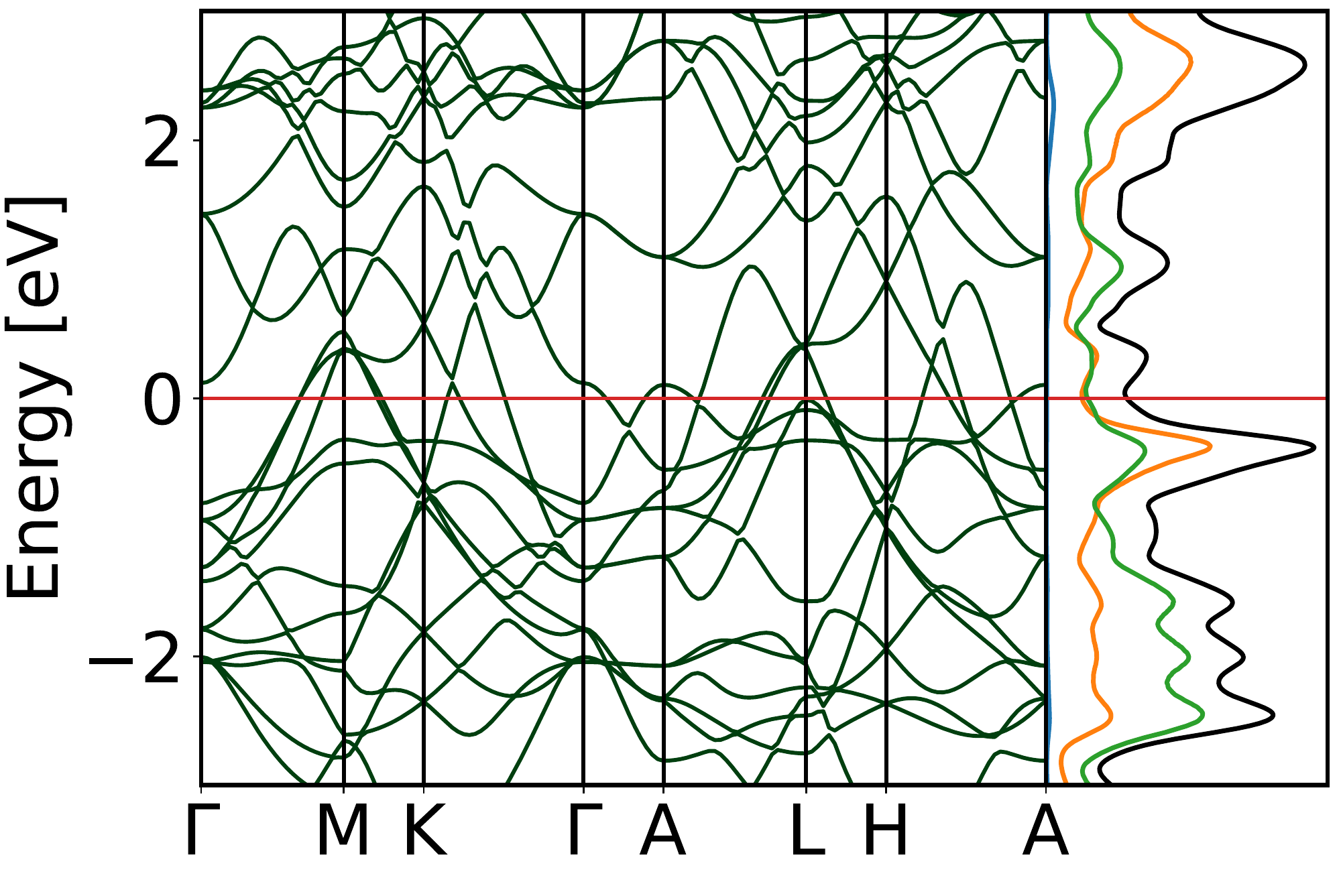}{6.11}{6.11}{9.63}    & \bandfigure{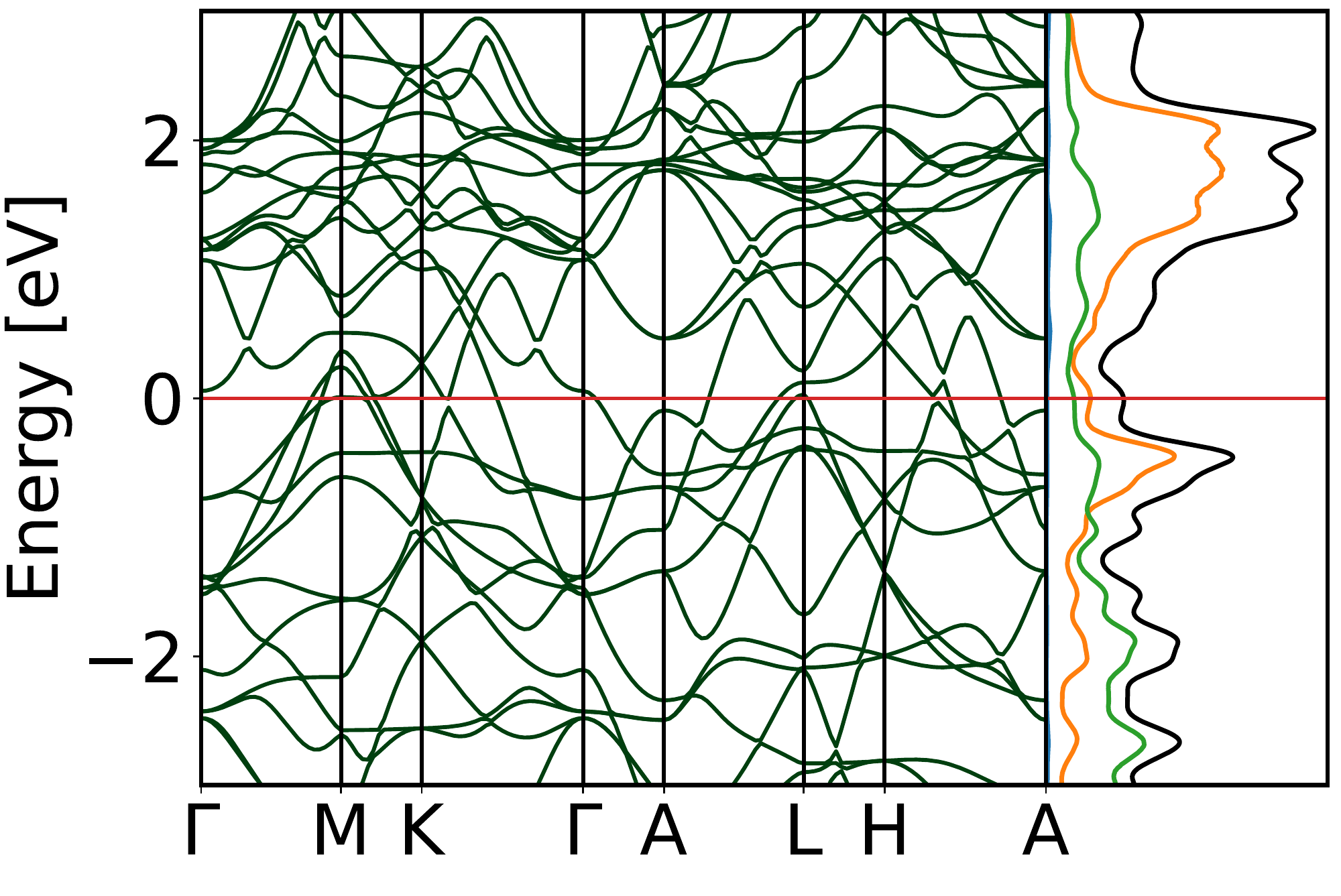}{5.85}{5.85}{8.89}    \\[5cm]

\end{tabular}
\end{figure*}

\begin{figure*}[ht]
\centering
\removetag
\caption{\large \textbf{\underline{Ce}}}
\begin{tabular}{cc}
  \bandfigure{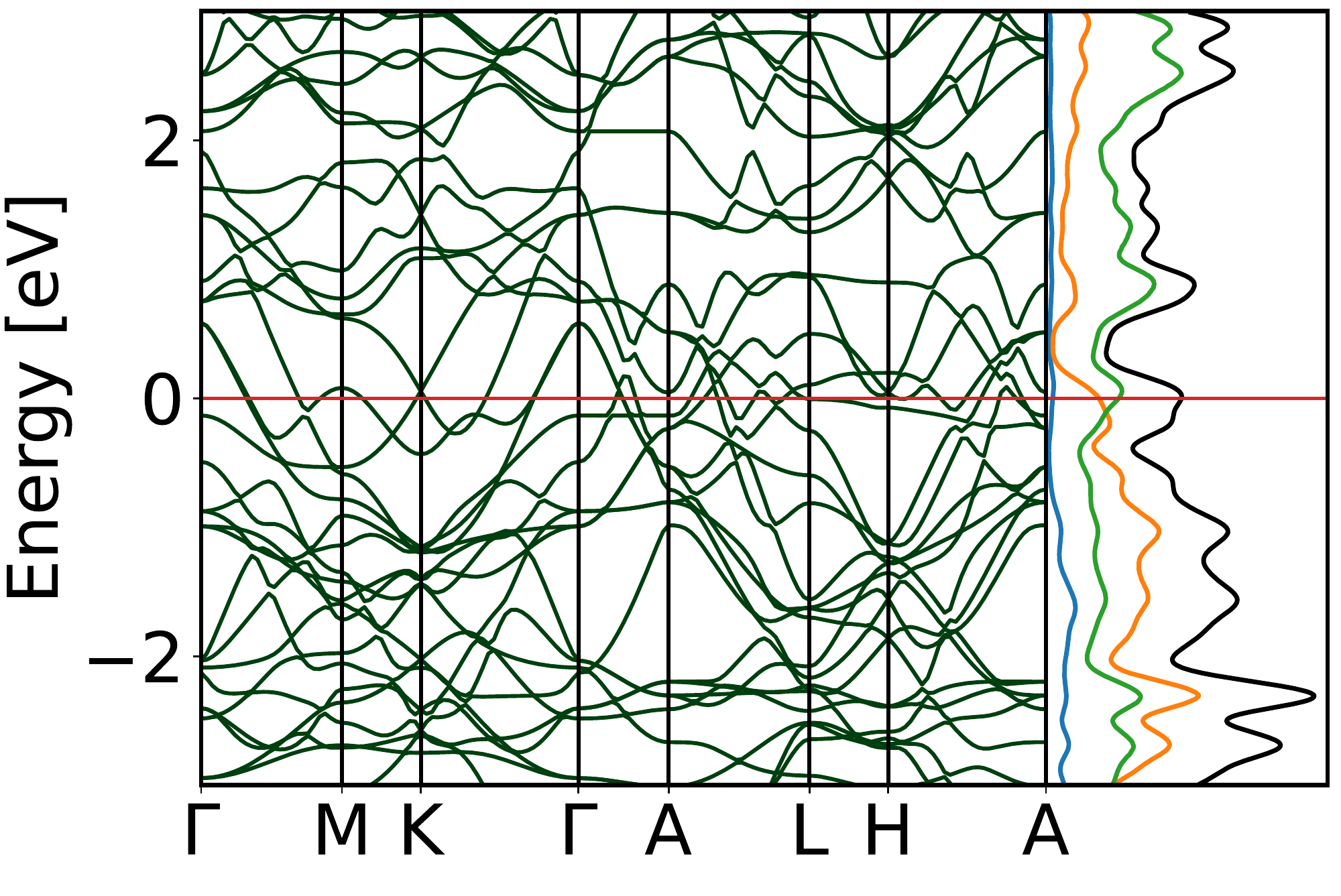}{5.87}{5.87}{7.41}    & \bandfigure{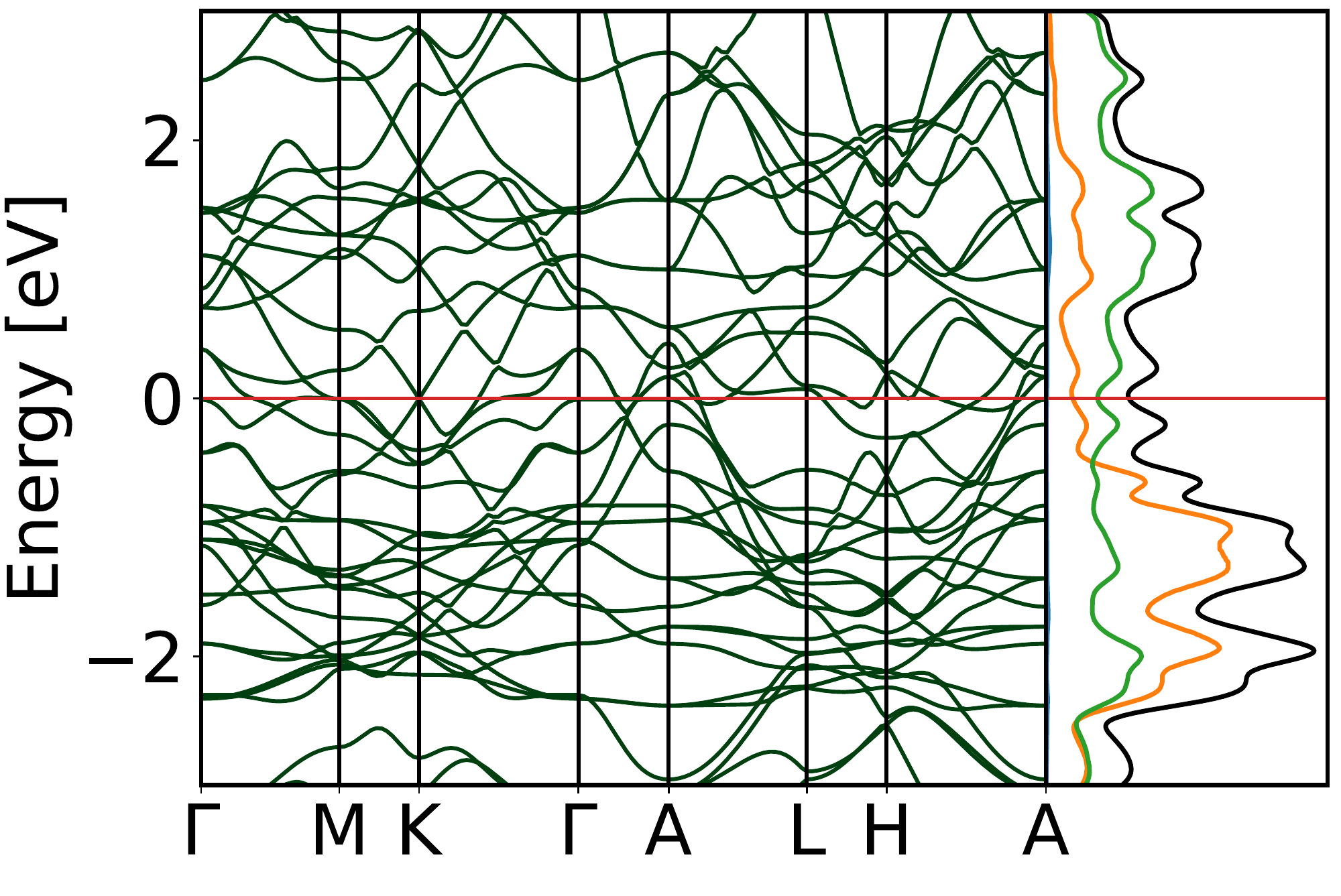}{5.46}{5.46}{7.34}    \\[5cm]
\end{tabular}
\end{figure*}

\begin{figure*}[ht]
\centering
\begin{tabular}{cc}
  \bandfigure{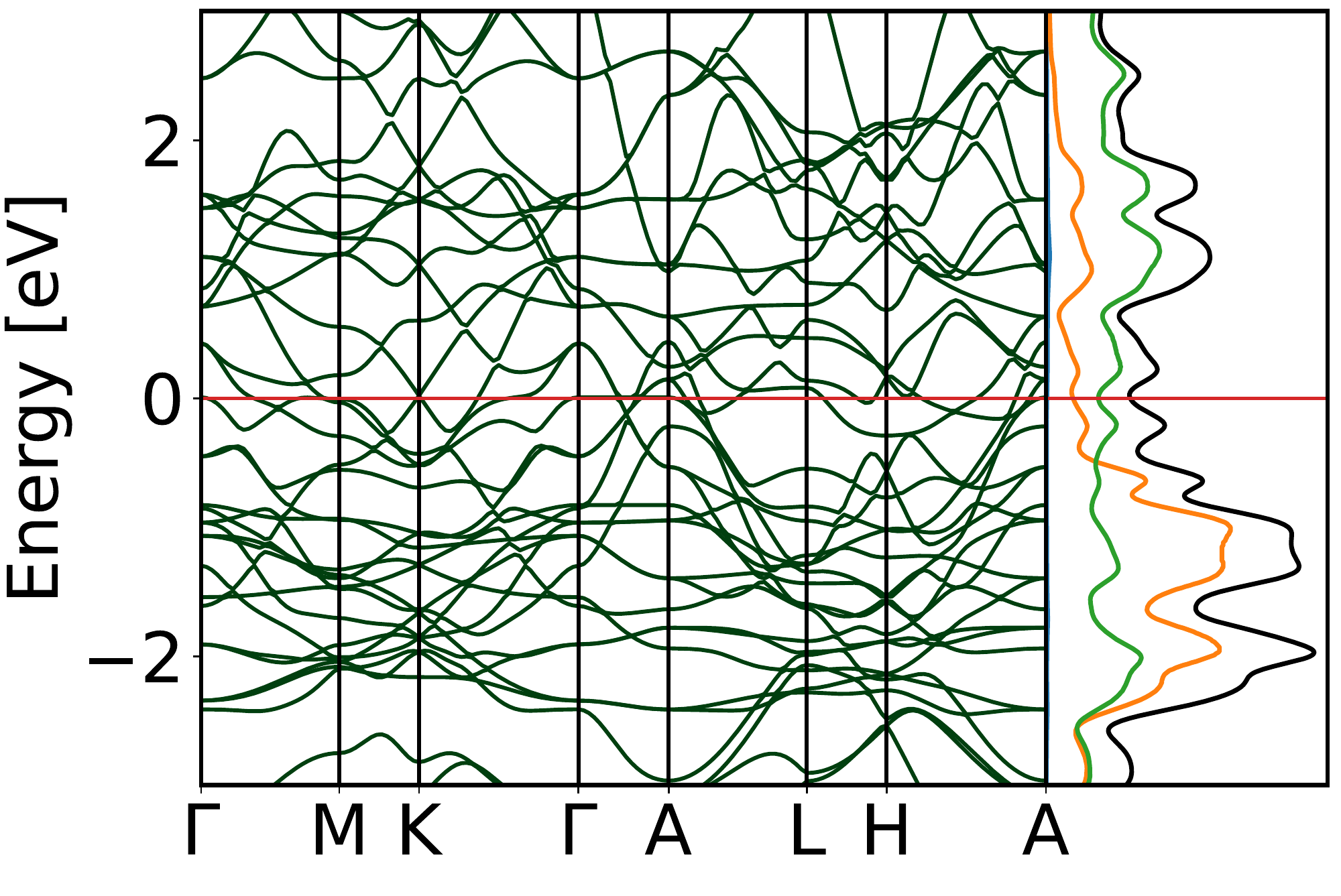}{5.45}{5.45}{7.34}    & \bandfigure{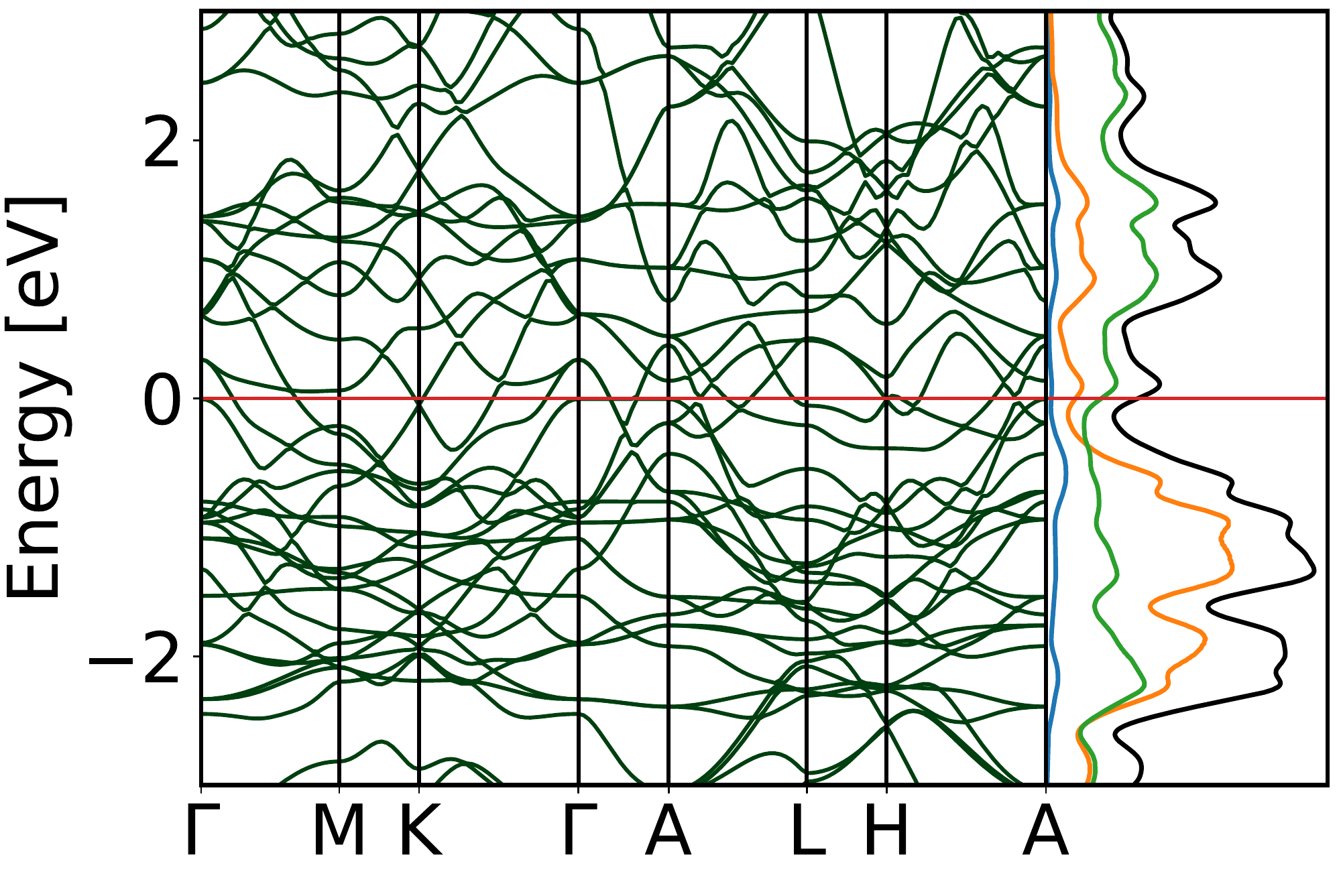}{5.49}{5.49}{7.32}    \\[5cm]
  \bandfigure{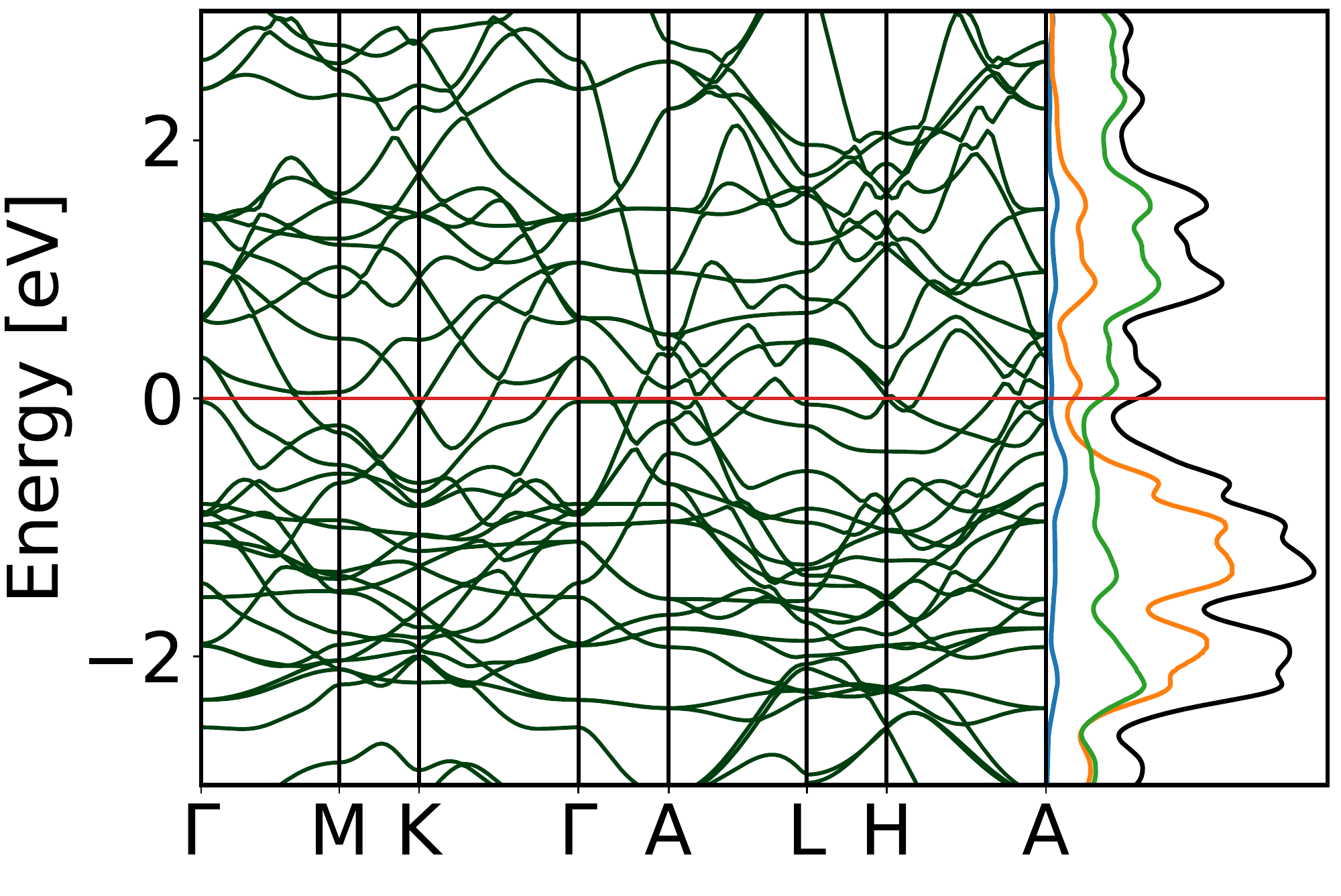}{5.51}{5.51}{7.32}    & \bandfigure{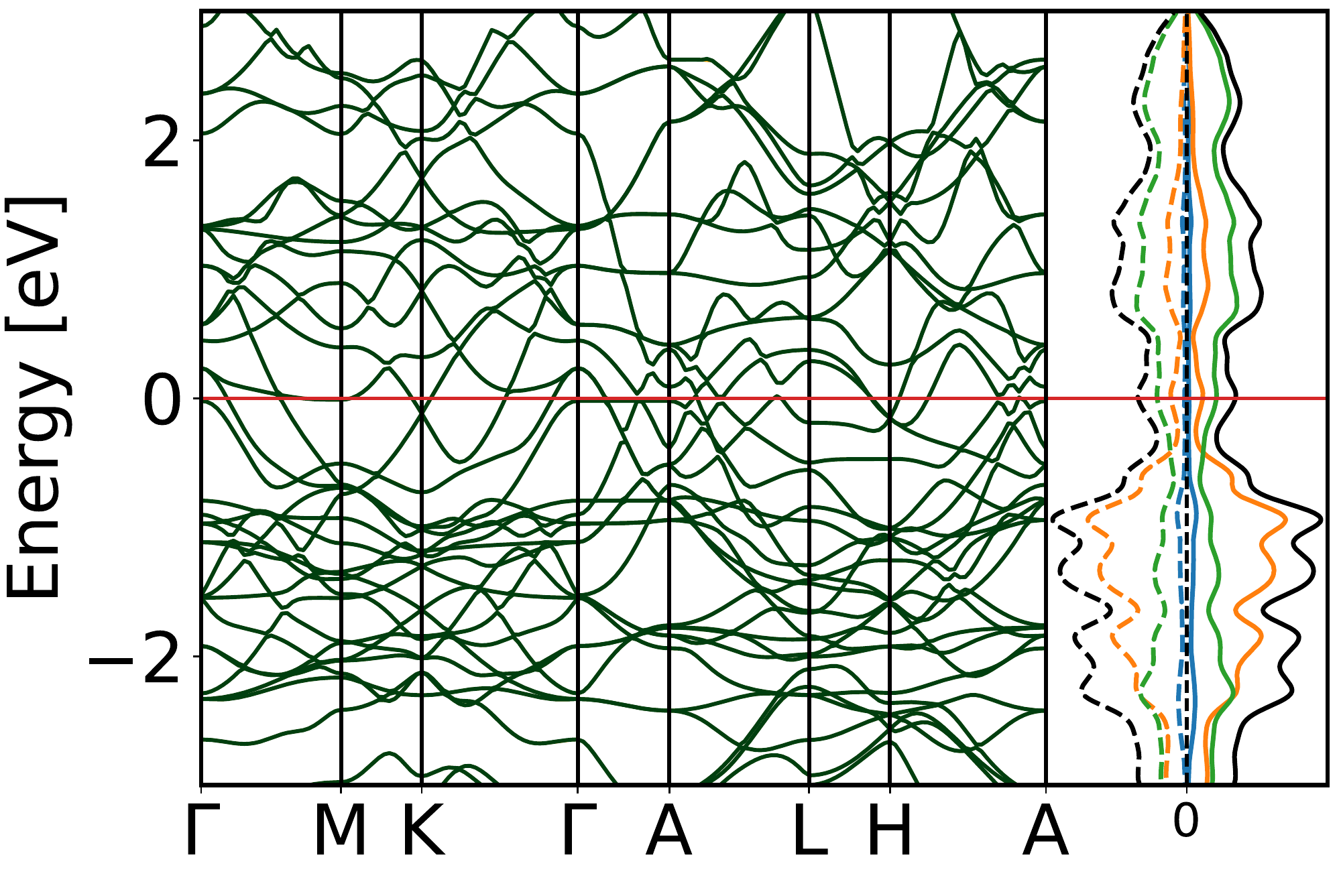}{5.56}{5.56}{7.28}    \\[5cm]
  \bandfigure{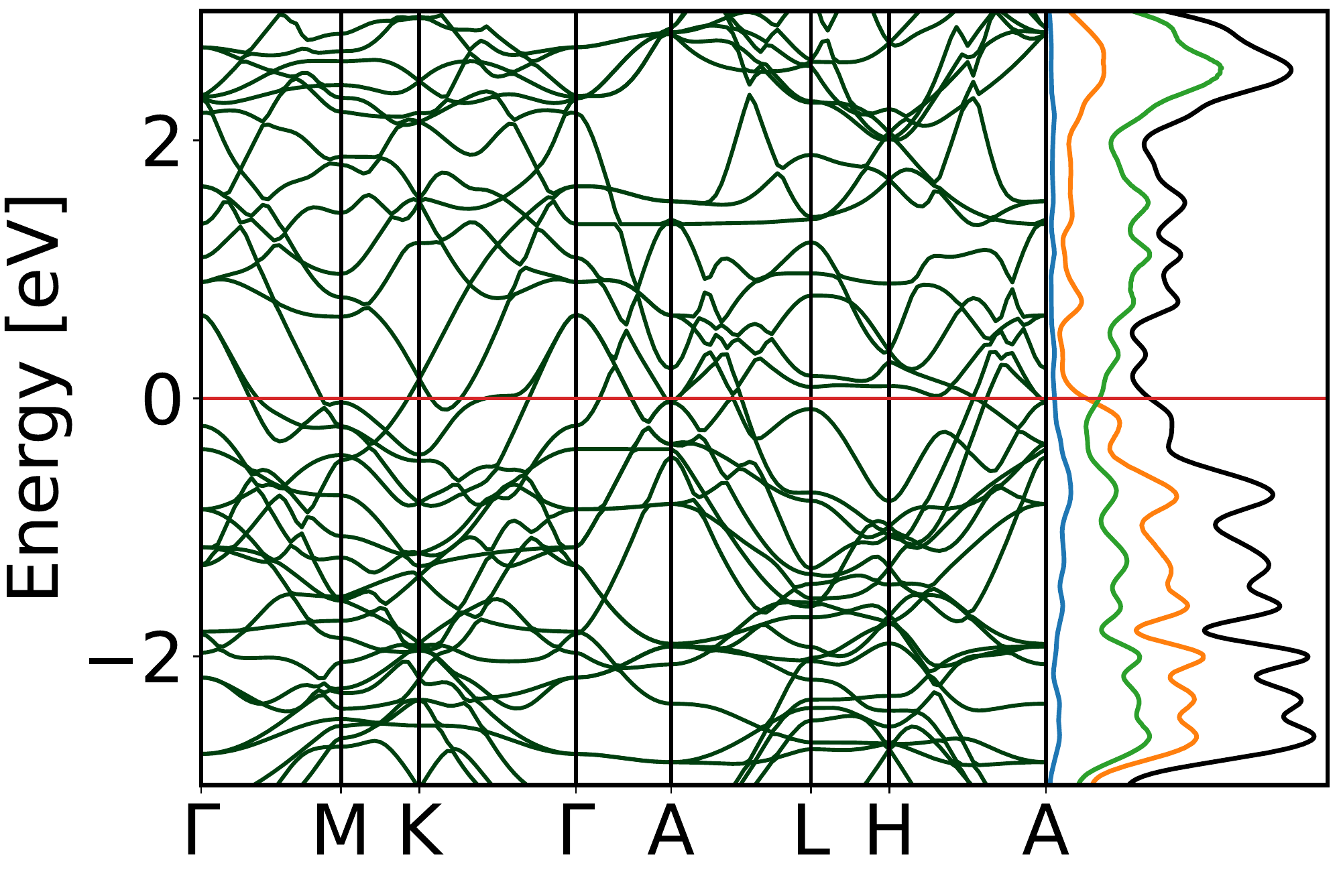}{5.84}{5.84}{7.34}    & \bandfigure{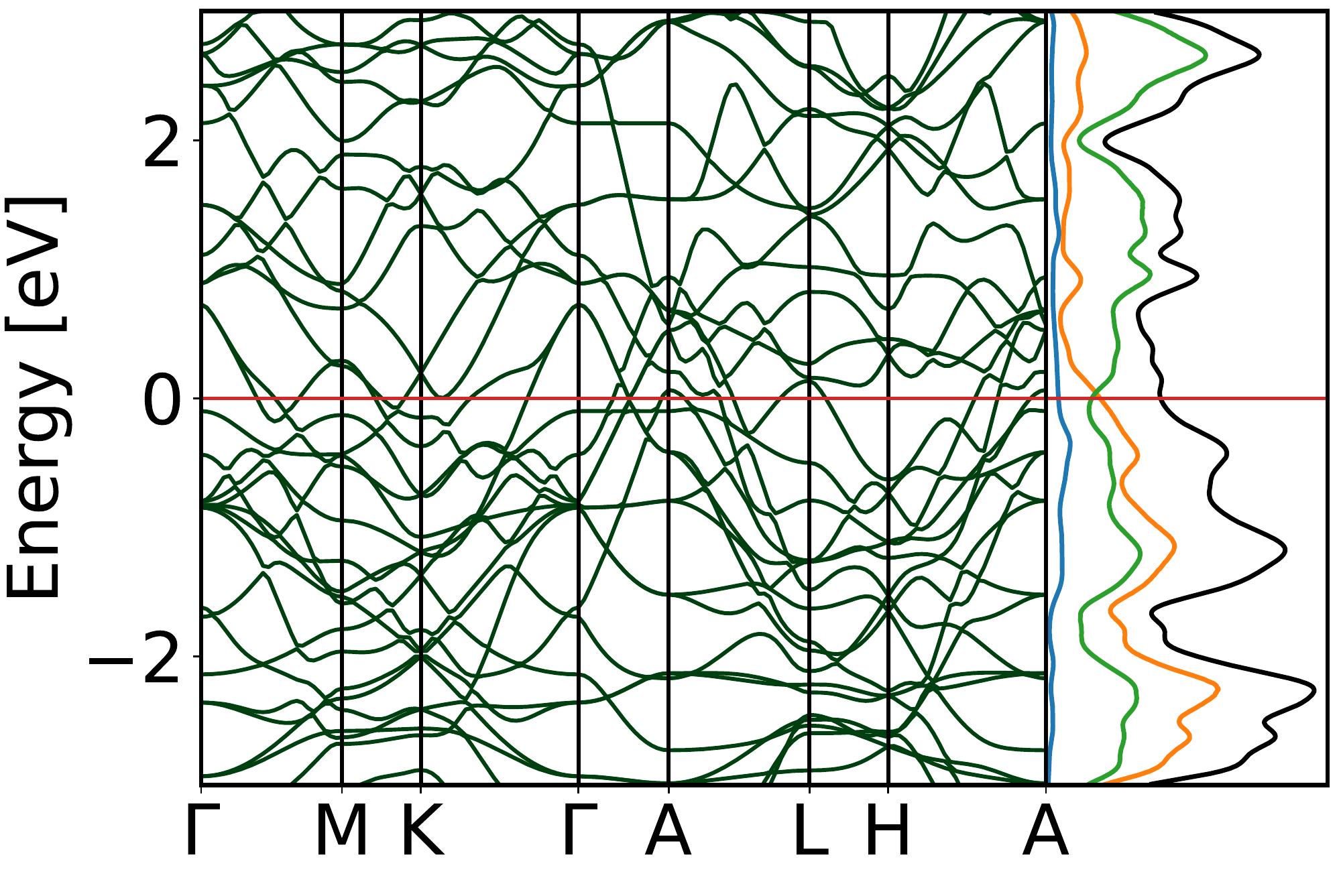}{5.81}{5.81}{7.45}    \\[5cm]
\end{tabular}
\end{figure*}

\begin{figure*}[ht]
\centering
\begin{tabular}{cc}
  \bandfigure{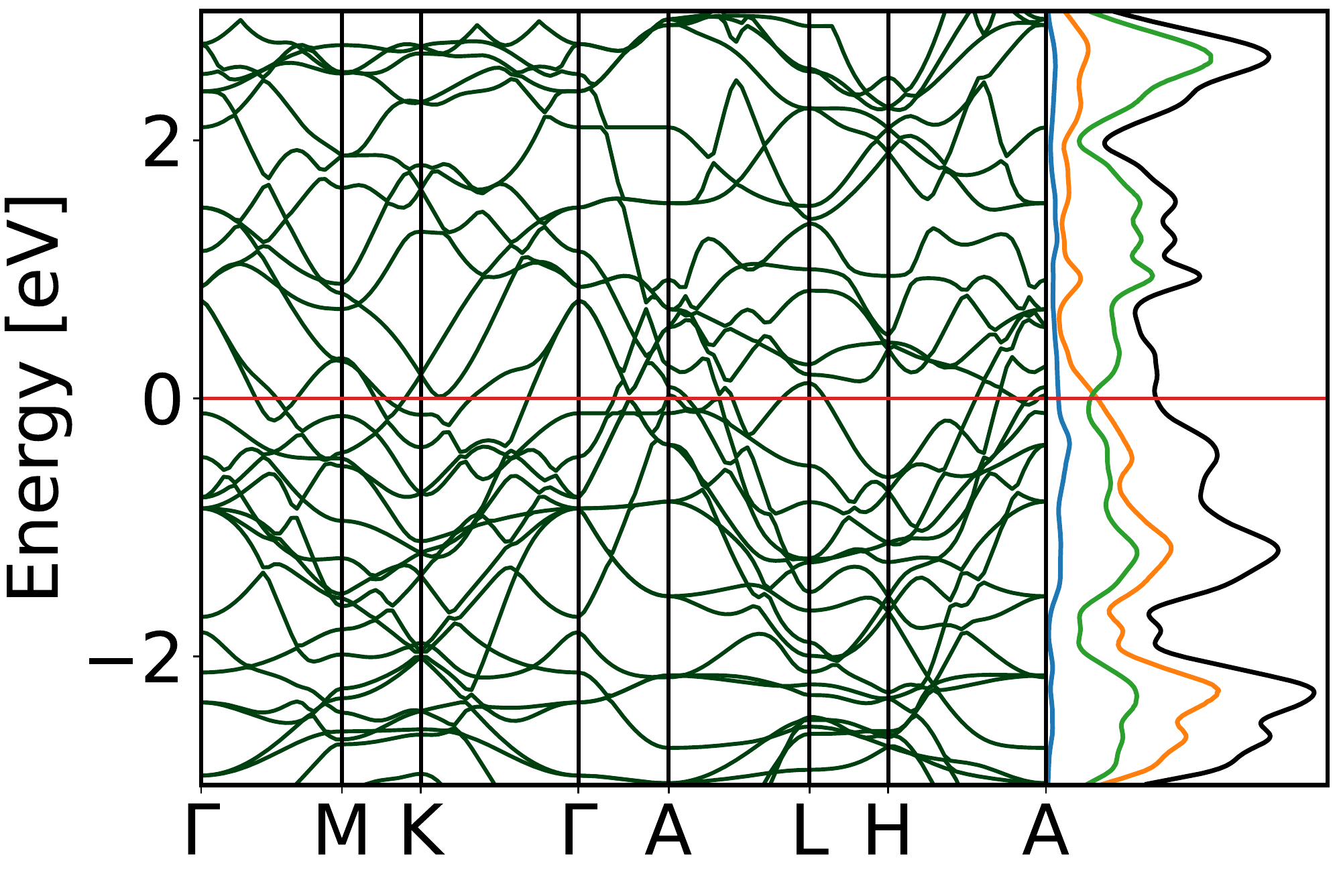}{5.82}{5.82}{7.44}    & \bandfigure{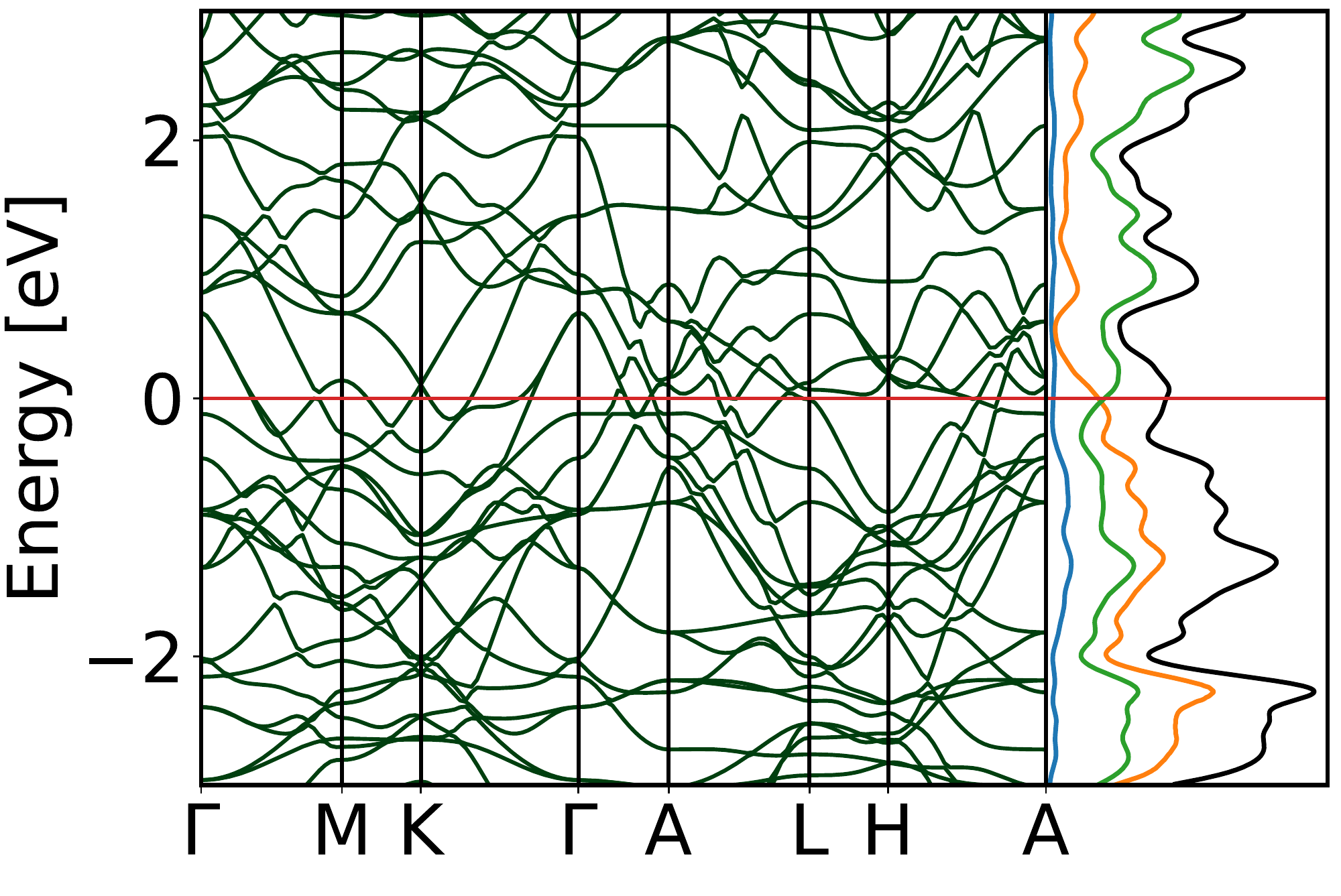}{5.84}{5.84}{7.44}    \\[5cm]
\end{tabular}
\end{figure*}

\begin{figure*}[ht]
\centering
\caption{\large Electron and phonon band structures of \ce{INb3Bi5}. The black curve represents the total density of states (DOS), while the projected atomic density of states is shown in blue (atom I), orange (atom Nb), and green (atom Bi).}

\begin{tabular}{c c }
    \includegraphics[height=.31\textwidth,trim=0.1cm 0.4cm 0 0cm, clip]{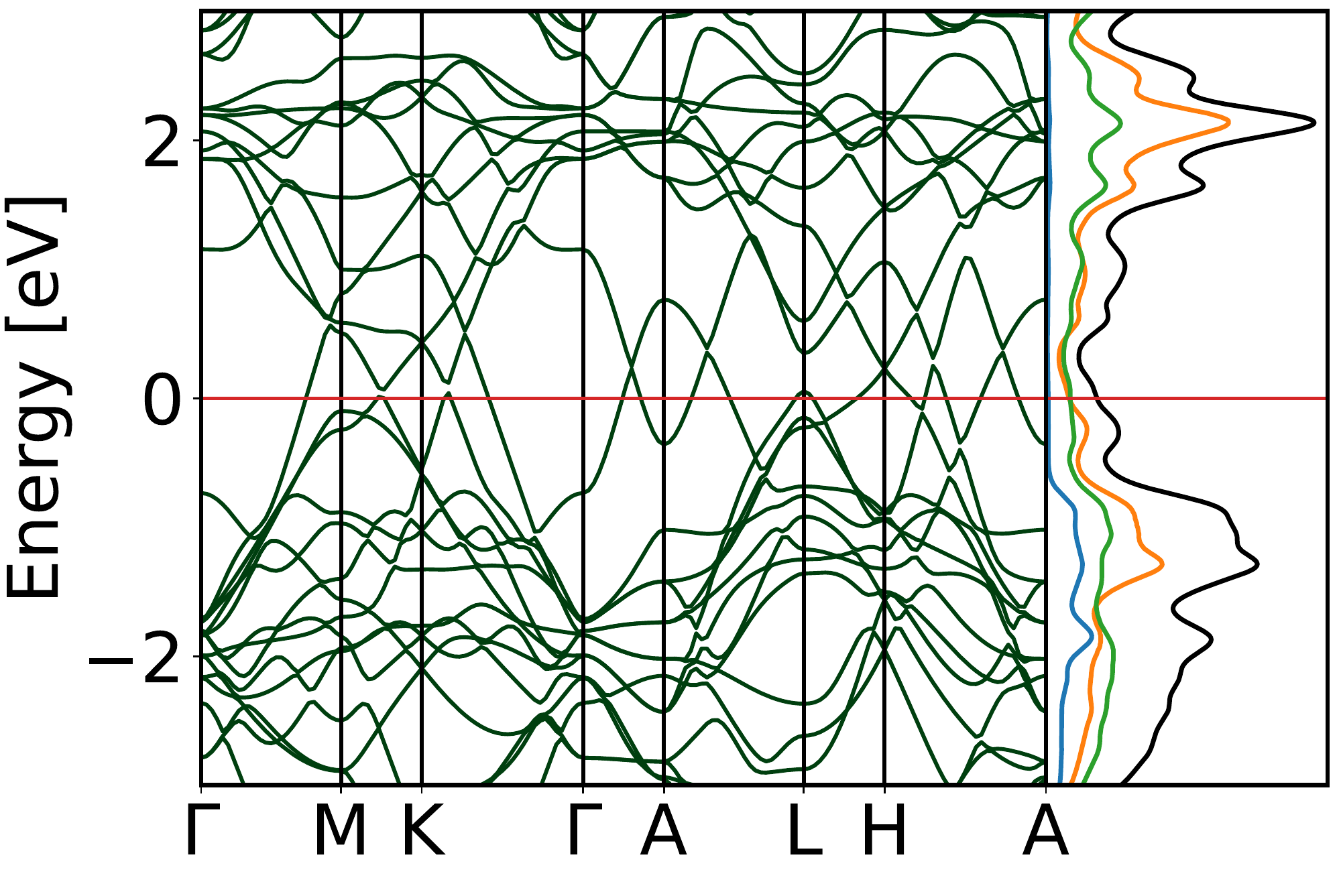} &
    \includegraphics[height=.31\textwidth,trim=1cm 0 6cm 1.8cm, clip] {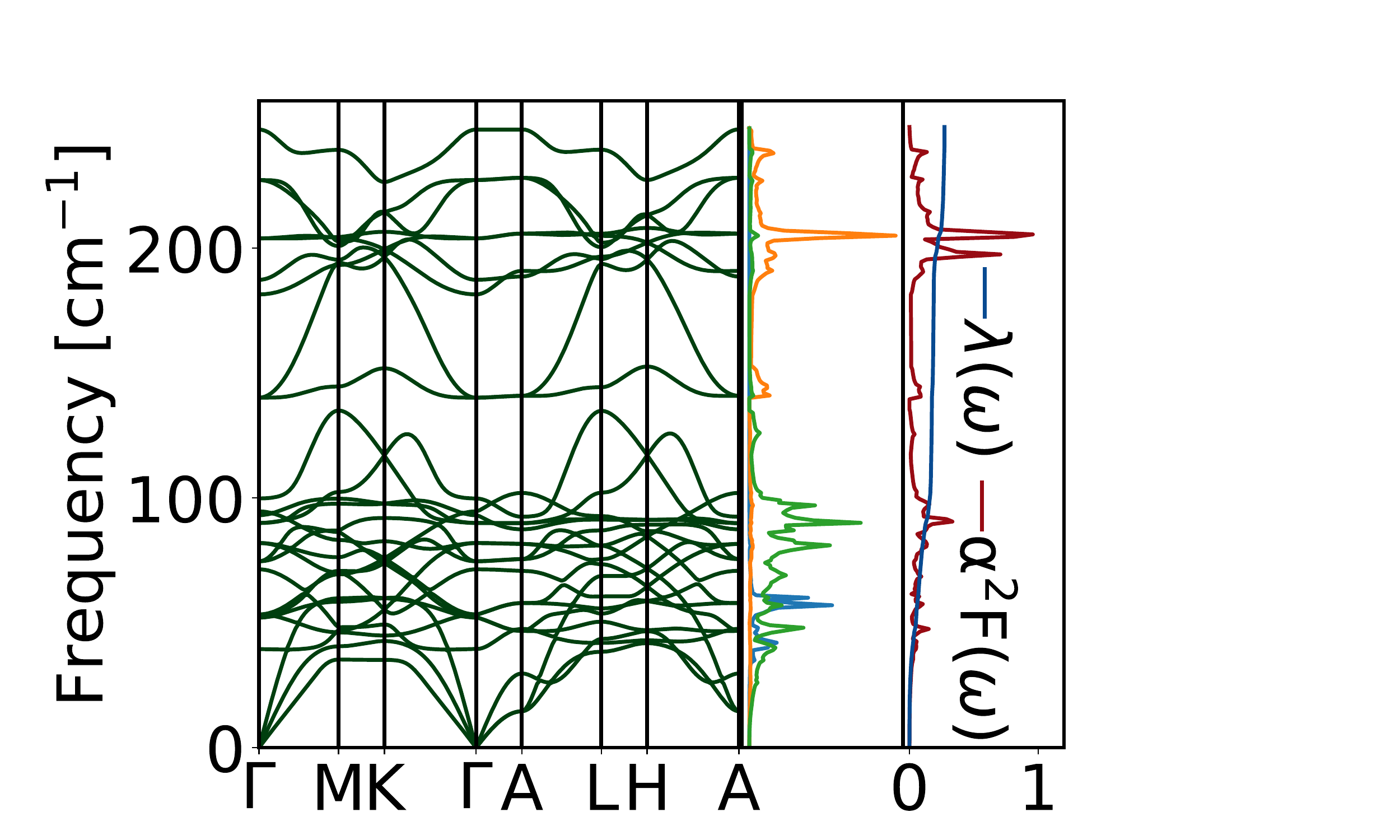} \\

    \color{white} \rule{5cm}{6cm} & {}
\end{tabular}
\label{fig:band_structures}
\end{figure*}

\clearpage
\bibliography{bib.bib, cal.bib}